# Correlated terahertz phonon-ion interactions control ion conduction in a solid electrolyte

Kim H. Pham[1*], Kiarash Gordiz[2*], Natan A. Spear[3*], Amy K. Lin[1], Jonathan M. Michelsen[1], Hanzhe Liu[1], Daniele Vivona[2], Geoffrey A. Blake[4], Yang Shao-Horn[2,5], Asegun Henry[2], Kimberly A. See[1], Scott K. Cushing[1¶]

[*]These authors contributed equally to this work, [¶]Corresponding author

Ionic conduction in solids that exceeds 1 mS/cm is predicted to involve coupled phonon-ion interactions in the crystal lattice. Here, we use theory and experiment to measure the possible contribution of coupled phonon-ion hopping modes which enhance $Li^+$ migration in $Li_{0.5}La_{0.5}TiO_3$ (LLTO). The *ab initio* calculations predict that the targeted excitation of individual $TiO_6$ rocking modes greatly increases the $Li^+$ jump rate as compared to the excitation of vibrational modes associated with heating. Experimentally, coherently driving $TiO_6$ rocking modes via terahertz (THz) illumination leads to a ten-fold decrease in the differential impedance compared to the excitation of acoustic and optical phonons. Additionally, we differentiate the ultrafast responses of LLTO due to ultrafast heating and THz-range vibrations using laser-driven spectroscopy (LUIS), finding a unique long-lived response for the THz-range excitation. These findings provide new insights into coupled ion migration mechanisms, indicating the important role of THz-range coupled phonon-ion hopping modes in enabling fast ion conduction at room temperature.

## Introduction

Solid-state ion conduction has applications in Li-ion batteries[1], biological membranes[2], solid oxide fuel cells[3], and more. Ongoing research in this field has led to the continual improvement of superionic conductors that can enable energy dense solid-state electrochemical systems while maintaining fast conductive properties akin to liquids[4–8]. Realizing an all-solid-state battery with a superionic solid-state electrolyte and a Li metal anode can meet the necessary energy and power requirements for developing long-range electric vehicles, and improve safety[9].

The fundamental mechanism of ion conduction ($\sigma_{ion}$) relies on the Arrhenius form equation as shown in **Eq. 1**, where $\sigma_0$ is the Arrhenius pre-factor, $T$ is temperature, $E_a$ is the activation energy, and $k_B$ is the Boltzmann constant[10].

$$\sigma_{ion} = \frac{\sigma_0}{T} * \exp\left(-\frac{E_a}{k_B T}\right) \qquad (1)$$

Researchers have sought out different strategies to increase $\sigma_{ion}$ by lowering the $E_a$ of the ion migration pathway by targeting crystal structures with corner sharing frameworks[6] and connected migration channels[9,11], employing highly polarizable polyanions in the sub-lattice[12,13], introducing disorder to cation sites[6,8], and leveraging polyanion-motion-mediated ion transport[14–16]. These principles have led to the discovery of several superionic conductors through aliovalent/isovalent substitution strategies[17] and a general search for ideal structural families[8,13,14,16]. $Li_{10}GeP_2S_{12}$[4,6,7], $Li_7P_3S_{11}$[6,9,14,15], Li-argyrodites[6,8,12], and Li-oxyhalides[18–20] have very high conductivity values, but the exact mechanisms supporting superionic conductivity remain debated[20,21]. In particular, the coupling of phonon modes with ion hopping has been suggested to play a significant role in the ionic conduction mechanism[13,22–26].

To understand phonon-ion coupling, we can expand the pre-factor $\sigma_0$ in **Eq. 1.** As shown in **Eq. 2**, $\sigma_{ion}$ depends on the entropy of migration ($\Delta S_m$) and enthalpy of migration ($\Delta H_m$)[13].

$$\sigma_{ion} \propto \exp\left(\frac{\Delta S_m}{k_B}\right) \exp\left(-\frac{\Delta H_m}{k_B T}\right) * \frac{1}{T} \qquad (2)$$

The value of $\Delta S_m$ is determined by collective phonon-ion interactions[27] which are predicted to correlate with and enhance ionic conductivity[13–17,22–25,27–29]. The value of $\Delta H_m$ is experimentally measured as $E_a$ (determined by **Eq. 1).**

Several experimental efforts have attempted to explore the phonon-ion couplings predicted to relate to $\Delta S_m$. Nuclear Magnetic Resonance (NMR) techniques have been used to probe phonon-ion coupling in Li-argyrodites[22,30] and other conductive Li-sulfides, -nitrides, and -oxides[31]. Neutron scattering techniques have been applied to Na-sulfides and -selenides[16,23,32,33], as well as Na-orthophosphates and -sulphates[16,34], specifically by characterizing the reorientation of anions as quasi-elastic structure factors[31,34].

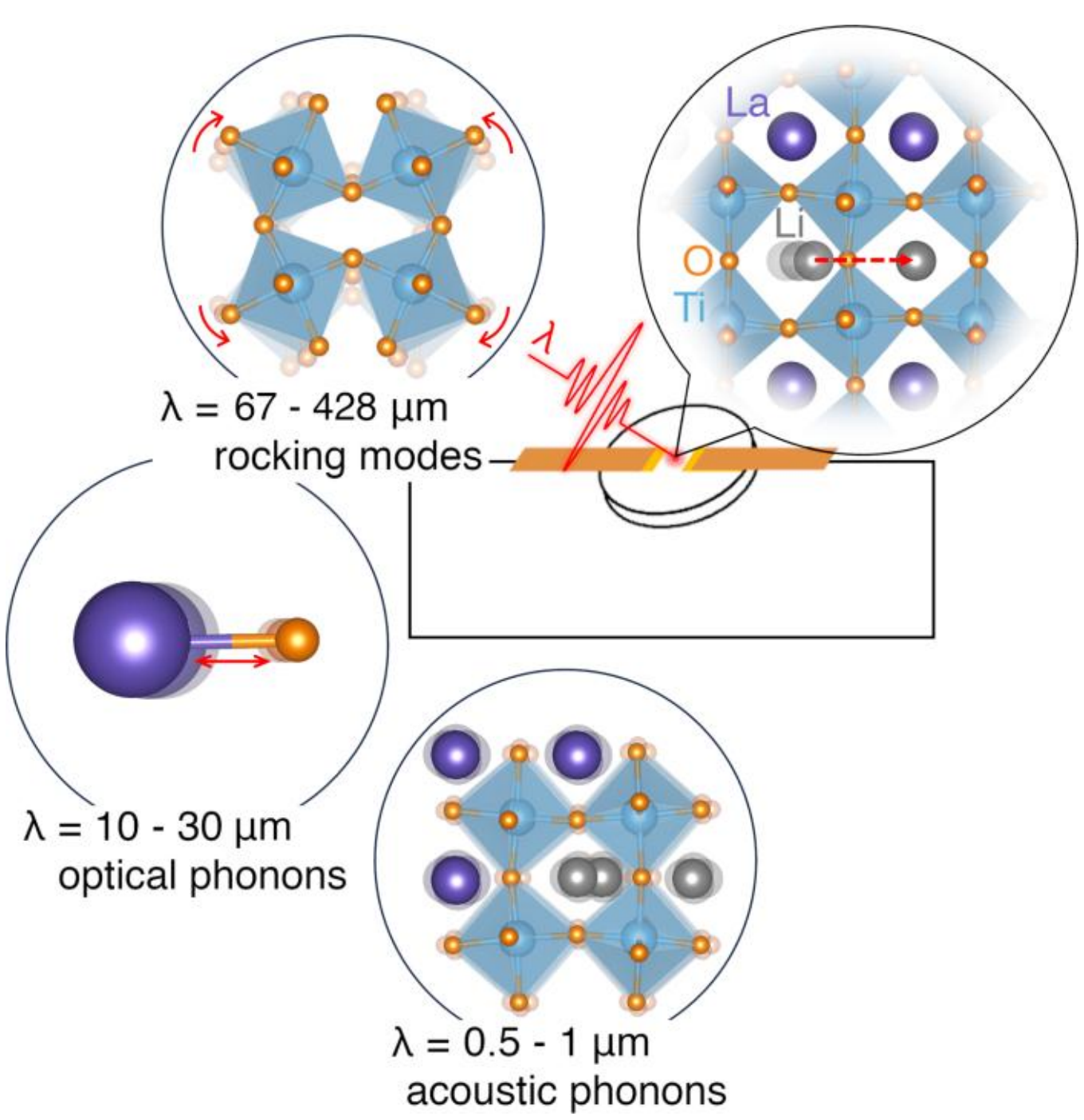

**Fig. 1 | Schematic of the laser-driven impedance method with LLTO.** An excitation source ($\lambda$) excites the sample between two in-plane Au electrodes while EIS measurements are collected. Depending on the wavelength of the excitation source, coupled phonon-ion rocking modes, optical phonons, and acoustic phonons can be excited to influence the $Li^+$ migration in LLTO.

Another approach to measuring how phonon-ion coupling enhances $Li^+$ hopping is to drive relevant modes with light and measure the resulting change in conductivity. The terahertz (THz) region, in particular, can selectively target polyanion modes theorized to mediate ion transport[29,35–38]. Specifically, modes that involve $(PS_4^{3-})$[14,23,26], $(SiS_4^{4-})$[14], $(PO_4^{3-})$[38], $(YBr_6^{3-})$[39], and $(ErCl_6^{3-})$[39] polyanion vibrations or rotations occur with vibrational frequencies at or below 10 THz[29,38]—similar to the picosecond timescales of ion hopping[16]. The same modes are theorized to be responsible for coupled hopping in LISICON-like conductors ($Li_{10}SiP_2S_{12}$[40], $Li_{10}GeP_2S_{12}$[4,29], $Li_3PO_4$[25,38], $Li_7P_3S_{11}$[14,29]), Li-argyrodites ($Li_6PS_5Cl$, $Li_6PS_5Br$)[12,22], and superionic Li-halides ($Li_3YBr_6$, $Li_3ErCl_6$)[29,39]. Similarly, octahedral rotations are also reported to promote fast ionic transport in several perovskite-related phases[24,41,42], making phonon-driven ion migration a promising approach for unveiling ion-coupled mechanisms that may lead to superionic conductivity.

While driving acoustic and optical modes to change electronic and magnetic properties is common in condensed matter physics[28,43,44], very few studies have studied the effects of directly driving phonon-ion couplings to promote ion transport[35,45]. In one report, Poletayev *et al.* employed a THz pump to induce a $Na^+$, $K^+$, and $Ag^+$ ion hop in $\beta$-$Al_2O_3$ and then measured the change in birefringence using an 800 nm probe. The THz Kerr Effect measured the anisotropy caused by the THz excitation, which they calculated to lead to a 20x increase in hop anisotropy compared to thermal conditions[45]. In addition, Gordiz *et al*. used Nudged-Elastic Band (NEB) calculations and Molecular Dynamic (MD) simulations to predict that over 87% of the lattice energy required for the ion hop in Ge-substituted $Li_3PO_4$ came from <10% of the modes in the system[38]. By computationally exciting highly contributing modes, they enhanced the ionic conductivity by orders of magnitude without increasing the bulk temperature.

In this paper, we use *ab initio* calculations, MD simulations, and experiments that combine laser-driving and electrochemical impedance spectroscopy (EIS) to determine the role of octahedral rocking $TiO_6$ phonon modes which couple with the hopping of $Li^+$ in LLTO. The bottleneck of $Li^+$ migration in LLTO (**Fig. S1**) is comprised of 4 oxygen atoms formed by 4 corner sharing $TiO_6$ octahedra[41,42] that are reported to "rock" in place at low THz frequencies. These "rocking modes" are depicted in **Fig. 1**. To determine how well the $TiO_6$ modes couple with the hopping of $Li^+$ ions, we first excite these modes with THz irradiation from 0.7 to 4.5 THz, or 67 - 428 μm (general schematic shown in **Fig. 1**). We compare the driving of targeted phonon modes predicted to couple with $Li^+$ conduction to laser-induced heating, which is performed via the indirect excitation of acoustic phonon modes with NIR light (800 nm or 0.8 μm). Acoustic phonons can mediate ion conduction via increased random thermal vibrations[46]. Lastly, we excite relevant optical phonon modes within the MIR light range. The La-O mode (11.6 μm) is predicted to weaken if $Li^+$ interactions with the $TiO_6$ octahedra are strong[42], so we test whether driving the La-O vibrations with MIR light (9.4 - 14.16 µm) can weaken $Li^+$ - $TiO_6$ - coupling and modulate $Li^+$ transport.

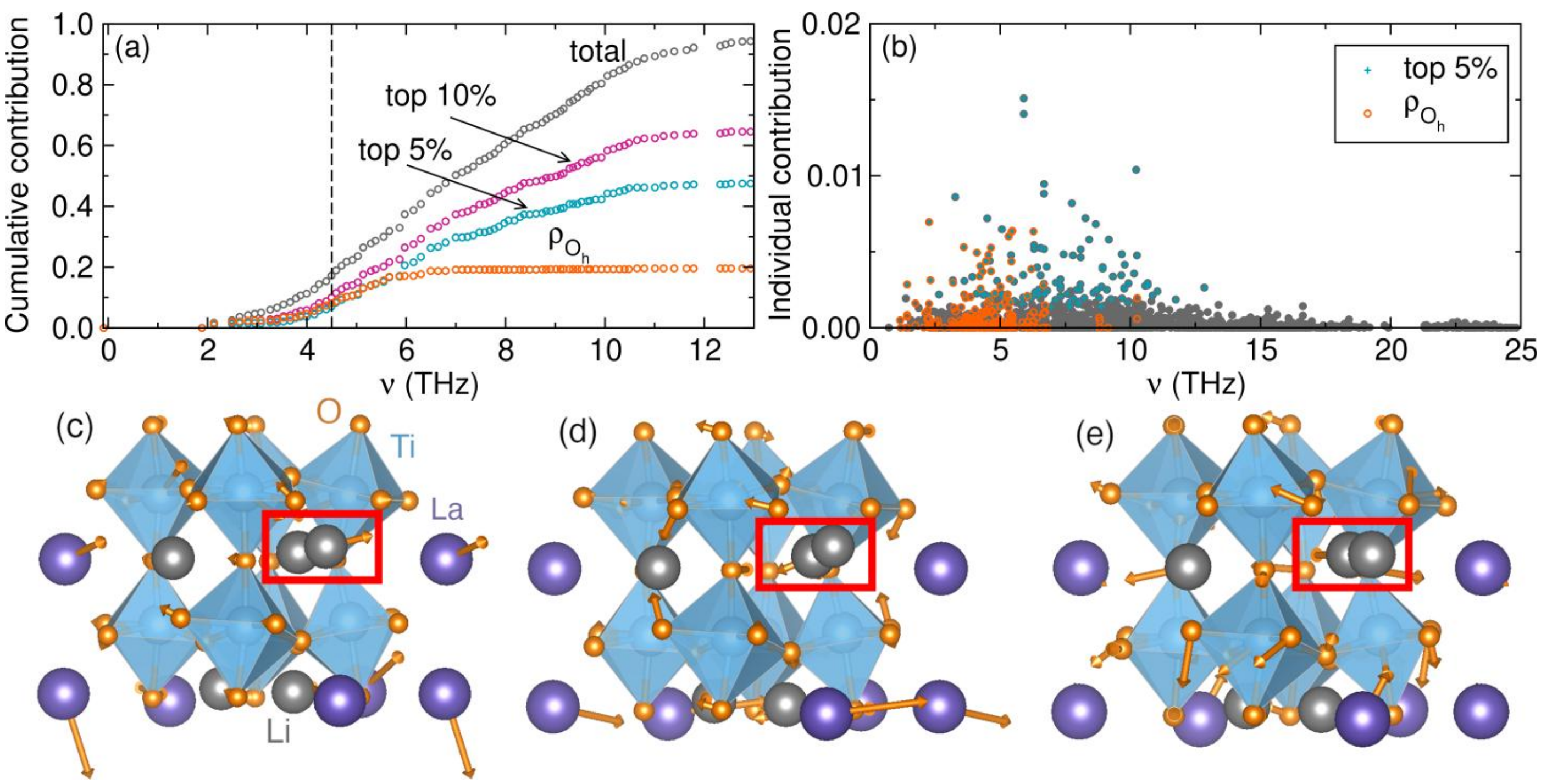

**Fig. 2 | Calculated phonon contributions to $Li^+$ hopping and representative LLTO phonon modes. a,** the cumulative contributions of phonon modes to the normalized energy required for ion hopping. The black dashed line marks the experimental limit for THz generation (4.5 THz). The cumulative effects of the highest 5% and 10% contributing modes, as well as the octahedral rocking modes are specifically shown. **b**, the individualized contributions of phonon modes to the normalized energy required for ion hopping where $\rho_{O_h}$ represents octahedral rocking modes. The data is normalized over all 22 $Li^+$ hop investigations. **c-e,** the eigenvectors of three vibrational modes show how the lattice displacements project along with the $Li^+$ hop, under **c**, a thermal mode (2.66 THz), **d**, the highest contributing non-rocking mode (4.29 THz), and **e**, the highest contributing rocking mode (3.46 THz). The red rectangle highlights the hopping $Li^+$ which is being investigated. The enhancement provided by each of these modes is outlined in **Table 1** and the hop to which these modes contribute is depicted in **Fig S5.1**.

We simultaneously use EIS to probe the change in impedance due to NIR – THz excitation, correlating these excitations with changes in the ion transport behavior. We quantify the role of the differing phonon couplings that mediate ion conduction by measuring shifts in impedance at each excitation wavelength and normalizing the measured enhancement by the power density of the incident laser beam. Additionally, we utilize LUIS[47,48] to differentiate the ultrafast responses of the NIR and THz excitations. We separately quantify the impacts of laser-induced heating on the sample, finding that long-lasting thermalization effects do not account for the differences in the measured enhancement, indicating that the targeted photoinduced lattice dynamics predominantly affect ion migration. We note that the aim of this study is not to significantly reduce the sample's resistance using the laser. Instead, we use a differential measurement to compare the impact of phonon excitations with that of laser-induced heating, providing an initial proof-of-concept demonstration that vibrational modes play a significant role in enabling fast ion conduction.

In qualitative agreement with the *ab initio* calculations, our measurements indicate that selectively exciting the rocking modes in the THz range which enable phonon-ion coupled $Li^+$ hopping leads to an order of magnitude decrease in the bulk resistance, or $R_{bulk}$, as compared to exciting acoustic or optical modes. Additionally, our LUIS measurements demonstrate a unique long-lived response time in the case of the THz excitation. Reminiscent of other metastable phases induced by light in inorganic-organic hybrid perovskites[49–52], the enhancement that we measured is persistent and reversible, even though the recorded changes are averaged over the 250 Hz repetition rate for the THz source and 1 kHz for the generated NIR-MIR light.

**Table 1** | MD-simulated enhancement in the ion hopping rate of one of the 22 investigated hops by targeted excitation of selected modes which have frequencies in the experimentally accessible frequency range. The visualization of this hop, which is a single hop in the disordered LLTO structure, is provided in **Fig. S5.1**. Details on the calculation are provided in the **Methods**.

| | | 400 K (natural MD simulation) | 700 K (natural MD simulation) | 400 K (with mode excited to 700 K) |
|---|---|---|---|---|
| Jump rate (#jumps/ps) | Thermal mode (2.66 THz) | 0.0011 | 1.45 | 0.0027 |
| | Highest contributing non-rocking-mode (4.29 THz) | | | 1.27 |
| | Highest contributing rocking-mode (3.46 THz) | | | 1.39 |

Our results provide initial evidence that laser driving phonon-ion coupled hopping modes can facilitate fast ion transport in a solid, suggesting that structures which have these modes in the THz range may make for fast ion conductors with ionic conductivities greater than 1 mS/cm.

## Results

Using *ab initio* calculations, the contributions of phonon-ion coupled hopping modes to the energy needed by $Li^+$ to reach its transition state (herein referred to as ion hopping) in LLTO were calculated based on the magnitude of the projection of the atomic displacement field obtained for the transition state on the eigenvectors of vibration for each specific phonon, as explained in detail in the **Methods**. The structural parameters for the calculations were obtained from quantitative Rietveld refinements of experimental synchrotron diffraction data (**Fig. S2)**. To efficiently sample different possible $Li^+$ hops and relevant contributing vibrational modes, 22 unique $Li^+$ hops based on two different hopping mechanisms (single and concerted) across three degrees of La - Li disorder in the LLTO crystal were chosen and analyzed (**Fig. S3**).

**Fig. 2a** shows the normalized cumulative contribution of different phonon modes to the $Li^+$ hopping process, here defined as the energy needed for a $Li^+$ to reach its transition state. **Fig. 2b** shows the individual breakdown of the total phonon mode contributions that sum to **Fig. 2a**. Within the experimentally excited region of <4.5 THz, marked by the dashed line in **Fig. 2a**, there are numerous phonon modes which couple strongly with the hopping of $Li^+$ ions. These modes are further separated based on their projection on different atomic species in **Fig. S4.** The modes below 4.5 THz contribute to 17.5% of all sampled hops, corroborating that low frequency phonon modes strongly couple with ion transport as studied in other theoretical reports[15,29].

The significance of the phonon contribution is further supported by the increase in ion hopping rate determined using MD simulations (see **Methods** and **Table S1**). The effects of targeted excitation of a thermal mode, the highest contributing non-rocking mode, and the highest contributing rocking mode are depicted in **Table 1**, showing that a three order of magnitude increase in the jump rate is found through targeted excitation of a phonon-ion coupled rocking mode relative to exciting a random phonon mode. The modes in **Table 1** are further examined in **Fig. 2c-e** where the random mode at 2.66 THz does not appear to align with the hopping path of the $Li^+$ ion (**Fig. 2c**), which is highlighted by the red box, while the highly contributing modes, both rocking and non-rocking, appear to have crystal lattice vibrations aligned with the $Li^+$ jump path (**Fig. 2d-e**), again highlighted in red. These figures show how increased alignment between the hopping $Li^+$ and the excited phonons lead to an enhanced ion hopping rate and increased phonon-ion coupling. The overall hopping path that these modes couple with is depicted in **Fig S5.1**. Similar hopping pathways are further examined in **Table S1** and **Figs. S5.2 and S5.3** indicating similar levels of enhancement due to the targeted excitations of coupled phonon-ion hopping modes which have a high degree of alignment between the phonon and the hopping ion. Due to the low frequency nature of modes below 4.5 THz, the inclusion of quantum effects changes their contribution by less than 1% (see **Methods and Fig. S6**), making the calculations relevant to the room temperature experiments.

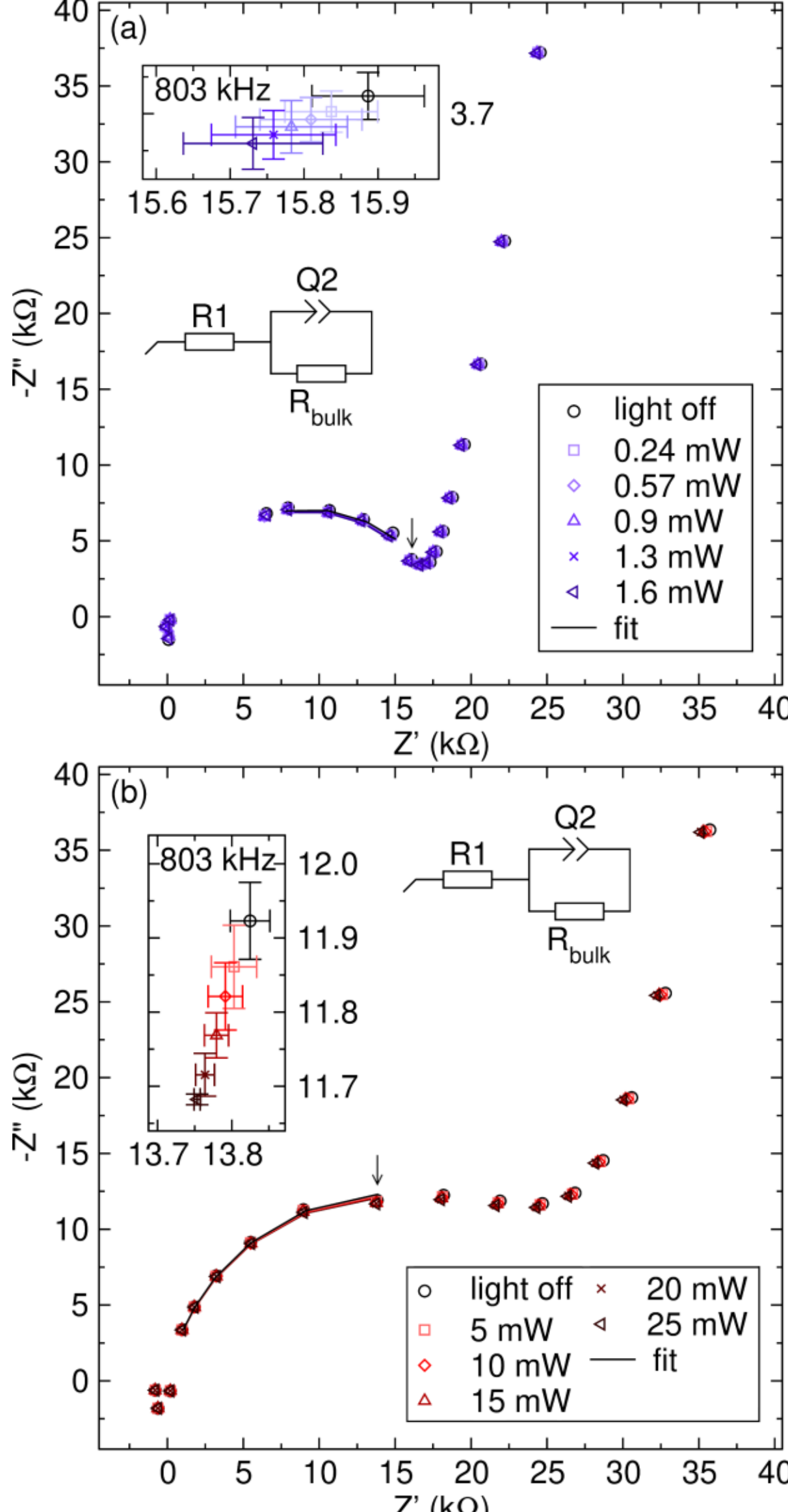

**Fig. 3 | Nyquist plots of LLTO under varied light intensities.** The zoomed-in Nyquist plots of the bulk ion migration feature measured over a total frequency range of 32 MHz to 1 Hz with a sinus amplitude of 50-100 mV to minimize electronic noise for **a**, THz light excitation powers varied between 0 and 1.6 mW and **b,** 800 nm light excitation varied between 0 and 25 mW. Because the absolute changes in impedance are difficult to discern, insets are shown for data points measured at 803 kHz, also marked by an arrow on the corresponding Nyquist plot. The average value of each data point under each light intensity condition across three samples and three trials each is plotted.

To experimentally test the role of phonons in coupling with $Li^+$ conduction in LLTO, we first synthesize the material in accordance with literature[53] and structurally characterize it using synchrotron X-ray diffraction, as described under **Methods**.

We then irradiate the LLTO using THz, NIR, and MIR light using an open electrochemical cell described in the **Methods** and described previously in literature[48]. We adopt an in-plane electrode geometry, as depicted in **Fig. 1**. This sample geometry allows for the irradiation of the sample but has irregular (i.e. nonplanar) surfaces with no accurately defined thickness, meaning the ionic conductivity of the laser-excited measurements cannot be calculated, requiring the use of $R_{bulk}$ to study the photoinduced changes in ion migration. Accordingly, we quantify the laser-induced enhancement on ion conduction through the decrease in $R_{bulk}$, which we determine by fitting the Nyquist plots (obtained with EIS) to a R1+$R_{bulk}$/Q2 circuit, shown in both plots of **Fig. 3**. The high frequency data points, which represent the bulk feature, between 32 MHz and 803 kHz, are partially affected by electronic induction which arises due to the wiring of the experimental setup not physical phenomena that we aim to probe in our samples. Consequently, we include only approximately 5 points in our fit. We are still confident, however, that we are probing the bulk region as EIS measurements on LLTO have long-reported low-impedance bulk and high-impedance grain features[54], which we have replicated in **Fig. S7**.

We first drive the THz rocking modes predicted to couple with $Li^+$ hopping (**see Figs. S8 and S9**) with 0.24-1.6 mW of THz power focused to a 250 μm beam diameter with a 250 Hz repetition rate (**Methods and Fig. S10**). The bandwidth and field strength of the THz radiation are determined by the generation process, described in more detail in the **Methods**. When the 0.7 – 4.5 THz modes are driven, a linear decrease in the Z' on the order of 100s of Ωs is measured, scaling with increased power (**Table S2**). We next measure the enhancement induced by heating using NIR light. We excite the sample with 5-25 mW of 800 nm pulsed light focused into a 250 μm beam diameter at a 1 kHz repetition rate to populate acoustic phonon baths associated with random thermal vibrations, or heat[33] (see **Methods**). The Nyquist plots for LLTO under 800 nm laser excitation are shown in **Fig. 3b**, with the replicable enhancements tabulated in **Table S3**. We additionally rule out any enhancement from structural changes in the material during and after excitation. We perform a damage test, as outlined in the **Methods**, and find that that the Raman spectra acquired at the laser spot before and after excitation are identical (see **Fig. S11**).

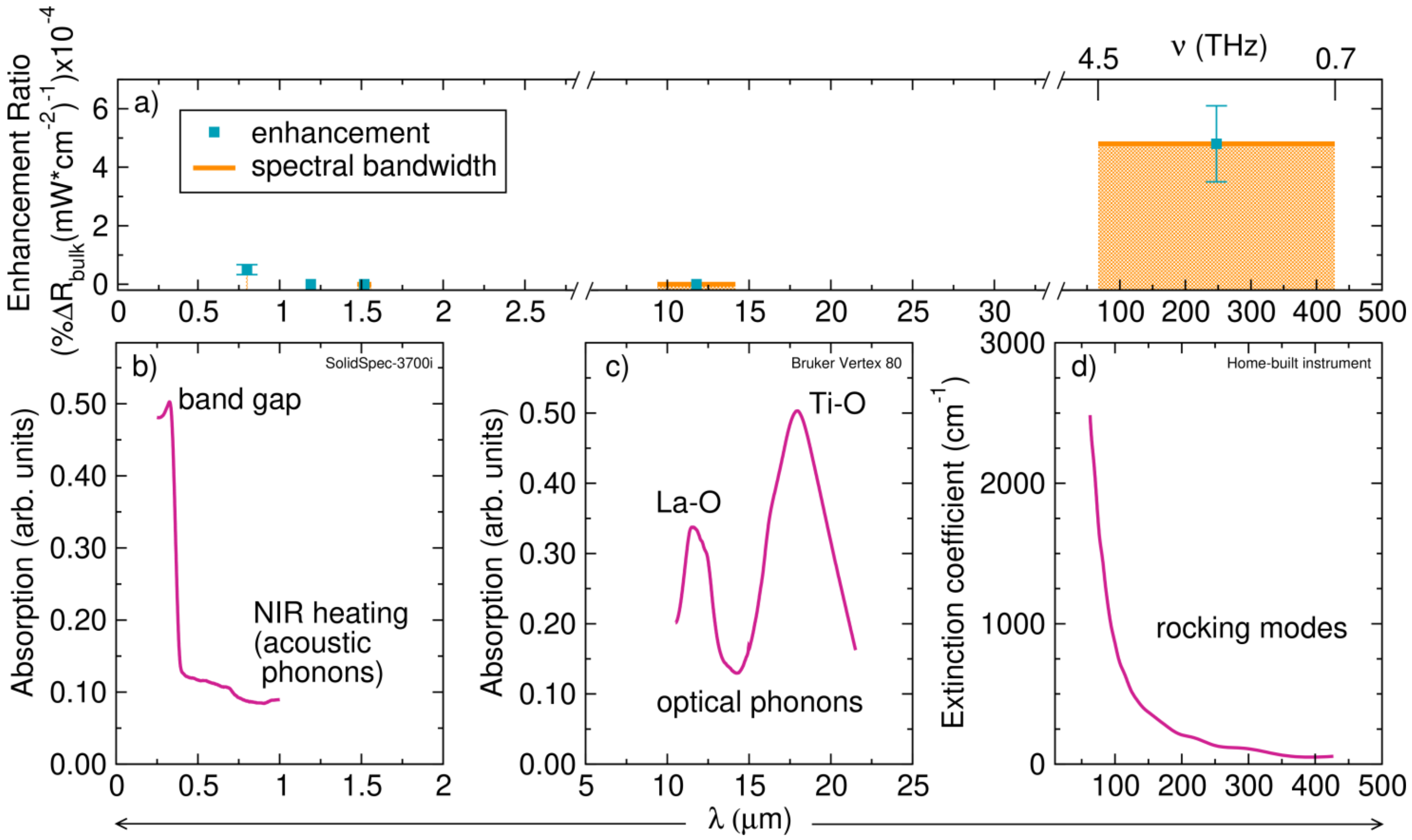

**Fig. 4 | Enhancement in bulk ion migration due to illumination across near-IR (NIR), mid-IR (MIR), and THz light. a,** Enhancement ratios (cyan squares) reported in $\%\Delta R_{bulk}(mW * cm^{-2})^{-1}$x $10^{-4}$, calculated from exciting LLTO between NIR – THz light. The width of the horizontal orange bar represents the spectral bandwidth of the excitation pulse, with the frequency range probed in the THz regime also noted. The NIR enhancement corresponds to incoherent heating of the acoustic phonon bath. The MIR excitation corresponds to coherent driving of optical phonon modes, which was inconclusive for the same excitation density as the THz modes. The THz light coherently drives highly contributing phonon-ion coupled hopping modes, showing the largest relative enhancement. **b-d,** Absorption spectra of LLTO (magenta) in the NIR, MIR, and THz regions obtained using UV-vis spectroscopy, FTIR, and THz transmission, respectively. The NIR absorption plot was measured on a Shimadzu SolidSpec-37000iDUV spectrophotometer, the MIR was measured on the Bruker Vertex 80, and the THz absorption was measured on a custom THz time-domain spectroscopy set up which is presented in more detail in the **Methods**. The use of different spectrometers necessitates the differing units on each plot.

Using the fitted Nyquist plots shown in **Fig. 3**, we compare the percent change in $R_{bulk}$ across different powers and excitation wavelengths (**Fig. S12)**. We find that incoherently heating the acoustic phonon bath with the 800 nm light induces a 0.10% decrease in $R_{bulk}$ per mW of power. Comparatively, a 0.78% decrease in $R_{bulk}$ per mW is measured when exciting phonon-ion coupled hopping modes with THz light, in line with the predictions from *ab initio* computational modeling which predict a significant enhancement associated with the excitement of single phonon modes. **Fig. S12** not only demonstrates the greater enhancement induced by the THz-range coupled phonon-ion modes, but also rules out any nonlinear effects, given the linear correlation seen between the excitation enhancement and power for both wavelengths.

In **Fig. 4**, we normalize the enhancement by the average power density of the beam used to excite the sample. Normalization by average power density is utilized because it allows for comparison across different excitation wavelengths which have varied spot sizes and repetition rates, as mentioned previously and outlined in the **Methods**. With this normalization, we find a $4.8*10^{-4}\%$ change in $\Delta R_{\mathrm{bulk}}(mW * cm^{-2})^{-1}$ for the broadband THz excitation of phonon-ion coupled hopping modes, while laser-induced heating gives a calculated $5*10^{-5}\%$ change in $\Delta R_{\mathrm{bulk}}(mW * cm^{-2})^{-1}$. We also attempt to excite optical phonon modes with lower energy MIR 1100 and 1500 nm light, finding no detectable change (**Fig. 4a**). The results are consistent with our theoretical calculations where the optical modes contribute significantly less to ion hopping above 10 THz (~30 μm) shown in **Fig. 2b,** which possibly justifies why no detectable changes are observed.

The small magnitude of the normalized enhancements in **Fig. 4a** satisfies our aim of utilizing the differential nature of the measurement to compare the impact of different excitations without damaging the sample (**Fig. S11**) and while avoiding any nonlinear effects (**Fig. S12**). Still, we acknowledge that the absorption depths of the 800 nm and THz excitations (7 and 6 μm, respectively) are roughly 1% of the total sample thickness and thus may diminish the measured enhancement. We chose this thickness due to the brittle nature of LLTO causing thinner pellets of this size to crack during measurements. Still, we aim to lessen this spatial mismatch by using an in-plane geometry, which is expected to primarily apply the EIS probe near the sample surface between the sputtered electrodes, where the laser most strongly interacts with the sample (**Fig. S13**). In **Fig. S14**, one can find a graphical depiction of the laser-driven EIS experiment. The entire sample holder is depicted in detail in previous work[47].

We next spectroscopically explore what types of modes explain the differential enhancement measured in **Fig. 4a** using UV-VIS, FTIR, THz transmission, and Raman measurements. In **Fig. 4b**, one can see there is low absorption in the NIR used for incoherent thermal laser heating. This indicates why higher powers of the incident laser are needed in measurements (**Fig. S12**). In **Fig. 4c**, one can see the vibrations of Ti-O and La-O vibrational modes[42]. Evidently, these modes do not affect $Li^+$ hopping within the experimental error bars. Finally, in **Fig. 4d**, we see high absorption coefficients, on the order of 1000 $cm^{-1}$ for THz light excitation. Given that there are many modes in this range, as indicated by

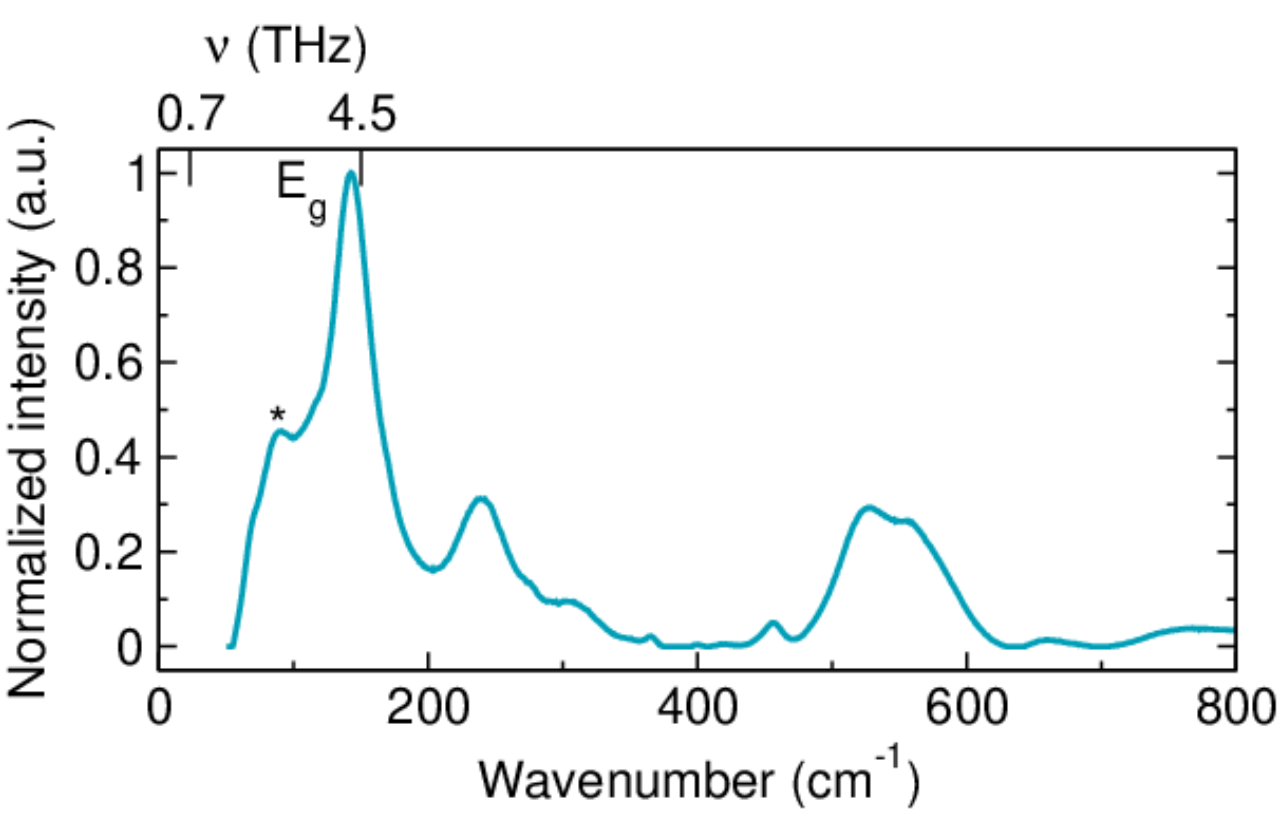

**Fig. 5 | Raman spectrum of LLTO.** The Raman spectrum of LLTO, demonstrating available phonon modes which can be accessed in the excitation range, from 0.7-4.5 THz, as highlighted in the figure.

**Fig. 2a-b**, we cannot discern individual modes via THz absorption measurements. However, in **Fig. 5**, we identify Raman-active modes that exist within the THz excitation energy range. The first of these is a shoulder peak <100 $cm^{-1}$, marked with an asterisk, which is hypothesized to be related to octahedral tilts and small oxygen shifts[55], very similar to the key modes which we believe couple strongly with $Li^+$ conduction. Additionally, the $E_g$ symmetry mode is associated with Ti vibrations, which affect the $TiO_6$ polyhedra[42], similar to the computationally identified $TiO_6$ rocking modes. While not all Raman-active phonon modes can be directly excited by THz irradiation, it has been proposed that THz excitation of IR-active modes can decay into available Raman modes via coupled pathways, which we hypothesize to be relevant in this study[56]. We further discuss the spectroscopic evidence for highly contributing modes in LLTO, particularly the THz absorption measurements in **Fig. 4d**, at the beginning of the Discussion.

Finally, we utilize LUIS to differentiate the response of LLTO when exciting THz-range phonon-ion coupled hopping modes, as compared to 800 nm NIR light, which is meant to represent the effects of laser-induced heating. In LUIS, as described in previous work[47,48], a single frequency AC field with similar amplitude to the laser-driven EIS experiments, is used to initiate ion hopping on site-to-site timescales of tens of picoseconds. Then, a laser excitation is used to induce a change in the crystal lattice, in this case specifical vibrational modes with THz light and laser-induced heating with 800 nm light. The corresponding impedance response is then measured with a high-speed oscilloscope. This technique is described in further detail in the SI, and is depicted in **Fig. S15**. We provide a summary comparing the laser excitation and AC probing field parameters for the laser-driven EIS and LUIS measurements in **Table S4**.

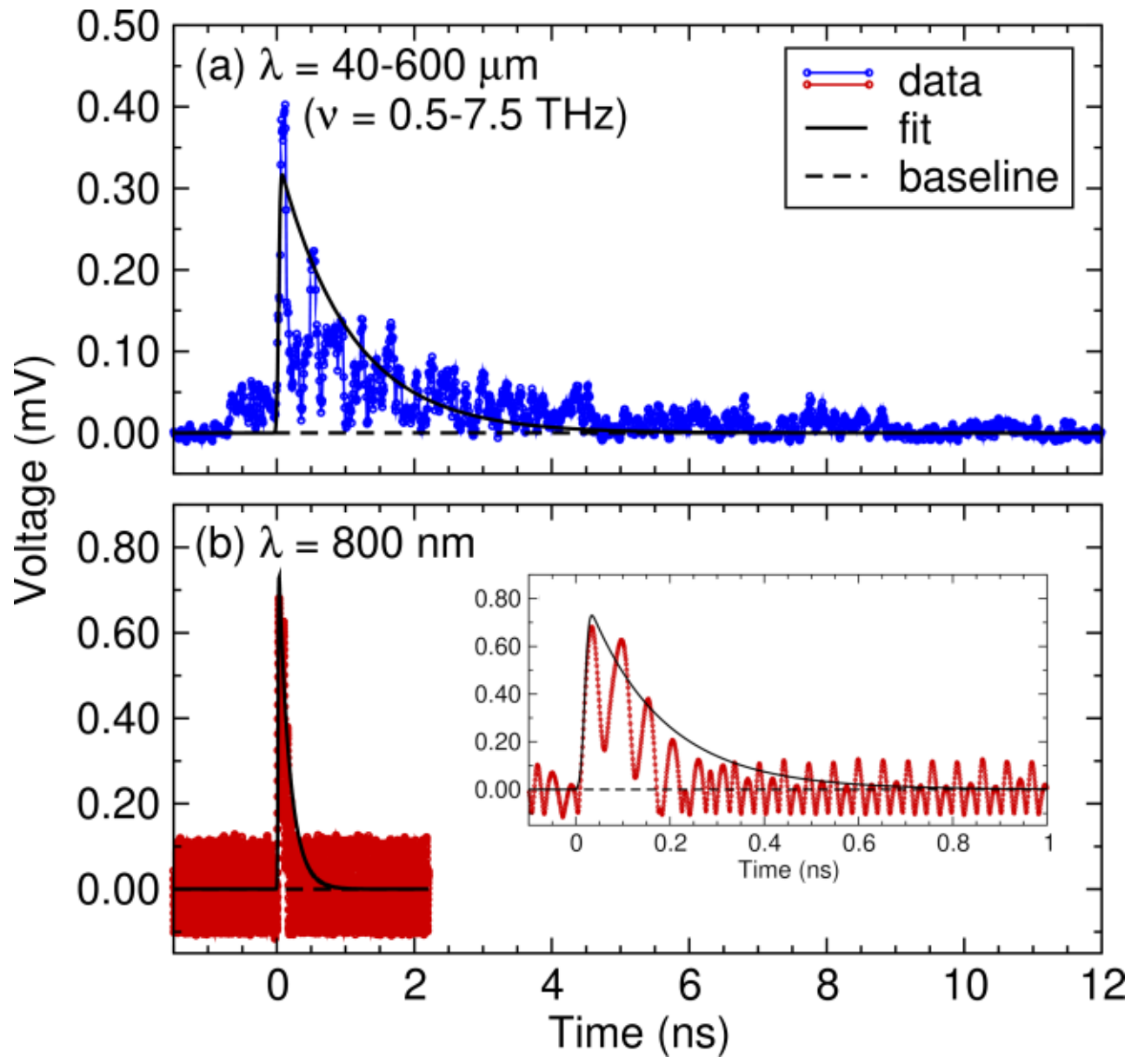

**Fig. 6 | Laser-Driven Ultrafast Impedance Response of LLTO.** The LUIS response of LLTO due to (a) broadband THz excitation, with a wavelength range of 54.5-599 um (0.5-5.5 THz) at a field strength of 744 kV/cm and (b) an 800 nm 8 mJ per pulse excitation. The red and blue signals represent the averaged amplitude demodulation response due to a 19 GHz, -5 dBm reflected voltage signal with a root-mean-squared envelope function applied to the response signal. The inset in (b) highlights the sub-nanosecond response of the 800 nm excitation. Both signals were fit to a gaussian convolved with an exponential decay, described in detail in the text, and denoted by the solid black lines. The dashed black lines at 0 mV demonstrates the baseline.

For these experiments, we utilize a 19 GHz, -5 dBm (178 mV sinus amplitude) alternating current (AC) field to maximize the signal-to-noise ratio. The probing depth of the 19 GHz AC field in LLTO is 1.15 cm[57], allowing for the full 450 $\mu$m-thick sample to be probed before reaching the metal short, which reflects the microwave signal. For these measurements, a thinner sample can be used as compared to the laser-driven EIS experiments. For the LUIS experiments, the pin used to transmit the AC signal is 350 $\mu$m in diameter, fully sampling the 6-7 $\mu$m laser absorption depth for the THz and 800 nm excitations, allowing for through-plane measurements, rather than the in-plane laser-driven EIS measurements. The signal is further processed using averaging and amplitude demodulation, as seen in **Figs. S16a-b and S17a-b**.

We fit the signal responses to a Gaussian response function convolved with an exponential decay, detailed in the **Methods**. The width of the gaussian is reported as the rise time, limited by the sampling rate of the oscilloscope, which is 7.8 ps. Further subsampling is possible when recording sub-nanosecond responses, leading to the differing sample rates seen in **Fig. 6a-b** and rise times reported in **Table S5**. We also report the exponential decays with a 95% confidence interval and attribute these responses to dynamics induced by each laser excitation. The deconvolution of these fits is seen in **Figs. S16c and S17c**.

We find, in **Fig. 6a**, that the THz excitation induces in a 1.03$\pm$0.03 ns decay time, which we hypothesize is due to increased ion hopping due to the excitation of strongly absorbing phonon-ion THz-range rocking modes (**Fig. 4d**), as well as potential decay pathways associated with the phonon-ion coupled hopping modes that are theorized to enhance ion hopping (**Fig. 2a-b**). As seen in **Fig. 6b**, the 800 nm excitation induces a much sharper response with a decay time of 0.15$\pm$0.01 ns, which we attribute to the excitation of an acoustic phonon bath associated with heating[48]. We summarize the fitted rise and decay times in **Table S5**. In short, the differing ultrafast responses of the two excitations demonstrate that the THz excitation induces a unique, long-lived response that is distinct from that due to laser-induced heating and is responsible for the outsized impact of phonon-ion coupled hopping modes in the THz energy range, as detailed throughout this work.

**Discussion**

The relative decrease in $R_{bulk}$ with THz light excitation of rocking modes is almost ten-fold that seen with 800 nm excitation, which is predicted to incoherently heat the sample. The significant difference in enhancement emphasizes the potential for theorized phonon-ion coupled hopping modes to promote ion migration at higher rates than uncorrelated thermal, acoustic phonon-induced hops.

At the macroscopic scales probed by the several hundred micron diameter spot size of the THz absorption data shown in **Fig. 4d**, the absorption coefficient of LLTO above 2-3 THz exceeds 1000 $cm^{-1}$, values in excess of those seen in high dipole moment liquids such as water and acetonitrile[58]. The multi-octave bandwidth of the THz pulse enables both resonant electric dipole-allowed transitions and sum frequency excitation via the polarizability operator, that is Raman active phonon modes[59]. The absolute values and smooth nature of the cross sections suggest that the transition dipoles are large, and that local variability in the Li/La ratio impacts the phonon mode frequencies, leading to a broad distribution of modes in this energy range, which cannot be cleanly decoupled in the THz absorption spectra (**Fig. 4d**). This is further corroborated by Raman measurements of $Li_{3x}La_{2/3-x}TiO_3$ ($0.04 \leq x \leq 0.15$), where the local distribution of $Li^+$ and $La^{3+}$ alters the spectroscopic signature of modes associated with the $TiO_6$ octahedra[42]. We hypothesize that the absorption coefficients at the level of **Fig. 4d** require significant coupling of octahedral rocking and ion displacement, likely with potential energy surfaces that are quite flat for small displacements. This suggests the coupling of relatively higher frequency phonons with $Li^+$ modes at relatively lower frequencies, as proposed by Xu et al[29]. Furthermore, recent work indicates that in LLTO, modes such as the key $TiO_6$ rocking modes we investigate here can distort Li sites, and the corresponding distortion can lead to a mechanism termed "ion-lattice-ion concerted diffusion"[60]. The activation of modes that give rise to this concerted mechanism upon THz excitation may be responsible for the long decay time found in the LUIS experiments and the strong relative enhancement observed in laser-driven EIS. Determining the fraction of THz modes that couple with actual ion hopping/conductivity requires careful comparison to the theory of the hopping mechanism and action spectroscopy to assess the role of THz excitation on ion hopping quantitively.

Having performed these measurements, we propose that the enhancement due to THz laser excitation can be attributed to phonon-ion couplings which enhance $Li^+$ hopping. However, the generation of electronic carriers and heating effects must also be considered. Any electronic contribution toward the total measured impedance, including the excitation of electronic carriers due to THz fields, are unlikely in LLTO, because (1) there is no change to the measured open circuit voltage during the EIS experiments, (2) the THz radiation is $<0.025$ eV[61] which lies sufficiently below the 3.3 eV band gap of LLTO[48], and (3) no nonlinear effects are observed (**Fig. S12**). Furthermore, we have previously demonstrated that, even in the case of a band gap excitation, electronic contributions to changes in the conductivity are negligible[48]. Therefore, the THz fields are unlikely to generate electronic carriers or have sufficient energies to promote enhanced electronic conductivity in LLTO. In the presented work, the measured changes in impedance should be attributed to changes in ion hopping. Additionally, no dependence on excitation near the contacts on the LLTO was found at the metal-material junction, ruling out the generation of carriers on the metal contacts.

Deconvoluting thermal enhancements of ion hopping from those associated with exciting targeted lattice-driven dynamics is non-trivial. A complete Nyquist plot that captures the bulk semi-circular feature requires 10s of seconds for each measurement, preventing direct probing of the "peak" temperature change induced by the laser which occurs in less than 100 ns (**Figs. S18** and **S19**). On the milliseconds-to-seconds timescales of the EIS measurements, we expect both excitations to have contributions from the decay of the initial phonon excitation into heat. However, if heating is the only mechanism by which the laser enhances ion hopping for both the 800 nm and THz light excitations, we would expect these excitations to cause similar levels of enhancement when normalized, given they have similar penetration depths and laser-sample interaction volumes. Instead, the THz light leads to an order of magnitude larger enhancement compared to the 800 nm light when normalized by power density (**Fig. 4a**). Additionally, the timescale of the ion hopping response due to the THz excitation is an order of magnitude longer, as measured by LUIS (**Fig. 6**). Such a discrepancy requires further investigation into the differences in laser-indued heating and longer-term photo-modulation effects between both excitation energies.

To more accurately compare the enhancements in ion migration between the laser-induced heating and THz-range phonon-ion coupled hopping modes and separate contributions due to heating and photo-modulated ion hopping at milliseconds-or-longer timescales, we adopt a finite-difference time domain (FDTD) method, detailed in the **Methods**. The method aims to capture resistor heating, while accounting for wavelength-dependent experimental parameters, outputting a single metric of heating normalized by the average laser power (**Fig. S20**). We can then separate the effects of heating from the different excited phonon baths at each wavelength.

We first identify the absorption depths of both 800 nm and THz light. Then, using the model, we produce cross-sectional heating maps of the sample, which show differences in the temperature gradient between the THz and 800 nm light (**Figs. S21 and S22**). Given the challenges in reconciling the region probed electrochemically with that heated by the laser, we aim to quantitatively determine an upper bound on the heating relative to the milliseconds- or longer timescales probed by EIS. We note that ultrafast laser pulses will induce peak temperatures well above the metric we define, but this heating is unlikely to be captured by the EIS measurement. As such, we instead capture the average maximum temperature in the sample at all time points during and after the final pulse, once the sample has reached a baseline thermal steady state, which occurs within 100 ms (**Figs. S23 and Fig. S24**). Using this metric, we find that 800 nm light induces a temperature change of 0.03 K/mW. In contrast, THz light induces a change of 0.02 K/mW. We confirm the accuracy of the method by calibrating it against measurements with an IR thermal gun for the 800 nm excitation (**Methods and Fig. S25**), and as described in previous work[48].

In **Table 2**, we use the thermal enhancement factors in K/mW to differentiate the computed thermal effects from the measured total enhancement, the slope in **Fig. S12**. To ascertain the ion migration enhancements associated with resistor heating, we utilize **Eq. 3.**

$$Thermal\ Effects = \frac{\Delta K}{mW} * \frac{\%\Delta \mathrm{R}}{\Delta \mathrm{K}} \qquad (3)$$

We measure the decrease in sample resistance per degree Kelvin, the slope in **Fig. S26**, and multiply this factor by the computed wavelength-dependent heating factors, discussed above, thus making up the right side of **Eq. 3.** The value provided by this equation is a quantitative measure of the upper limit for resistor-like thermal-induced heating effects on timescales relevant to EIS, in units of %ΔR/mW, as depicted in the second column of **Table 2**. Finally, we compare the computed thermal effects to the slopes of **Fig. S12**, the third column in **Table 2**.

**Table 2.** Deconvolution of Thermal-Induced Ion Enhancement from the Total Measured Enhancement of Ion Migration

| Excitation Source | Calculated Thermal Effects (%ΔR/mW) | Measured Total Enhancement (%ΔR/mW) |
| --- | --- | --- |
| 800 nm | 0.07±0.01 | 0.10±0.01 |
| 0.7-4.5 THz | 0.04±0.01 | 0.78±0.04 |

From **Table 2**, we can determine the fraction of total enhancement attributable to resistor-like heating relative to the total measured enhancement, finding that 60%-80% of the enhancement at 800 nm is associated with thermal heating. We note that the approximately 30% discrepancy likely indicates that laser and resistor heating affect ion hopping differently. However, the model still allows us to attribute most of the 800 nm excitation enhancement to purely thermal effects .Given this, we believe that this wavelength measures the enhancement associated with populating the acoustic phonon bath, or heating the sample[48]. For the THz excitation, we can again find the fraction of the total enhancement which is associated with resistor-like heating effects. We find that approximately 5% of the THz-induced enhancement is attributable to heating. In comparison, the remaining 95% should be attributed to additional photo-modulation effects associated with the population of phonon-ion coupled hopping modes, in qualitative agreement with the computational work presented in this paper. By separating the thermal contributions from the total measured enhancement, we demonstrate that we are indeed probing the photo-modulation of ion hopping rather than simply the effects of laser-induced heating.

We further aim to account for grain boundary scattering by varying the thermal conductivity of the sample in the FDTD model. Grain boundaries can suppress phonon scattering and thus reduce the thermal conductivity of the sample[62]. We find that altering thermal conductivity does not meaningfully change the wavelength-dependent enhancement, as shown in **Tables S6 and S7**. This model does not explicitly account for wavelength-dependent phonon scattering and instead generalizes these effects in the remainder of the total enhancement, beyond the calculated resistor-like heating contribution. This remainder not only includes any directly excited dynamics in the crystal lattice, but also the associated decay pathways, as seen in **Fig. 6a-b**. Based on the previously presented computational results, we hypothesize that the modes seen in **Fig. 5** and the highly absorbing nature of the phonons found in **Fig. 4d** are part of the THz-specific decay pathway and affect the $TiO_6$ octahedra in ways that are theorized to couple with ion hopping.

In addition to ruling out electronic carriers and heat, we also aim to address discrepancies between the theoretical and experimental data. The results in **Table 1**, which demonstrate a three order of magnitude increase in ionic conduction, represent limiting cases where energy can be deposited into three well-defined comparative cases: the excitation of the full phonon bath, individual highly contributing modes, and non-contributing modes. In comparison, the THz excitation excites all phonon modes which can absorb THz light which does not match the computationally excited cases. Rather, we must use THz absorption and Raman spectroscopy to gain insight into what our light excites. As shown in **Fig. 4d**, we observe strong but smooth absorption across the THz spectrum, preventing the identification of individual modal components via deconvolution. This smooth absorption is due to local variability in Li/La ratios throughout LLTO. However, when investigating what modes exist in the low-energy THz regime, we can consider the Raman spectrum (**Fig. 5**), where we identify modes that may couple strongly to hopping $Li^+$-ions and represent plausible decay pathways[56]. The optical excitation of these modes leads to a factor of ten increase in the differential enhancement as compared to laser-induced heating, which quantitatively differs from **Table 1** by two orders of magnitude. We believe that this disagreement can mainly be attributed to the differences in the computational and experimental excitations discussed above.

Still, we consider alternative explanations below. One can rule out the mismatch in EIS probe and laser absorption depth because both the 800 nm and THz light have similar absorption depths, so the relative enhancement ratios should be maintained, given that the laser pumps the same sample volumes. We also note the mismatch between the timescales of the computational and experimental results. While the computed enhancement is reported in hops/ps, the 803 kHz EIS probe used in **Fig. 4a** probes processes occurring on microsecond timescales, meaning that we may not perfectly probe bulk properties, but instead may be probing local structural effects. La/Li ordering[63], impurities inherent in LLTO[64], and domain boundaries[65] may be captured by EIS but are not captured computationally, even when considering both ordered and disordered structures (**Fig. S3**). As discussed, we attribute quantitative discrepancies to differences between computational and optical excitations, as well as to differences between computational and experimental probes, while emphasizing the qualitative agreement across the various portions of this work. This, when combined with the long-lived response measured in the LUIS experiments, paints an intriguing picture regarding the important role of THz-range vibrations on enhancing ionic conduction.

## Conclusion

In conclusion, we compare the relative role of acoustic, optical, and THz phonon modes in ion migration through laser-frequency-selective perturbation of EIS and LUIS measurements. By directly driving highly contributing phonon-ion coupled hopping modes, which are most likely rocking modes, in the 0.7 – 4.5 THz region, a ten-fold enhancement in ion migration is measured relative to incoherent heating. Using an FDTD model, we attribute most of this enhancement to laser-driven photo-modulation effects and find that these excitations lead to notably different responses on ultrafast timescales. The anomalous contribution of rocking modes to ion hopping in the <4.5 THz region is qualitatively consistent with ab initio calculations, validating the use of laser-driven methods to probe the mechanism of phonon-ion-coupled interactions.

Furthermore, the findings herein can aid in the design of future solid-state ion conductors by emphasizing the design of structures with a large population of phonon modes that couple to ion hopping and are inherently active at room temperature (<6 THz). More specifically, our study emphasizes the role of complex phonon-ion interactions, indicating that, at least in perovskite structures, engineering structures with highly active THz rocking modes may increase ionic conductivity by coupling lattice dynamics with the hopping ion. Additionally, THz-range phonon-ion coupling has been found to be relevant in polymer electrolytes, potentially allowing for new design rules which emphasize key vibrational modes which can enhance ion hopping rates[66,67]. Similarly, in biological systems, THz light may enable the modulation and control of ion channels[68]. This work provides an initial foray into the direct experimental measurement of how THz-range vibrational modes affect ion hopping. We emphasize this exploratory motivation, which still offers intriguing results, demonstrating the outsized role of low-energy phonon-ion coupled hopping modes in enhancing ionic conduction in a solid-state ion conductor.

## Methods

**General DFT setup.** $Li^+$ ion hop simulations in this study were conducted under the spin-polarized density functional theory (DFT) approximation using the Vienna Ab initio Simulation package (VASP)[69] within the projector augmented-wave approach. Perdew–Burke–Ernzerhof (PBE)[70] generalized-gradient approximation (GGA) functionals, electronic KPOINTs equal to 4x4x4, an electronic energy cutoff of 500 eV, and Gaussian smearing with sigma of 0.05 eV were employed for the calculations. The valence electrons for lanthanum, lithium, titanium, and oxygen atoms were generated in the $5s^2$ $5p^6$ $5d^1$ $6s^2$, $1s^2$ $2s^1$, $3p^6$ $3d^2$ $4s^2$, and $2s^2$ $2p^4$ configurations, respectively.

**Details for NEB calculations.** The minimum energy pathway for $Li^+$ ion hops in the lattice were calculated using the nudged elastic band (NEB) methodology following the formulations by Henkelman et al.[71] to ensure the correct determination of the saddle point as implemented in VASP. Three middle images were considered for the NEB calculations. Structures at the end of the hops were relaxed to lower the forces and variations in energy below $10^{-4}$ eV/ Å and $10^{-5}$ eV, respectively. Obtaining these end-point relaxed structures are critical in a successful NEB calculation to obtain meaningful migration barriers (MB). Because of the challenges in the convergence of DFT relaxation simulations, which is mainly caused by the multitude of relaxation sites available in the lattice of highly $Li^+$ conductive compounds,[72] different optimization algorithms were tested for the convergence of both local relaxed structures (corresponding to the beginning and end of the hops) and NEB calculations (reaction coordinates). The structure relaxation and NEB calculation were initially tested for convergence using conjugate-gradient energy-based optimization method (IBRION=2). If no convergence was achieved using energy-based optimizers, force-based optimizers were then employed. Two force-based optimizers[73] were identified to be effective in helping the convergence: (*i*) QM=Quick-Min (IOPT = 3, IBRION = 3, POTIM = 0), and (*ii*) FIRE = Fast Inertial Relaxation Engine (IOPT = 7, IBRION = 3, POTIM = 0). We reported the relaxed structure, the migration barrier (MB), the reaction coordinates using whichever method that converged. Afterwards, we utilized the reaction coordinates for the phonon modal analysis of the respective ion hop.

**Choosing the initial structure for computations.** The same experimental structure identified from synchrotron X-ray diffraction of $La_{0.5}Li_{0.5}TiO_3$ in this study is used for the *ab initio* NEB and phononic calculations. For ease of calculations and better visualizations, the experimentally determined $R\bar{3}c$ structure ($a = b = 5.4711Å, c = 13.4040Å, \alpha = \beta = 90°, \gamma = 120°$) is transformed to its pseudo-cubic (perovskite-looking) counterpart[74] ($a = b = c = 7.7378Å, \alpha = \beta = \gamma = 90°$) through the rotation/transformation of the lattice based on the **Eq. 4** where, $\alpha = 0°$, $\gamma = -45°$, and $\beta = -35°$ correspond to yaw, roll, and pitch angles, respectively. We considered 40 atoms in this pseudo-cubic unit cell, and we used it for the rest of our simulations. After the transformation, only internal atomic coordinates were relaxed during the calculations.

$$R = R_z(\alpha)R_x(\gamma)Ry(\beta) = \begin{bmatrix} \cos(\alpha) & -\sin(\alpha) & 0 \\ \sin(\alpha) & \cos(\alpha) & 0 \\ 0 & 0 & 1 \end{bmatrix} \begin{bmatrix} 1 & 0 & 0 \\ 0 & \cos(\gamma) & -\sin(\gamma) \\ 0 & \sin(\gamma) & \cos(\gamma) \end{bmatrix} \begin{bmatrix} \cos(\beta) & 0 & \sin(\beta) \\ 0 & 1 & 0 \\ -\sin(\beta) & 0 & \cos(\beta) \end{bmatrix} \quad (4)$$

**Maximizing the sampling of $Li^+$ hop in the LLTO lattice through different choices of lattice disorder, $Li^+$ hopping mechanisms, and hopping pathways.** The experimentally determined coordinates for La and Li contain partial occupancies and need to be substituted with atoms in order to perform atomistic simulations. It is important to maintain the exact experimental stoichiometry

($La_{0.5}Li_{0.5}TiO_3$) during this process. Depending on which La and Li partial occupancy sites are selected for atom placement in the simulation, locally different structures of LLTO can be formed. Experimental[75] and computational[76–78] studies have shown that the various structures of LLTO can be classified into three categories based on the degree of La|Li disorder in the lattice (Li|La orderings): (*i*) fully ordered structure, (*ii*) partially ordered structure, and (*iii*) fully disordered structure. However, due to the limited size of the supercell used in our simulations (40 atoms), it is difficult to distinguish between the partially ordered and fully disordered structures. Therefore, to maximize our sampling of $Li^+$ ion hops in the LLTO lattice within the constraints of our supercell size, we included three different La|Li orderings in our analysis of $Li^+$ ion hops, as illustrated in **Fig. S3**. These chosen structures included one structure with ordered Li|La layers, as well as two structures with disordered Li|La layers.

Since our calculations do not involve any hopping for La atoms, the position of La atoms is the primary factor that distinguishes these three structures with different La|Li orderings. Therefore, these structures can be regarded as parent structures, under which we have defined several substructures by shuffling $Li^+$ ions among unoccupied sites in the LLTO lattice. Although our simulation cell sizes may not be large enough to fully capture the exact differences between these parent structures with different La|Li orderings, we believe that incorporating these different ordered and disordered parent structures, along with their substructures and the different $Li^+$ hopping mechanisms included in these substructures, has increased our sampling of different local chemical environments in the LLTO lattice in our calculations. The schematics of the parent structures with different Li|La ordering layers, the number of $Li^+$ hops calculated for different substructures under these parent structures obtained by shuffling $Li^+$ ions in the lattice or along different hopping pathways, and the different considered hopping mechanisms are illustrated in **Fig. S3**.

In addition to the different degrees of disorder in the LLTO lattice, we also considered two different hopping mechanisms: (*i*) single and (*ii*) concerted ion hops. For the single ion hop mechanism, the hop occurs through a vacancy. However, there are no vacancy sites in the stoichiometry considered in this study (i.e., the nominal number of vacancies in the stoichiometric $La_{2/3-x/3}Li_x\square_{1/3-2x/3}TiO_3$ for x=1/2 is zero). To allow the occurrence of the vacancy-mediated mechanism, a La cage must be doubly occupied by two $Li^+$ ions. This results in an unoccupied La cage that was previously occupied, which is now available for other $Li^+$ ions to hop into. Thus, the choice of the doubly occupied La cage and the resulting hopping pathway will be the determining factor for performing our NEB calculations. We combined different (*i*) orderings of Li|La, (*ii*) choices of which La cage to doubly occupy, and (*iii*) hopping mechanisms to choose 22 different $Li^+$ hopping events to maximize our sampling efficiency of the LLTO lattice. The distribution of the different hopping mechanisms (single and concerted) across the chosen structures with different degrees of disorder is illustrated in **Fig. S3**.

**DFPT setup for modal calculations.** To obtain the vibrational modes (phonons) for each of the structures considered in this study, we first calculated the Hessian matrix using density-functional-perturbation theory (DFPT) with the VASP software and the Phonopy package[79] to capture the matrix. Next, we calculated the phonon frequencies and eigenvectors using a custom lattice-dynamics Python code[79,80]. It was important to obtain these values for the exact supercell containing 40 atoms, since the NEB calculations were also performed on cells of this size.

**Phonon modal analysis of ion hop.** To identify the contribution of different phonons to the $Li^+$ hop in the lattice, we used the methodology recently proposed based on combining the lattice dynamics and NEB calculations[38]. Following this method, after the minimum energy pathway for the $Li^+$ hop is identified using the NEB calculation, we project the displacement field obtained from the transition state (saddle point) on the eigenvectors of vibration belonging to the structure at the beginning of the hop. The magnitude of projection shows the degree to which each normal mode is contributing to that displacement field, which can be quantified using the following expression[81,82],

$$Q_n = \sum_{i=1}^{N} \sqrt{m_i}\, \boldsymbol{e}_{i,n}^{*} \cdot \boldsymbol{u}_i \qquad (5)$$

where $\boldsymbol{u}_i$ is the displacement of atom $i$ from its equilibrium position in the configuration at the hopping origin; $\boldsymbol{e}_{i,n}$ is the eigenvector for mode $n$, assigning the direction and displacement magnitude of atom $i$ obtained from the lattice dynamics calculation[81,83,84] for the structure at the hopping origin; $*$ denotes the complex conjugate operator; and $m_i$ is the mass of atom $i$. $Q_n$ is the modal displacement coordinate, the square of which is proportional to the mode potential energy $E_n$ according to the following equation[82],

$$E_n = \frac{1}{2}\omega_n^2 Q_n^2 \tag{6}$$

The total energy of the system $E$ is equal to the summation over all modal energy values ($E = \sum_n E_n$); Here, $E_n$ is the contribution by mode $n$ to the potential energy of the displaced lattice during the ion migration, which can be interpreted as the contribution by mode $n$ to the ion hop along its migration pathway. In addition, $\omega_n$ is the frequency of vibration of mode $n$. $E_n$ can be divided by $E$ to calculate the normalized contribution of that phonon to the $Li^+$ hop in that specific ion hop event. Moreover, by subsequently normalizing $E_n$ through division by the total number of ion hopping events considered in the study, the resultant normalized $E_n$ can be regarded as the normalized contribution of mode n to the $Li^+$ hopping events in the LLTO compound under investigation. The code used to identify the modal contributions of $Li^+$ hop in the LLTO lattice, following the outlined methodology, was written in Python and is available as a GitHub repository[81].

**Algorithm for identifying the rocking modes.** In this study, we identified the octahedral modes of vibration by quantifying the degree of circulation the eigenvectors of vibration impose on the octahedron units ($TiO_6$) in the lattice. To this end, the eigenvectors of vibration for the 6 oxygen atoms belonging to the $TiO_6$ octahedron unit were utilized to quantify the circulation of the unit according to the definition of circulation ($\Gamma$), e.g., $\Gamma_{\mathrm{n}}^{i} = \oint \vec{v} \cdot \overrightarrow{dl}$, where $\Gamma_{\mathrm{n}}^{i}$ is the circulation belonging to the octahedron unit $i$ (in our case we have 8 octahedron units in our simulation cells, thus $i = \{1, 2, 3, \ldots, 8\}$) coming from mode of vibration $n$. In addition, $\overrightarrow{dl}$ is the path vector along which the circulation is being evaluated, and $\vec{v}$ is the eigenvector of vibration belonging to the specific atoms along the $\overrightarrow{dl}$ direction. In practicality, to simplify the evaluation of the circulation formula above, we considered the perfect representation of a perovskite structure (e.g., with untitled octahedron units) and evaluated $\Gamma_{\mathrm{n}}^{i}$ along the circulation pathways normal to $x$, $y$, and z directions, which results in $\Gamma_{\mathrm{n}}^{i,\alpha}\{\alpha = x, y, z\}$. By averaging the obtained values of $\Gamma_{\mathrm{n}}^{i,\alpha}$ over $\alpha$, a representative circulation value is found for octahedron unit number $i$ $\left(\Gamma_{\mathrm{n}}^{i}\right)$. By averaging again over all the octahedron units in the simulation cell, $i$, a representative value of circulation is obtained for all the octahedron units in the simulation cell ($\Gamma_n$). If this value is larger than a specified criterion, the mode of vibration number $n$ is identified as an octahedral mode of vibration. Example MATLAB code block below demonstrates this calculation:

```
% mondenum = 3*Natoms % number of modes of vibration
in the system
for n=1:modenum
    vortTi = zeros(8, 3);
    % TiO6 unit 1 (atomids listed below)
    ix1 = 23;
    ix2 = 26;
    iy1 = 21;
    iy2 = 27;
    iz1 = 34;
    iz2 = 36;
    vortTi(1,1) = (ev((iy2-1)*3+3, n) - ev((iy1-
1)*3+3, n)) - (ev((iz2-1)*3+2, n) - ev((iz1-1)*3+2,
n));
    vortTi(1,2) = (ev((iz2-1)*3+1, n) - ev((iz1-
1)*3+1, n)) - (ev((ix2-1)*3+3, n) - ev((ix1-3)*3+3,
n));
    vortTi(1,3) = (ev((ix2-1)*3+2, n) - ev((ix1-
1)*3+2, n)) - (ev((iy2-1)*3+1, n) - ev((iy1-1)*3+1,
n));
    .
    .
    .
    for ii=1:8
        vortTi_mag(ii) = (vortTi(ii,1)^2 +
vortTi(ii,2)^2 +  vortTi(ii,3)^2)^.5;
    end
    vortTi_ave(n) = mean(vortTi_mag);
    vortTi_mag_allmodes(n, :) = vortTi_mag;
end
tagOctRot(1:modenum, i) = 0;
for n=1:modenum
    vorticity_criteria = 0.25;
    if vortTi_ave(n)^2 > vorticity_criteria
        tagOctRot(n) = 1;
    end
end
```

**Calculating the imparted energy on different atomic species by different modes of vibration.** To calculate the energy that each mode of vibration imposes on the hopping $Li^+$, we leverage the following identity equation existing from lattice dynamics theory[82], $\sum_n \vec{v}_n^{\,i} \cdot \vec{v}_n^{\,i} = 1$, where $\vec{v}_n^{\,i}$ is the eigenvector of vibration belonging to mode $n$, imposing vibrations on atom $i$. The vibrations of atoms in the crystal are governed by the eigenvectors of vibration, and the imparted vibrations on an atom will add up to give an energy equal to $k_BT$ to that atom. In this view, the $\vec{v}_n^i \cdot \vec{v}_n^i$ term in the summation above can be considered to be proportional to $k_BT$. This methodology can be extended to the eigenvectors of vibration on the hopping Li$^+$, where they can be projected along the hopping direction of the Li$^+$ to give the energy imparted on Li+ along the hopping pathway, e.g., $E_{Li^+}^{hopping\ direction} = \vec{v}_n^{Li^+} \cdot \vec{u}_{hopping\ direction}$. Similarly, the $\vec{v}_n^{Li^+} \cdot \vec{u}_{hopping\ direction}$ term can be considered to be proportional to $k_BT$.

**Quantifying the change in O-4 bottleneck area due to phonon modes in the structure.** To evaluate the impact of each phonon on the O-4 bottleneck area during ion hopping, we applied a displacement to the lattice along the eigenvectors associated with that phonon. This displacement can be considered as an induced excitation of the lattice. The extent of this excitation can be calculated for each vibrational mode using the following equations.

The potential energy for an oscillator can be written as arising from the sum of harmonic and anharmonic contributions[82],

$$\langle H_{pot} \rangle = \langle H_{harmonic} \rangle + \langle H_{anharmonic} \rangle \quad (7)$$

where $\langle ... \rangle$ represents the ensemble average[85]. The harmonic portion has been shown to dominate the potential energy[86]. Thus, we will also leverage the harmonic potential energy to evaluate the mode amplitude that is needed to calculate the degree of displacement field in the structure. Based on the equipartition theorem, the average harmonic energy of a classical oscillator is equal to[87]

$$\langle H_{harmonic} \rangle = \frac{1}{2} \mathrm{k_B T} \quad (8)$$

Where $k_B$ is the Boltzmann constant and $T$ is the temperature of the system. The potential energy of a harmonic oscillator can also be calculated from the normal mode amplitude analysis via **Eq. 9.**[82]

$$\langle H_{harmonic} \rangle = \tfrac{1}{2} \mathrm{Q_n^2 \omega_n^2} \quad (9)$$

where $Q_n$ is the modal displacement coordinate (**Eq. 6**), and $\omega_n$ is the frequency of the eigen mode. It should be noted that calculating the harmonic energy of an eigen mode using the knowledge of force constant matrix and the respective atomic displacements for an eigen mode is equal to the approach based on the normal mode amplitude analysis (**Eq. 10**). The modal displacement coordinate can be explicitly obtained from **Eq. 3**, which is rewritten here[77].

$$Q_n = \sum_{i=1}^{N} \sqrt{m_i} \boldsymbol{e}_{i,n}^* \cdot \boldsymbol{u}_i \quad (10)$$

where $\boldsymbol{u}_i$ is the displacement of atom $i$ from its equilibrium position in the configuration at the hopping origin; $\boldsymbol{e}_{i,n}$ is the eigenvector for mode $n$, assigning the direction and displacement magnitude of atom $i$ obtained from the lattice dynamics calculation[81,83,84] for the structure at the hopping origin; $*$ denotes the complex conjugate operator; and $m_i$ is the mass of atom $i$. By combining **Eq. 8** and **Eq. 9** we then have,

$$Q_n^2 \omega^2 = k_B T \quad (11)$$

To determine the distribution of the harmonic energy for eigen mode amongst the atoms in the system, one can envision a state of the system whereby all of the atoms in the system are displaced from equilibrium in the direction of their respective eigen vectors from mode **n** (i.e., $\boldsymbol{e}_{i,n}$). In this view, the attributed displacement to an atom would be equal to,

$$\boldsymbol{u}_i = \mathrm{\alpha_n} \boldsymbol{e}_{i,n} \quad (12)$$

Where $\alpha_n$ is a scaling factor that associates a certain degree of displacement with the mode's amplitude at a given temperature. The exact value of the scaling factor can then be calculated from the combination of **Eq. 6**, **Eq. 11**, and **Eq. 12**. Replacing for the atomic displacement in **Eq. 6** with the definition in **Eq. 12**, we would have,

$$Q_n = \sum_{i=1}^{N} \sqrt{m_i} \boldsymbol{e}_{i,n}^* \cdot \alpha_n \boldsymbol{e}_{i,n} = \alpha_n \sum_{i=1}^{N} \sqrt{m_i} \boldsymbol{e}_{i,n}^* \cdot \boldsymbol{e}_{i,n} \quad (13)$$

which by the substitution of $Q_{i,n} = \sqrt{m_i} \boldsymbol{e}_{i,n}^* \cdot \boldsymbol{e}_{i,n}$, yields a simpler form,

$$Q_n = \alpha_n \left( \sum_{i=1}^{N} Q_{i,n} \right) \quad (14)$$

Using **Eq. 11** we then have $Q_n \omega_n = \sqrt{k_B T}$ and by incorporating **Eq. 14**, we can calculate the scaling factor as,

$$\alpha_n = \frac{\sqrt{k_B T}}{\left( \sum_{i=1}^{N} Q_{i,n} \right) \cdot \omega_n} \quad (15)$$

By using the obtained $\alpha_n$ in **Eq. 15**, the displacement field is obtained. The O-4 bottleneck area will then be calculated for both the equilibrium (no displacement) lattice and the lattice for which the atomic positions have been displaced according to **Eq. 15** to assess the capability of each phonon in increasing the O-4 bottleneck area against the ion hop. It should be noted that the four oxygen atoms forming the bottleneck area are not necessarily on the same plane in space, making it impossible to calculate the planar area connecting them. Therefore, four smaller triangular areas inside the bottleneck area are defined (as shown in **Fig. S1**), and their areas are calculated and averaged to approximate the bottleneck area in this situation, as demonstrated in **Fig. S1**. The percent change between the O-4 bottleneck area in the equilibrium lattice and the one in the displaced lattice is then calculated and reported as the change in the ability of that specific phonon in altering the bottleneck area against the hop of $Li^+$ ion.

**Quantum correcting the calculated modal contributions to the ion hop.** To include the quantum effects in our reported modal contributions to the $Li^+$ hop in LLTO, we multiply our obtained modal contribution by mode $n$ ($E_n$, obtained from **Eq. 6**) with frequency $\omega$ by a correction coefficient that is based on the Bose-Einstein distribution function at temperature T $\left(\mathrm{f_Q}(\omega, T)\right)$ based on the following formula to obtain the quantum corrected contribution of that mode $n$ to the hop in the lattice ($E_n^{qc}$)[81],

$$E_n^{qc}(\omega) = \mathrm{f_Q}(\omega, T)\mathrm{E_n}(\omega) \quad (16)$$

The function $f_Q$ represents the ratio of quantum to classical specific heat for mode $n$. It restricts the contributions of the high frequency modes at low temperatures and modulates the modal contributions obtained from the NEB-based ion hop simulations. The quantum expression of volumetric specific heat, based on Bose-Einstein statistics, is given by

$$C_q(\omega, T) = \frac{k_B x^2}{V}\frac{e^x}{(e^x - 1)^2}; x = \frac{\hbar\omega}{k_B T}, \quad (17)$$

and the classical volumetric specific heat is given by $C_c = \frac{k_B}{V}$, where $V$ is the volume of the supercell under study. Thus, the quantum heat capacity correction factor which is the ratio of $C_q$ and $C_c$ is

$$f_Q(\omega, T) = \frac{C_q(\omega, T)}{C_c} = \frac{x^2 e^x}{(e^x - 1)^2} \quad (18)$$

**Enhancing ion hopping rate by targeted excitation of phonons in MD simulations.** To assess if any increase in the hopping rate of $Li^+$ in LLTO lattice can happen by exciting the identified contributing phonons to the ion hop, we chose three hops from the 22 investigated hops in this study. We then excited the highly contributing phonons in their respective structures. We performed the following three tests: (*i*) exciting the top contributing (non-rocking) mode overall, (*ii*) exciting the top contributing rocking mode, and (*iii*) exciting a random mode.

To perform the molecular dynamics simulations, we utilized classical molecular dynamics simulations based on the Morse and Buckingham form potentials as proposed by Cormack and coworkers[88,89] and successfully utilized by Chen and Du[90] to investigate the diffusion of $Li^+$ diffusion in the LLTO lattice. The simulations were performed using the same DFT simulation cells using the large-scale atomic/molecular massively parallel simulator (LAMMPS)[91] package. The modes that were excited were the same modes that were detected from the DFT calculations. The goal of these experiments was to only assess if any extra energy (excitation) given to these highly contributing modes will result in any enhancement in the hopping rate of the ion in the lattice.

First, the structure was relaxed under the Nosé-Hoover canonical ensemble for 100 ps at T = 400 K; then, the structure was simulated until a $Li^+$ hops in the structure. Once the hop gets detected through continuous examination of equilibrium MD positions during the MD simulation[38], the simulation gets interrupted and restarted with a new initial condition for atomic velocities, which results in exploring a new region of the phase space. This process gets repeated until a total of 50 ns has been simulated. Then, the total number of detected ion hops will be compared based on the group of phonons that were excited during the simulations. The time step was chosen to be 1 fs.

Because ion diffusivity increases with temperature, we ensured that the bulk lattice temperature remained constant at a low temperature (T = 400 K), so that any change in the diffusivity could be attributed only to the excitation of the targeted modes and not to the bulk temperature. To do so, we kept the total kinetic energy of the system constant via a velocity-rescaling scheme, in which, the addition of energy to the intended modes was complimented by a uniform reduction in the kinetic energy of all other modes in the system. To change the temperature of mode $n$ to a desired temperature $T_d$, we modified the atomic velocities in the system according to the following formula:

$$\boldsymbol{v}_i = \boldsymbol{v}_i + \frac{1}{\sqrt{m_i}}\left[\sqrt{2k_B T_d} - \dot{Q}_n(t)\right]\mathbf{e}_{\mathrm{i,n}} \quad (19)$$

where $\boldsymbol{v}_i$ is the velocity of atom $i$, and $\dot{Q}_n$ is the modal velocity coordinate of mode $n$ defined by[78],

$$\dot{Q}_n = \sum_{i=1}^{N} \sqrt{m_i}\, \mathbf{e}^{*}_{\mathrm{i,n}} \cdot \boldsymbol{v}_i \quad (20)$$

In this study, the above-mentioned rescaling procedure was applied every five time steps (every 5 fs). Although a velocity-perturbation scheme has been used in this study, other methods for modal excitation, such as atomic position perturbation[92] or simultaneous perturbation of atomic positions and velocities[93], have also been employed in previous studies, particularly those with the goal of investigating the phonon-phonon interactions in MD simulations. However, the chosen method for modal excitation should not change the reported observations in this study because modal excitation (increasing the energy of a mode) can be achieved either by increasing the potential (position perturbation) or the kinetic (velocity perturbation) energy of the mode; the choice of which should not matter.

The results for the excitations are shown in **Table S1**. By targeted excitation of the detected contributing phonons in the LLTO structure (including the contributing rocking modes), the hopping rate of the $Li^+$ increases to the same order magnitude of increasing the hopping rate by increasing the temperature, but this time, by keeping the lattice at the base (T = 400 K) temperature.

**Development of finite-difference time-domain methods to deconvolute thermally induced impacts on ionic conduction.** The heat accumulation and subsequent change in temperature of the samples was simulated using a finite-difference time-domain method which has previously been developed to reproduce the results of a three-dimensional two-temperature model[94,95]. This method treats the laser pulse as instantaneous given that the timesteps in the simulation are significantly longer than the length of the pulse. For this paper, 100 ns timesteps were taken for a total simulation time of 1 s. A subsection of the sample with a 1.5 mm radius and 0.3 mm depth was simulated. The total simulation time was chosen because it allowed the sample to reach a clear steady state in its baseline temperature. Additionally, the timesteps and total volume chosen allowed heat to effectively diffuse through the sample, while also balancing the large computational cost of simulating for long periods of time.

For this model, the thermal conductivity and heat capacity of LLTO were estimated to be similar to other oxide solid-state electrolytes[96,97]. Given the measured porosity of our synthesized LLTO (~0.264), the thermal conductivity was then adjusted using previous relations developed for other oxide materials[98]. While these properties fluctuate greatly with temperature, given that most laser-induced heating effects from ultrafast lasers dissipate in far less than 1 ms, the time between the 800 nm laser pulses, these variations are likely to be very short-lived and have minimal impact on the final temperature values.

The penetration depth of the 800 nm light was estimated to be 7 $\mu$m using the following equation[99]:

$$\delta = \frac{c}{2\kappa\omega} \quad (21)$$

where c is the speed of light, $\kappa$ is the imaginary part of the refractive index, and $\omega$ is the angular frequency of light. For this calculation, the imaginary part of the refractive index was taken from a previous study on the optical properties of LLTO[100], while the absorption depth of THz light was found to be 6 $\mu$m via broadband THz absorption spectroscopy on LLTO.

The simulations produce a micron-resolution heating map, as seen in **Figs. S21 and S22**, showing the spatial variation of temperature both on the surface and into the depth of the sample. It is difficult to ascertain the exact thermal effects of the laser since there is a fast rise and decay of the lattice temperature caused by each pulse, pictured for the final pulse in **Figs. S18 and S19**. However, as seen in **Fig. S25**, the peak temperature in the sample was found to be between 0.4 and 0.8°C for a 20 mw 800 nm beam, and ~0.5°C from simulation. It was not possible to take a similar measurement for the THz light due to instrument constraints. As such, using this value as a calibration, we take the average of the sample's peak temperature, during and after the final pulse, as an approximate temperature in the region we are probing, allowing us to separate thermal effects due to laser illumination. As seen in **Figs. S23 and S24**, by the time 1 s has passed, a baseline steady state has been reached, making this average temperature a reasonable approximation for the temperature in the sample. This characteristic temperature increase is found to scale linearly with the average laser power as seen in **Fig. S20**. Taking the slope of these plots, one finds that, as discussed in the main text, the 800 nm light induces approximately twice as much thermal heating, when accounting for differences in the repetition rate, absorption depth, and spot size compared to THz light.

**Synthesis.** $Li_{0.5}La_{0.5}TiO_3$ (LLTO) was synthesized according to literature[101]. A stoichiometric amount of $La_2O_3$, $Li_2CO_3$, and $TiO_2$ were mixed in an agate mortar and pressed into pellets under 100 MPa of pressure. The pellets were placed on a bed of sacrificial powder and calcined at 800℃ for 4 h then 1200℃ for 12 h at a ramp rate of 1℃ / min. The following material was routinely characterized using a Rigaku X-ray diffractometer with CuKα radiation and scanned from 10 to 70 2θ at a scan rate of 0.4˚ per second. The following x-ray diffraction pattern was fitted using a Rietveld refinement with the GSAS II software. High resolution synchrotron powder diffraction data was collected using beamline 11- BM at the Advanced Photon Source (APS), Argonne National Laboratory.

**Sample preparation.** The resulting powder was pressed into a pellet with a diameter of 9 mm and a thickness of 0.7 mm under 2 5 minute cycles of 2 tons of pressure, yielding a 75-76% pellet density. The pellet was subsequently annealed at 1100℃ for 6 h at a ramp rate of 2℃ / min over a bed of sacrificial powder. A 1.6 mm thick strip of Au was sputtered onto one side of the pellet with a 1 mm gap in the center for the light source to excite the LLTO to eliminate possible effects on the impedance from illuminating the Au. Cu contacts were used to contact the sputtered Au to the 1260A Solartron impedance analyzer for EIS measurements. An open cell set-up was employed to allow THz irradiation to excite the sample without compromising average power.

The conductivity was then measured by an AC Impedance method over a frequency range of 32 MHz to 1 Hz with an applied 100 mV sinusoidal amplitude with and without the excitation source. We confirm the linearity of this 100 mV sinus amplitude by comparing results for $R_{bulk}$ at 20, 50, and 100 mV in a swagelok cell with Au-sputtered electrodes via EIS measurements between 3 MHz and 1 Hz. The results are seen in **Fig. S27** and **Table S8**. A copper mesh faraday cage was custom-made with a 1.4 mm copper wire spacing to reduce the noise in EIS measurements due to electromagnetic interference from the environment.

**Heating cell set up.** A heating cell was custom made to heat the sample between 298 K – 333 K. The cell temperature was controlled using a TC-48-20 OEM temperature controller, 12 V power supply, and corresponding TC-48-20 OEM software and thermocouple. The heating cell was placed inside a Faraday cage for all experiments to reduce noise from electromagnetic interference. This set up was used to acquire the data in **Fig. S26** and is depicted in previous work[47].

**Raman characterization.** Raman measurements were performed on a Horiba Instruments PLUS Raman spectrometer with a 532 nm laser focused on uncompressed LLTO powder on a glass microscope slide. The signal was acquired by averaging 200 acquisitions lasting 1 s each with a 50 $\mu$m slit and 500 $\mu$m hole and 10% average power.

**Damage testing**. Pressed pellets of LLTO were characterized using Raman, as described above. Then, we focused 25 mW of the 800 nm laser at a spot on the sample for 30 minutes on and 10 minutes off for 6 rounds. We then characterized this spot using Raman after the 800 nm excitation to ascertain if there was any damage. We note that the peaks in **Fig. S11** are much sharper than in **Fig. 5** because pressed pellets increase the intensity of otherwise hard-to-resolve peaks in the Raman spectra[102].

**Terahertz Kerr Effect (TKE) Setup.** The full THz set-up is pictured in **Fig. S28**. A 1 kHz, 38 fs, 13 mJ pulse centered at 800 nm is produced via regenerative amplification in a Ti:Sapphire laser (Coherent Legend Elite Duo). The output beam was split by a 25:75 beam splitter, with the more intense ~ 10 mJ beam used to pump an optical parametric amplifier (OPA, TOPAS, Light Conversion Inc.), generating a 2.1 W, 1440 nm output signal beam. The signal beam travelled through a mechanical delay stage and a chopper set at 250 Hz to reduce the average power to 1.1 W to not burn the DAST (4-*N*,*N*-dimethylamino-4′-*N*′-methyl-stilbazolium tosylate) crystal, which was purchased from Swiss Terahertz. The μW output of the DAST crystal, was focused through three OAPs into a ~250 μm beam spot, providing a peak field intensity in the 100 kV/cm range. The chamber was purged under inert gas ($N_2$) to minimize water absorption and to increase the field strength of the THz source. The THz field was characterized using electro-optical detection in 30 μm GaP[103]. The time-domain trace from this generation set-up is found in **Fig. S10**. We note that there is a water signal in the time-domain trace due to the ~10% relative humidity inside of the purge box. The 10 dB bandwidth of the generated light is 0.7 - 4.5 THz[104], as noted in **Fig. 5** in the main manuscript.

For the LUIS measurements, the same beam path to **Fig. S28** was used, however, the chopper was set to 500 Hz, and the nonlinear organic crystal used was 4-N,N-dimethylamino-4'-N'-methyl-stilbazolium 2,4,6-trimethylbenzenesulfonate (DSTMS). The DSTMS crystal was pumped with 820 mW of 1300 nm light (410 mW after chopping) generated by an OPA, and the signal was charcterized using electro-optic sampling, with 100 $\mu$m GaP. The time trace and frequency character of the signal are seen in **Fig. S29**. The peak field strength of the THz field for these measurements was 744 kV/cm.

**IR power to THz field strength.** Terahertz pulse energies were measured using a calibrated pyroelectric THz joulemeter (Genctec EO SDX 1211). The subsequent THz field strength values were used to calculate an average power for comparison to the non-resonant heating.

**THz average power calculation.**

$$\text{Pulse energy (J)} = \frac{\text{Voltage output (V)}}{614000\ (\frac{\text{V}}{\text{J}})} \quad (22)$$

$$\text{Avg. power (W)} = \text{J} * \text{Rep. rate (Hz)} \quad (23)$$

Using a pyroelectric THz joulemeter, a measured voltage was used to calculate the pulse energy of the THz field using **Eq. 22** where 614000 V/J is the conversion factor according to the manufacturer specifications. The average power of the laser was then calculated using **Eq. 23** where J is the pulse energy and the repetition rate of the pulses after passing the chopper is 250 Hz. The conversion between THz field strength and laser power (in mW) is seen in **Table S9**.

**Near-IR (NIR) to Mid-IR (MIR) light set-up.** The NIR and MIR light sources were generated using a regenerative Ti:Sapphire laser amplifier operating at 1 kHz. The 800 nm output was sent through an optical parametric amplifier for the NIR light and a difference frequency generation unit for the MIR light.

**THz Absorption Spectrometer.** The Terahertz (THz) absorption spectrum shown in **Fig. 5d** was acquired using a home-built transmission-geometry THz time domain spectrometer (THz-TDS) using optical light (Coherent Astrella 800 nm, 35 fs pulse width, 1 kHz rep-rate Ti:Sapphire regenerative amplifier) to illuminate a spintronic THz emitter (TeraSpinTech T-Spin2) to generate p-polarized THz radiation, which is passed through a wire grid polarizer to assure polarization purity. Generated THz light is subsequently collimated and refocused down onto the sample position using a train of off-axis parabolic mirrors (OAPs). In the sample position, we mount the sample to a translation stage to alternate the acquisition of reference (dry nitrogen) spectra and sample spectra by moving the sample in and out of the sample position to account for any systematic drift through the data-acquisition process. After passing through the sample position, THz light is once again collimated and focused onto a GaP THz detector crystal (Swiss THz, 100 μm thick active cut optically contacted onto 1 mm thick inactive cut) where it is profiled by electro-optic sampling. In this process, the arrival time of a linearly polarized optical probe (800 nm, 35 fs) is scanned using a mechanical delay stage. As the arrival time of the two pulses is scanned, the polarization of the optical probe is rotated due to the transient birefringence induced in the detector crystal by the THz pulse. This polarization shift is then detected in a pair of balanced photodiodes[105]. The combination of the THz emitter and detector detailed here results in a calculated field strength of ~15 kV/cm, and a 10 dB bandwidth of 0.5 - 7.3 THz[104]. We present the absorption spectrum in the range of 0.7 - 4.5 THz in **Fig. 5d** to match the bandwidth of the higher field strength radiation that was used to drive the rocking modes in LLTO. Finally, the THz-TDS spectrometer is fully enclosed in a dry nitrogen purge box and is monitored with a hygrometer to assure that all spectra are taken in a dry environment.

**Interpretation of THz-TDS.** Upon capturing the time trace of the THz pulses for our reference (dry nitrogen) and sample (pressed LLTO:PTFE pellet), we take the fast Fourier transform (FFT) to get the complex frequency spectra $E_{ref}(\omega)$ and $E_s(\omega)$, respectively. By dividing the sample spectrum by the reference, we arrive at the transfer function $H(\omega)$ defined below[106]:

$$\frac{E_s(\omega)}{E_{ref}(\omega)} = H(\omega) = \frac{4n_s(\omega)}{[n_s(\omega)+1]^2} * e^{-k_s(\omega)\frac{\omega l}{c}} e^{-j[n_s(\omega)-1]\frac{\omega l}{c}} \quad (24)$$

Where $l$ is the sample thickness, c is the speed of light, and $n_s(\omega)$ and $k_s(\omega)$ are the real and imaginary refractive index of the sample, respectively. By taking the argument and logarithm of the transfer function, we arrive at the following two equations:

$$\angle H(\omega) = -[n_s(\omega)-1]\frac{\omega l}{c} \quad (25)$$

$$|H(\omega)| = ln\left[\frac{4n_s(\omega)}{[n_s(\omega)+1]^2}\right] \quad (26)$$

From these, we then extract the optical constants and related absorption coefficient directly from the transfer function as shown below[106]:

$$n_s(\omega) = 1 - c\angle H(\omega) \quad (27)$$

$$k_s(\omega) = cl\left\{ln\left[\frac{4n_s(\omega)}{[n_s(\omega)+1]^2}\right] - \ln|H(\omega)|\right\} \quad (28)$$

$$\alpha_s(\omega) = \frac{2\omega k_s(\omega)}{c} \quad (29)$$

By comparing the pure absorbance spectra of ball-milled PTFE with that of the dilute LLTO in PTFE mixed pellets and assuming linear contributions to absorption by concentration, it was possible to extract the pure LLTO absorbance spectra.

**THz-TDS Sample Preparation.** To prepare the pellet samples for THz-transmission data, LLTO powder was loaded into a ball-mill grinder alongside the THz transparent polymer PTFE (teflon) in a known concentration and allowed to mix for 20 minutes. PTFE was chosen as the dilutant polymer due to its low porosity in pressed pellets, high tensile strength, and low propensity to capping defects at high pressure[107]. Following the mixing process, the PTFE and LLTO grains shared an average sub-wavelength, size of ~1 μm, measured with scanning electron microscopy. Following mixing in the ball-mill, pellets were pressed for 1 minute under 1 ton of pressure**.** To account for scattering effects, multiple pellets of varying concentrations of LLTO in PTFE, ranging from 0% to 7.5%, were analyzed.

**LUIS.** In LUIS, a 19 GHz, -5 dBm AC carrier is utilized to initiate ion hopping under a fast-oscillating electric field. Then, an ultrafast laser, as discussed in the TKE Setup and NIR-MIR light sections of the **Methods**, is used to excite the sample, inducing a change in ion hopping. Due to the impedance mismatch at the LLTO sample, there is a modulated reflection of the AC carrier signal due to the photoexcitation. The GHz signal is carried via coaxial cables and routed to the sample and the oscilloscope, using a hybrid coupler. Averaging, amplitude demodulation, and a root-mean-square (RMS) envelope are used to extract the signal (**Figs. S16a-b and S17a-b**). This modulated response is captured using a real-time oscilloscope, which samples at a rate of 128 gigasamples per second (every 7.8 ps), and, at shorter timescales, applies further subsampling. This setup is depicted in **Fig. S15** and the sample holder is detailed in previous work[47]. We fit the response to a gaussian convolved with an exponential decay, where the width of the gaussian represents our overall response time and the exponential decay characterizes the laser-induced dynamics that we measure. This is outlined below:

$$\mathrm{S(t)} = \int_{-\infty}^{t} \mathrm{G(t')E(t-t')dt'} \quad (30)$$

$$\mathrm{G(t)} = \mathrm{A} * \exp\left(-\frac{(t'-t_0)^2}{2\sigma^2}\right) \quad (31)$$

$$\mathrm{E(t)} = \mathrm{B} * \mathrm{H(t-t')} * \exp\left(-\frac{(t-t')}{v}\right) \quad (32)$$

$$\mathrm{S(t)} = \mathrm{C} * \exp\left(\frac{\sigma^2}{2{\tau_d}^2} - \frac{(t-t_0)}{2\tau_d}\right) * \left(1 + \mathrm{erf}\left(\frac{t - t_0 - \frac{\sigma^2}{2\tau_d}}{\sqrt{2}\sigma}\right)\right) \quad (33)$$

Where C is the fitted amplitude, $t_0$ is the time at which the laser excites the sample, $\sigma$ is the system response and width of the gaussian, and $\tau_d$ represents the characteristic decay time of the laser-induced dynamics.

**Acknowledgements**

FT-IR data was collected at the Laser Resource Center in the Beckman Institute of the California Institute of Technology with the assistance of Dr. Jay Winkler. Solid-state UV-VIS data was collected at the Earle M. Jorgenson Laboratory of the California Institute of Technology with the assistance of Dr. Weilai Yu. We thank Prof. David Hsieh and Dr. Omar Mehio for assistance in using the difference frequency generation unit for IR light. We thank Ricardo Zarazua at the Chemistry and Chemical Engineering Machine Shop for machining the custom heating cell in this work. We thank Dr. Zachery W. B. Iton for assistance with the characterization of LLTO. We thank Jadon Bienz for assistance with analysis of the XRD data including multi-phase Rietveld refinements of LLTO. We thank Jax Dallas for generating the THz light for the THz absorption and ultrafast impedance measurements. Use of the Advanced Photon Source at Argonne National Laboratory was supported by the U. S. Department of Energy, Office of Science, Office of Basic Energy Sciences, under Contract No.DE-AC02-06CH1135. The finite-difference time-domain heating model computations were conducted in the Resnick High Performance Computing Center at the Resnick Sustainability Institute at the California Institute of Technology.

**Funding**

KHP acknowledges financial support from the National Science Foundation, Air Force Office of Science & Research, and David & Lucile Packard Foundation. AKL acknowledges support by the California Institute of Technology through the Beckman-Gray Graduate Fellowship and the Natural Sciences and Engineering Research Council of Canada Graduate Research Scholarship Doctoral Program. NAS acknowledges support by the U.S. Army Research Office grant number W911NF-23-1-0001 and the National Defense Science and Engineering Graduate Fellowship.

**Author Contributions**

S.K.C. and K.A.S. supervised the project. S.K.C. and K.A.S. conceived the work. S.K.C., K.A.S., and K.H.P. designed the experiments. K.H.P. developed and performed the laser-driven EIS experiments. K.H.P., A.K.L., and N.A.S. synthesized and characterized the material. J.M.M. and H.L. generated the THz radiation, characterized the THz spectrum, and supported the integration of THz with the impedance technique. N.A.S. and A.K.L, performed the THz absorption and ultrafast impedance measurements. K.G. and D.V. performed the atomistic simulations. N.A.S. performed the heating model simulations. K.H.P., K.G., and N.A.S., wrote the manuscript. S.K.C., K.A.S., G.B., A.K.L., D.V., A.H., and Y.S.H. edited the manuscript.

**Data Availability Statement**

All data supporting the findings of this work are available in the Supplementary Information. The computational data that supports the findings of this study can be obtained from the corresponding author on request.

**Code Availability Statement**

The code is available under the MIT license and can be found at https://github.com/kgordiz/modalNEB.

**Ethics Declaration**

Competing interests

The authors declare no competing interests

Supporting Information for

# Correlated terahertz phonon-ion interactions control ion conduction in a solid electrolyte

Kim H. Pham[1*], Kiarash Gordiz[2*], Natan A. Spear[3*], Amy K. Lin[1], Jonathan M. Michelsen[1], Hanzhe Liu[1], Daniele Vivona[2], Geoffrey A. Blake[4], Yang Shao-Horn[2,5], Asegun Henry[2], Kimberly A. See[1], Scott K. Cushing[1¶]

[*]These authors contributed equally to this work, [¶]Corresponding author

[1]Division of Chemistry and Chemical Engineering, California Institute of Technology, Pasadena, CA, USA.
[2]Department of Mechanical Engineering, Massachusetts Institute of Technology, Cambridge, MA, USA.
[3]Department of Applied Physics and Materials Science, California Institute of Technology, Pasadena, CA, USA. [4]Division of Geological and Planetary Sciences, California Institute of Technology, Pasadena, CA
[5]Research Laboratory of Electronics, Massachusetts Institute of Technology, Cambridge, MA, USA.
*Correspondence to: scushing@caltech.edu

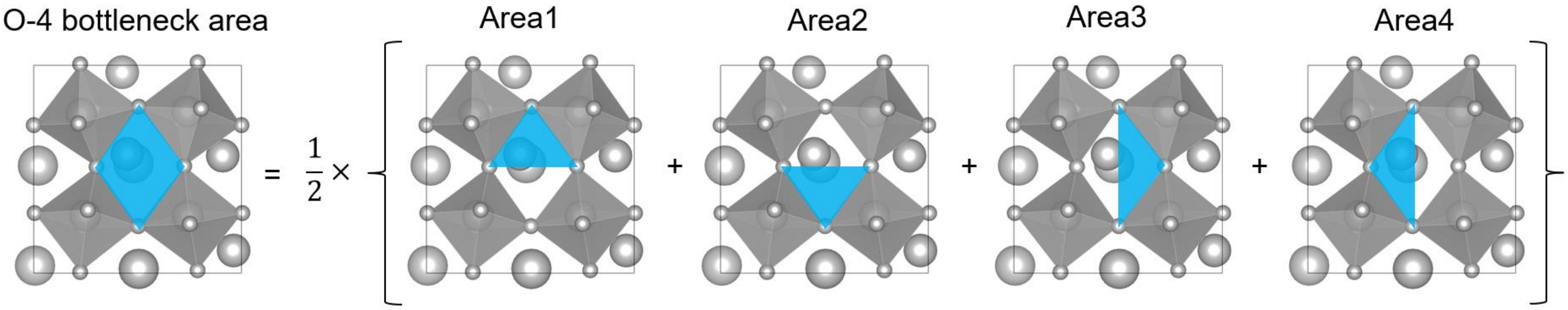

**Figure S1.** The O-4 bottleneck area in space is approximated by averaging over four smaller triangular areas (shown by blue).

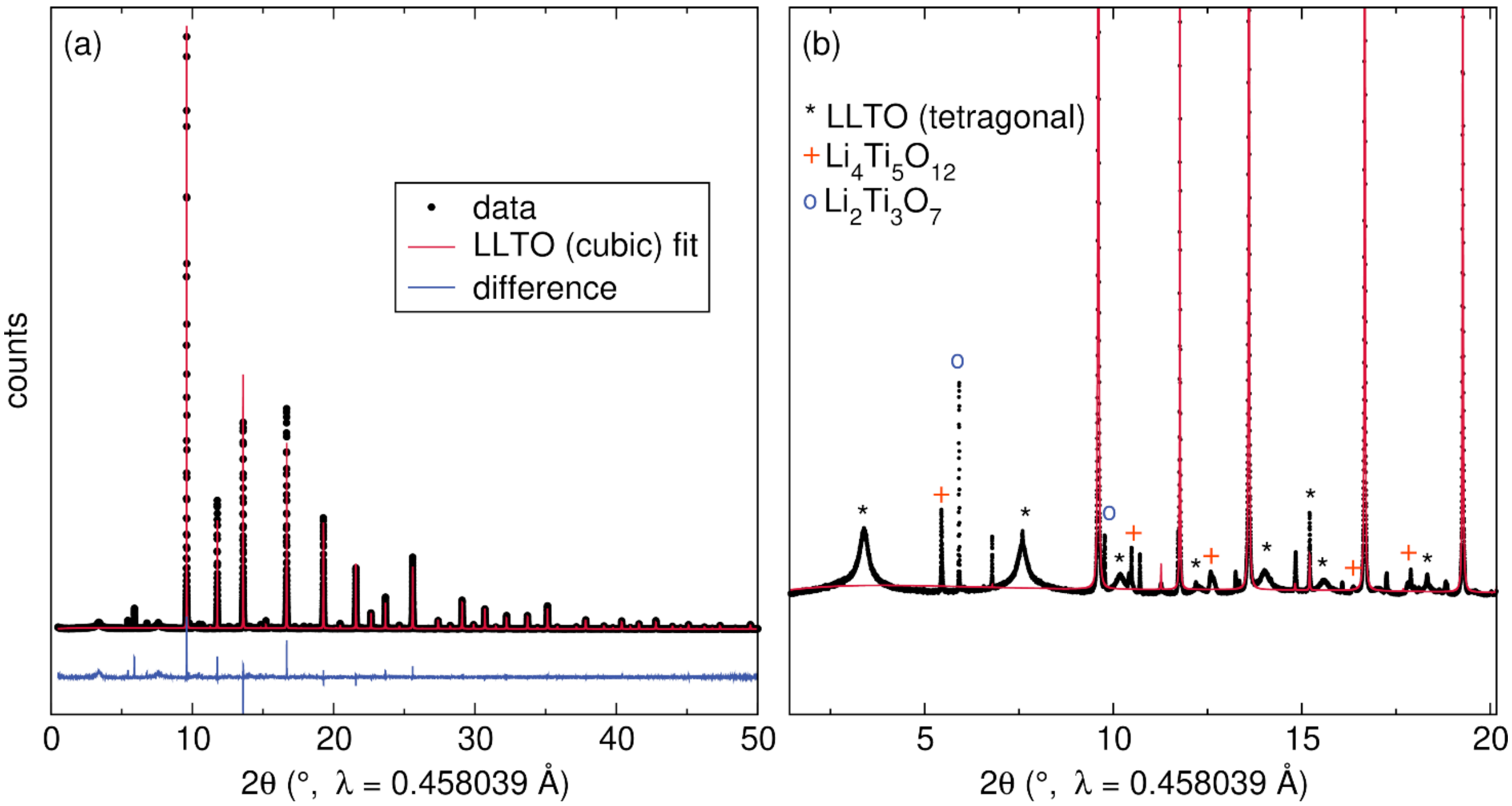

**Figure S2.** Synchrotron X-ray diffraction of LLTO. (a) Quantitative Rietveld refinement to the cubic phase of LLTO describes the data well. (b) The small peaks at low angle are associated with both the tetragonal phase of LLTO, which includes ordered La – a common phenomenon[1] – and some Li-Ti-O impurity phases. A four-phase fit to the data (not shown) suggests the impurity phases are much less than 8 wt% of the material. Additionally, no impurity phases are visible in the Raman spectrum (**Fig. 5**).

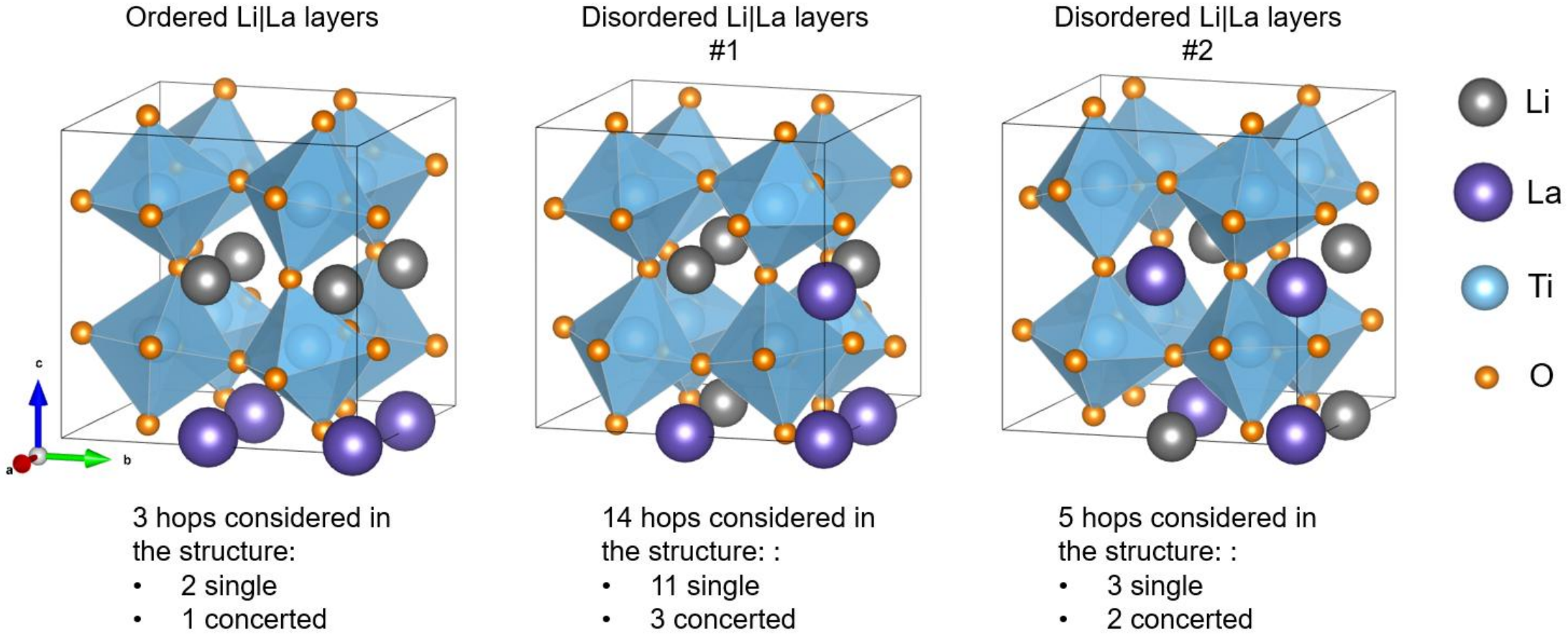

**Figure S3.** Schematics of three different Li|La orderings with respect to the c-axis in the $Li_{0.5}La_{0.5}TiO_3$ lattice included in our calculations. The chosen number of $Li^+$ ion hops with different hopping mechanisms are also included in the bottom of the schematics.

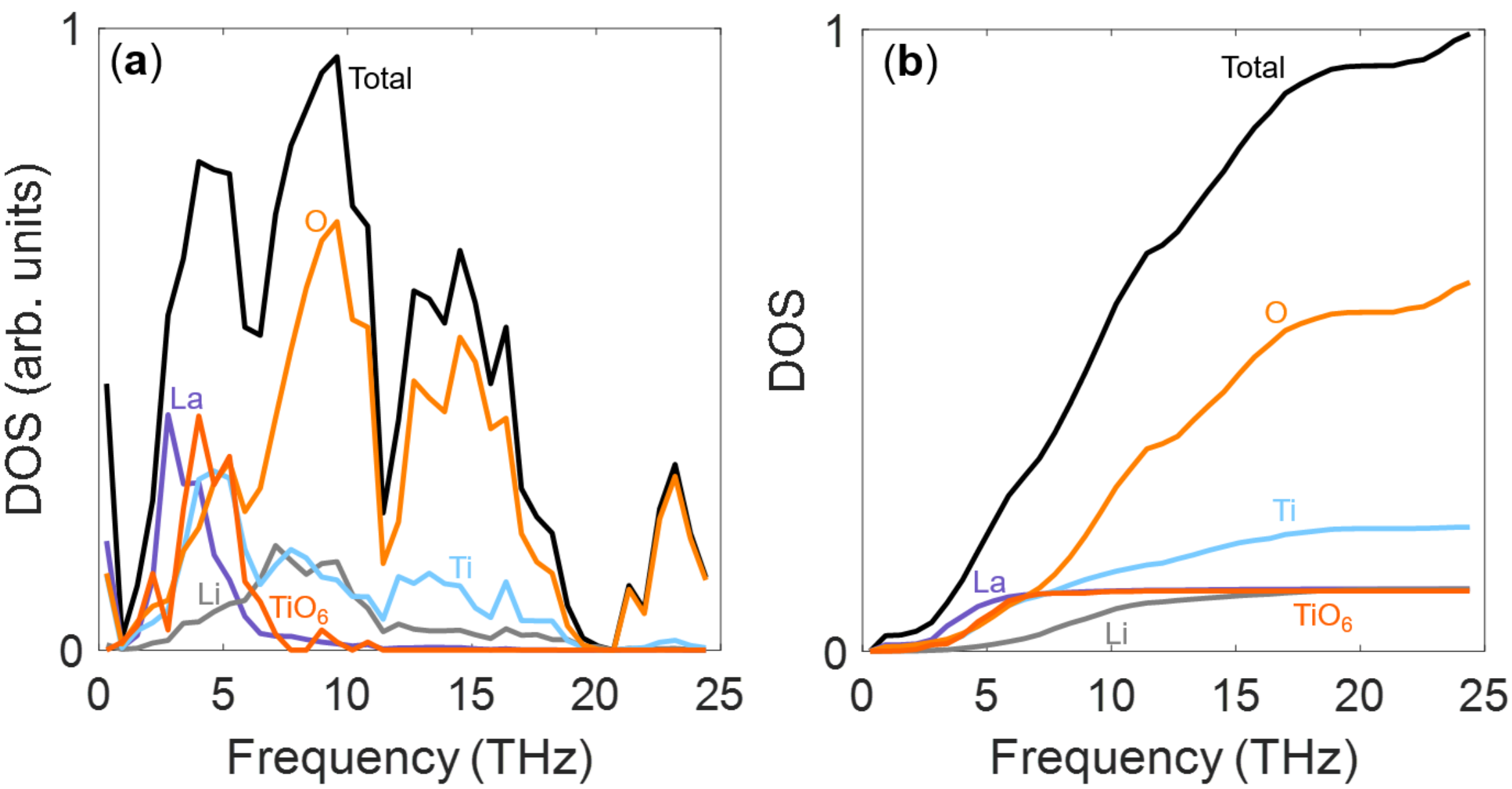

**Figure S4.** Total and partial phonon density of states in (a) non-accumulation and (b) accumulation format for different atomic species and rocking modes ($TiO_6$ units), showing that that 10 % of the total modes are of $TiO_6$ rocking nature, and in the <6THz region, 36% of the modes are of $TiO_6$ rocking nature.

**Table S1.** Enhancement in the ion hopping rate by targeted excitation of the highest contributing non-rocking mode, highest contributing rocking mode, and one random mode, that have frequencies in the experimentally accessible frequency range, during MD simulations. The targeted excitation of phonons during MD simulation is explained in the **Methods** and the method is applied to three hops among the hops investigated in this study. The visualization of these hops is provided in **Fig. S5**. The numbers included in the table are ion hopping rates (#jumps/ps). Following the sorted phonon contributions shown in **Fig. S7**, almost 50% of the existing phonons in the structure do not have any noticeable contribution to the $Li^+$ ion hop in the lattice, based on which, they can be categized/labeled as “random” as opposed to their “highly contributing” counterparts.

| | | | 400 K (natural MD simulation) | 700 K (natural MD simulation) | 400 K (with mode excited to 700 K) |
|---|---|---|---|---|---|
| Hop #1 | Jump rate (#jumps/ps) | Highest non-rocking-mode contributing mode (4.29 THz) | 0.0011 | 1.45 | 1.27 |
| | | Highest rocking-mode contributing mode (3.46 THz) | | | 1.39 |
| | | Random mode (2.66 THz) | | | 0.0027 |
| Hop #2 | Jump rate (#jumps/ps) | Highest non-rocking-mode contributing mode (4.50 THz) | 0.0024 | 2.07 | 3.61 |
| | | Highest rocking-mode contributing mode (1.84 THz) | | | 3.23 |
| | | Random mode (3.47 THz) | | | 0.0004 |
| Hop #3 | Jump rate (#jumps/ps) | Highest non-rocking-mode contributing mode (5.10 THz) | 0.0017 | 1.44 | 2.09 |
| | | Highest rocking-mode contributing mode (1.94 THz) | | | 1.94 |
| | | Random mode (4.43 THz) | | | 0.0011 |

**Figure S5.** The schematic of the three hops discussed in **Table S1**, with structures at the beginning, during, and end of the hop displayed in both standard view (upper panel) and top view (bottom panel). The “during the hop” representation depicts the hopping atom's different positions during the hop. The bottleneck of the hop is formed by four O atoms (O-4), indicated by black lines in the “during the hop” representation, and an assisted view is provided for better visualization of the bottleneck area. The relative energies corresponding to the minimum energy pathway are displayed in the bottom-right panel.

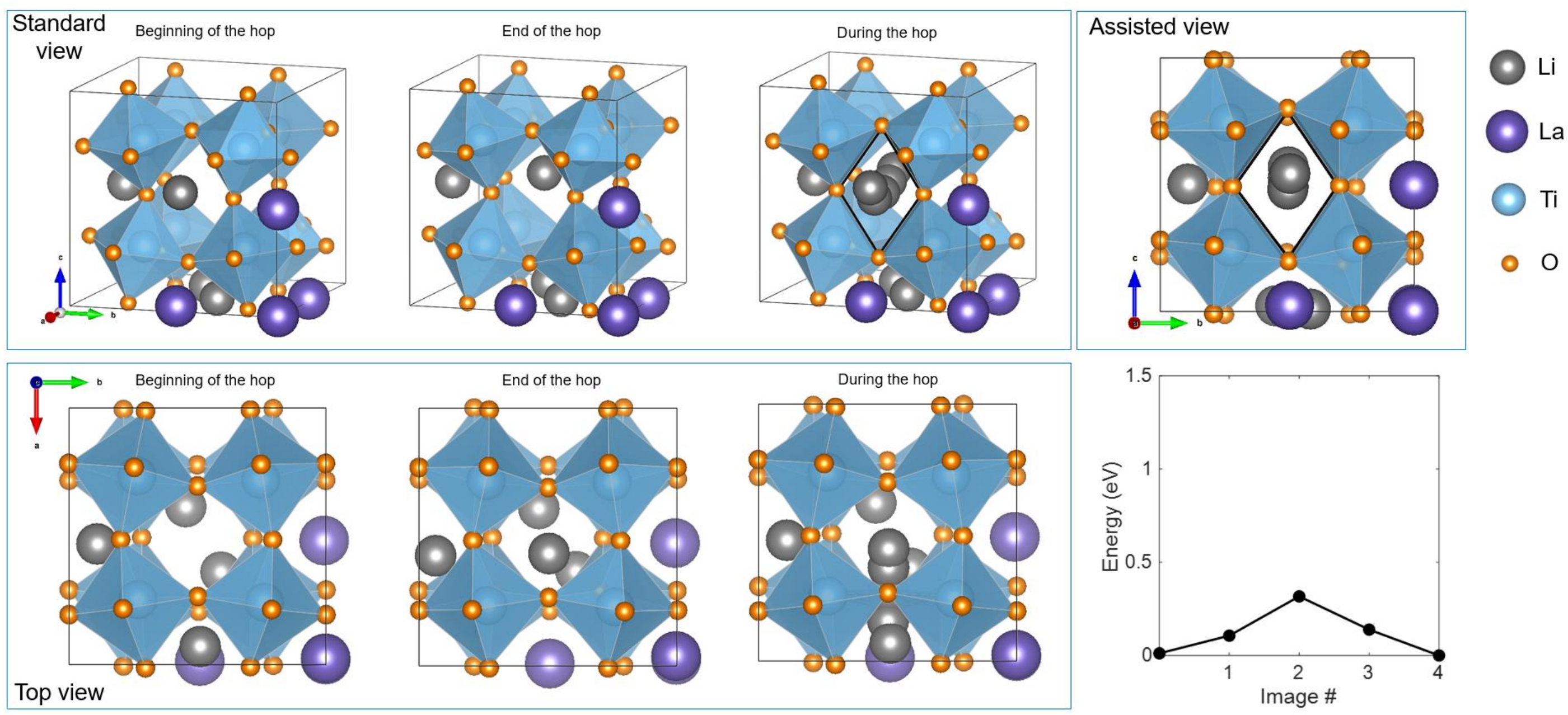

**Figure S5.1.** Representation of Hop #1. The shown hop is a single ion hop, and the structure is derived by repositioning of the Li atoms in the parent disordered structure #1 shown in **Fig. S3**. This hop is also described in **Fig. 2** and **Table 1** in the main manuscript.

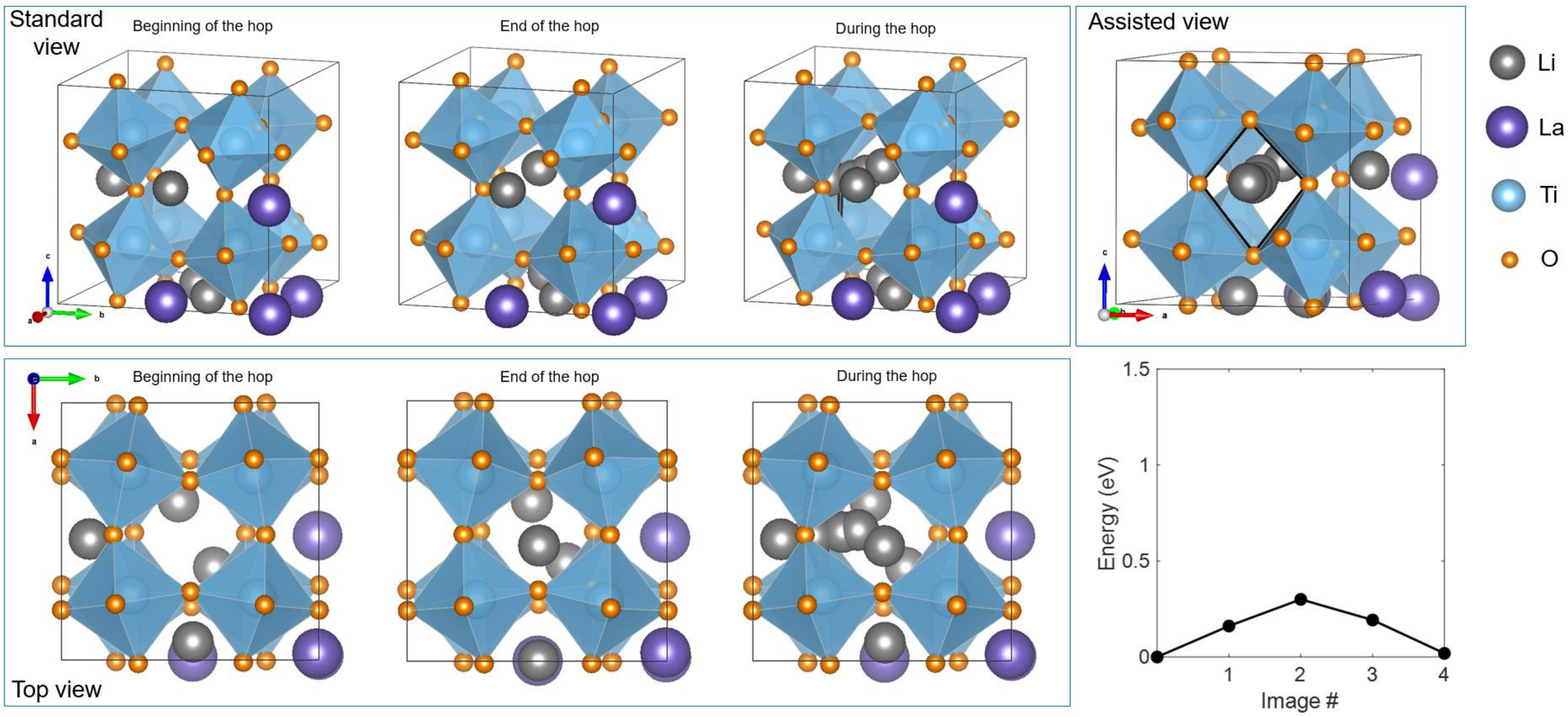

**Figure S5.2.** Representation of Hop #2. The shown hop is a single ion hop, and the structure is derived by repositioning of the Li atoms in the parent disordered structure #1 shown in **Fig. S3**.

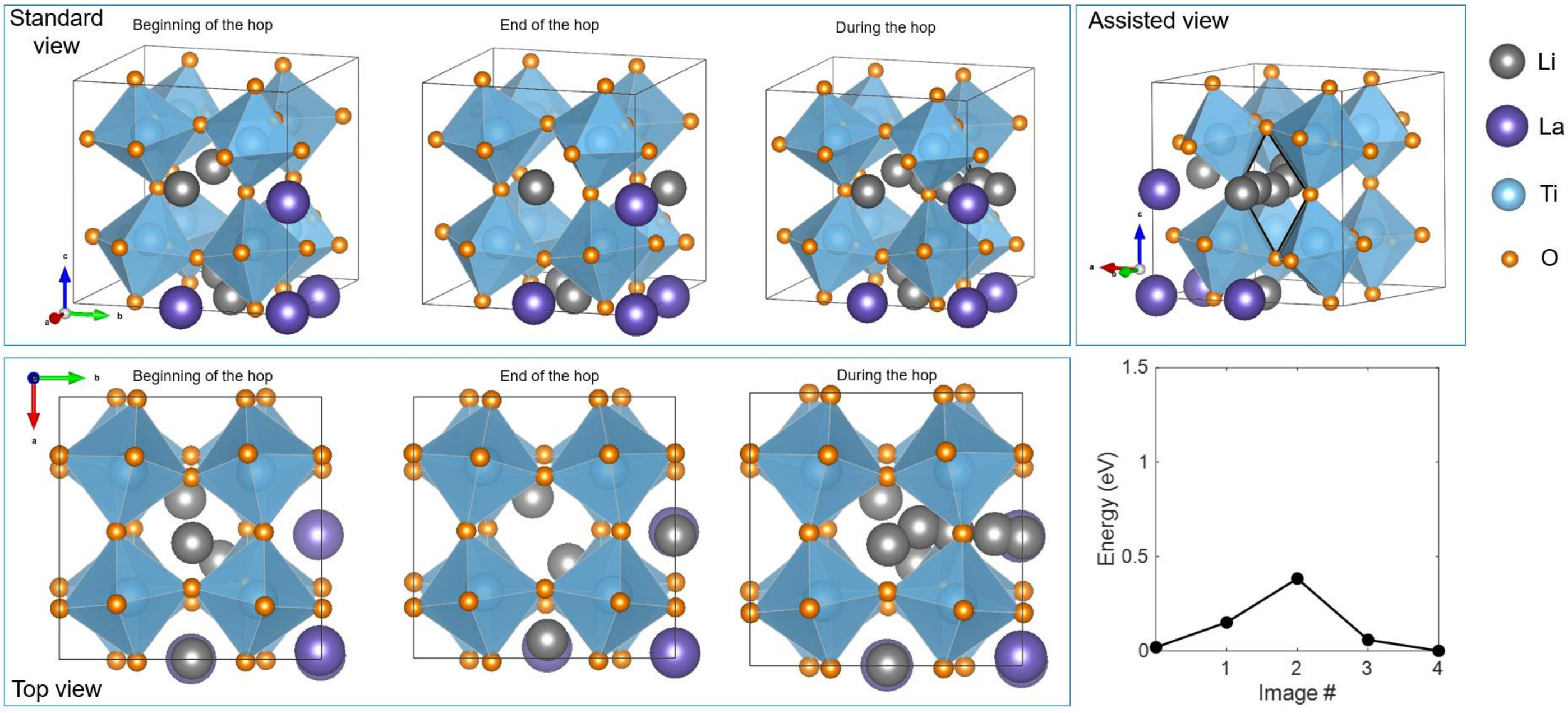

**Figure S5.3.** Representation of Hop #3. The shown hop is a single ion hop, and the structure is derived by repositioning of the Li atoms in the parent disordered structure #1 shown in **Fig. S3**.

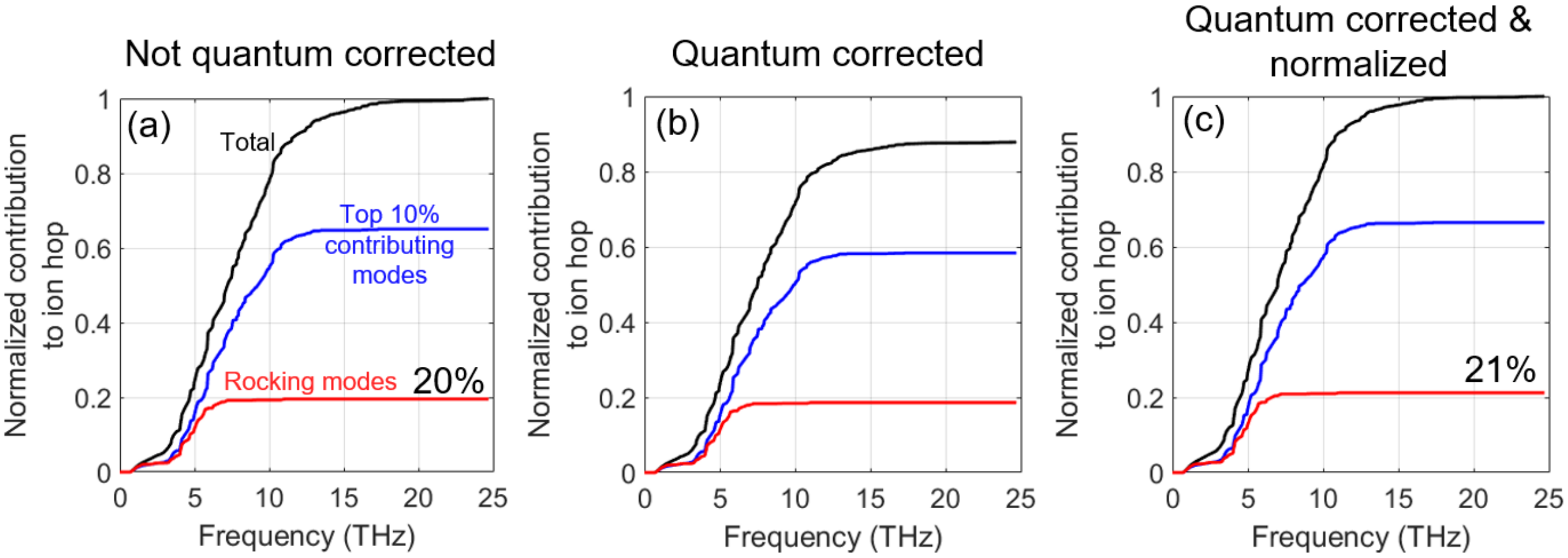

**Figure S6.** Effect of quantum correction on the calculated modal contributions to $Li^+$ hop in the LLTO lattice. The accumulative modal contributions (a) before, and (b) after quantum correction. (c) Shows the accumulative modal contributions after quantum correction, when the calculated modal contributions are normalized. It can be seen that, since most of the $TiO_6$ rocking modes have low frequencies (< 6 THz), correcting their contribution based on Bose Einstein distribution only slightly increases their overall percentage of contribution to $Li^+$ in the LLTO lattice (from 20% to 21%).

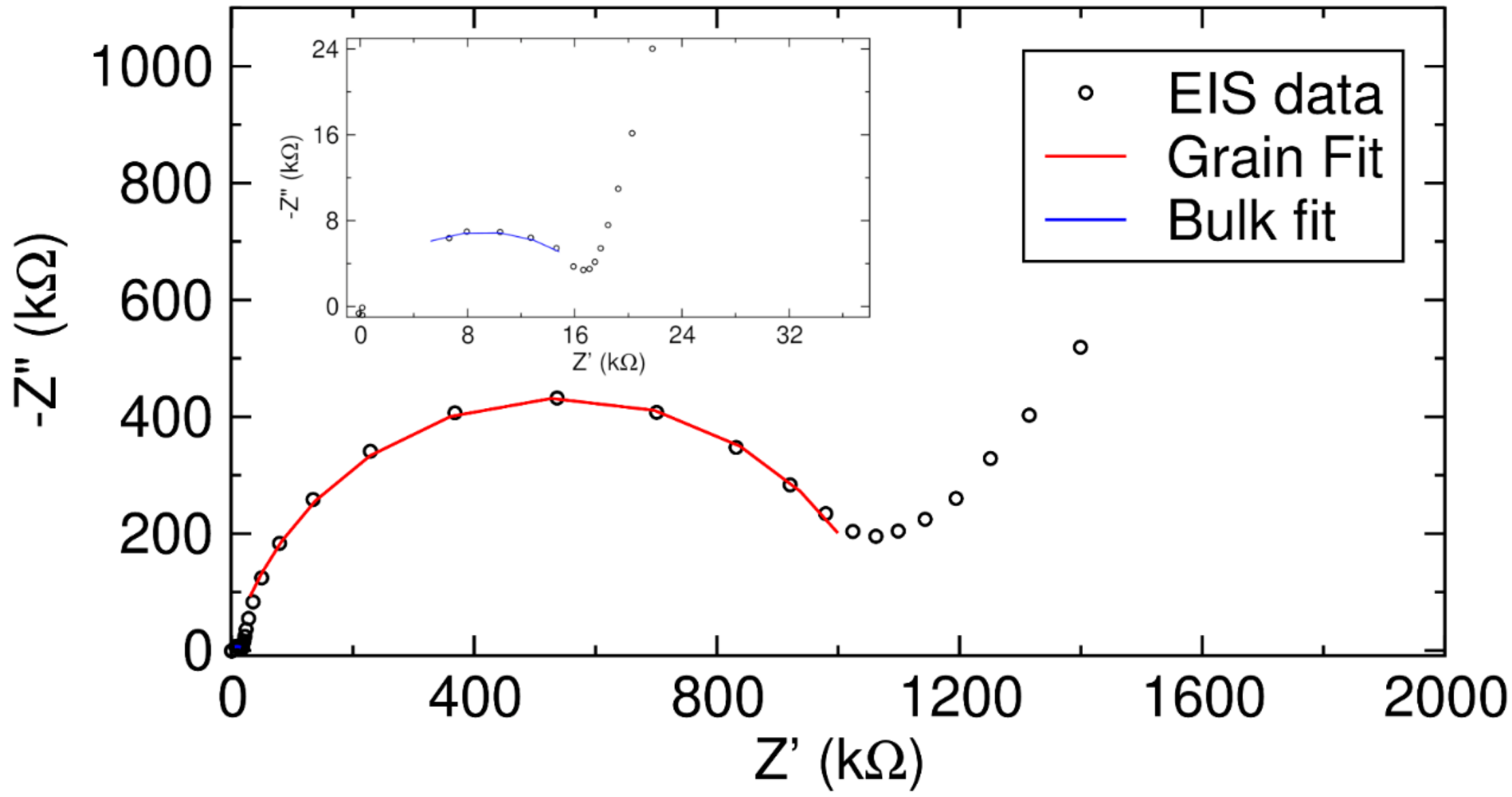

**Figure S7**. Full EIS spectra of LLTO with bulk and grain fits showing the bulk and grain features which are characteristic of LLTO samples, measured between 32 MHz and 1 Hz.

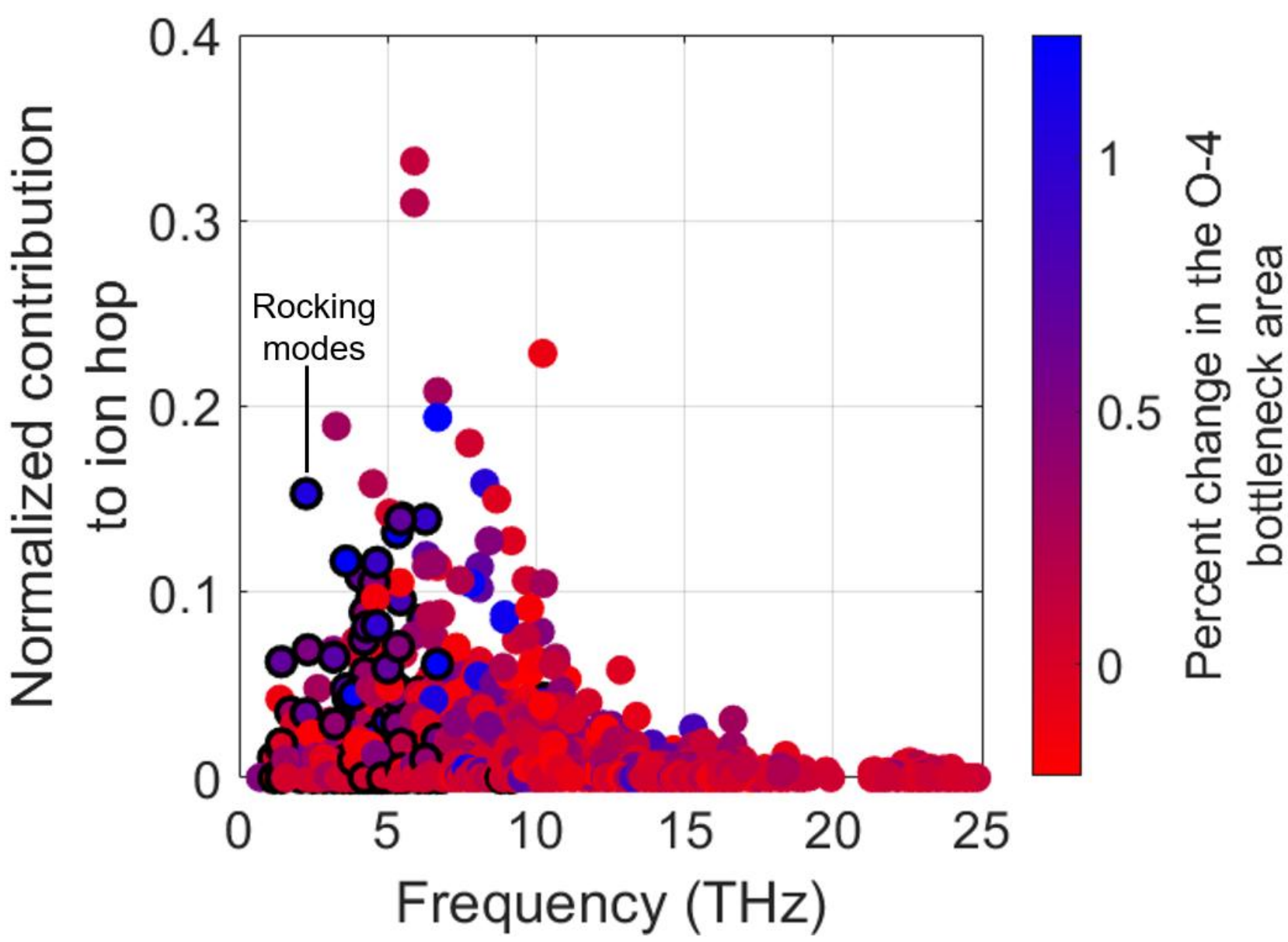

**Figure S8.** Scatter plot displaying the normalized individual contributions of phonons to the $Li^+$ ion hop in the LLTO lattice. The color of each data point represents the extent to which the corresponding phonon can modify the O-4 bottleneck area against the hopping of $Li^+$ ions in the lattice, with the color bar indicating the percentage of change in the bottleneck area, calculated using the methodology explained in **Section S1**. Positive percent changes refer to widening of the bottleneck while negative percent changes refer to shrinkage of the bottleneck. Rocking modes, indicated by black rings surrounding their data points, exhibit a comparatively higher ability to modify the O-4 bottleneck area in comparison to non-rocking modes.

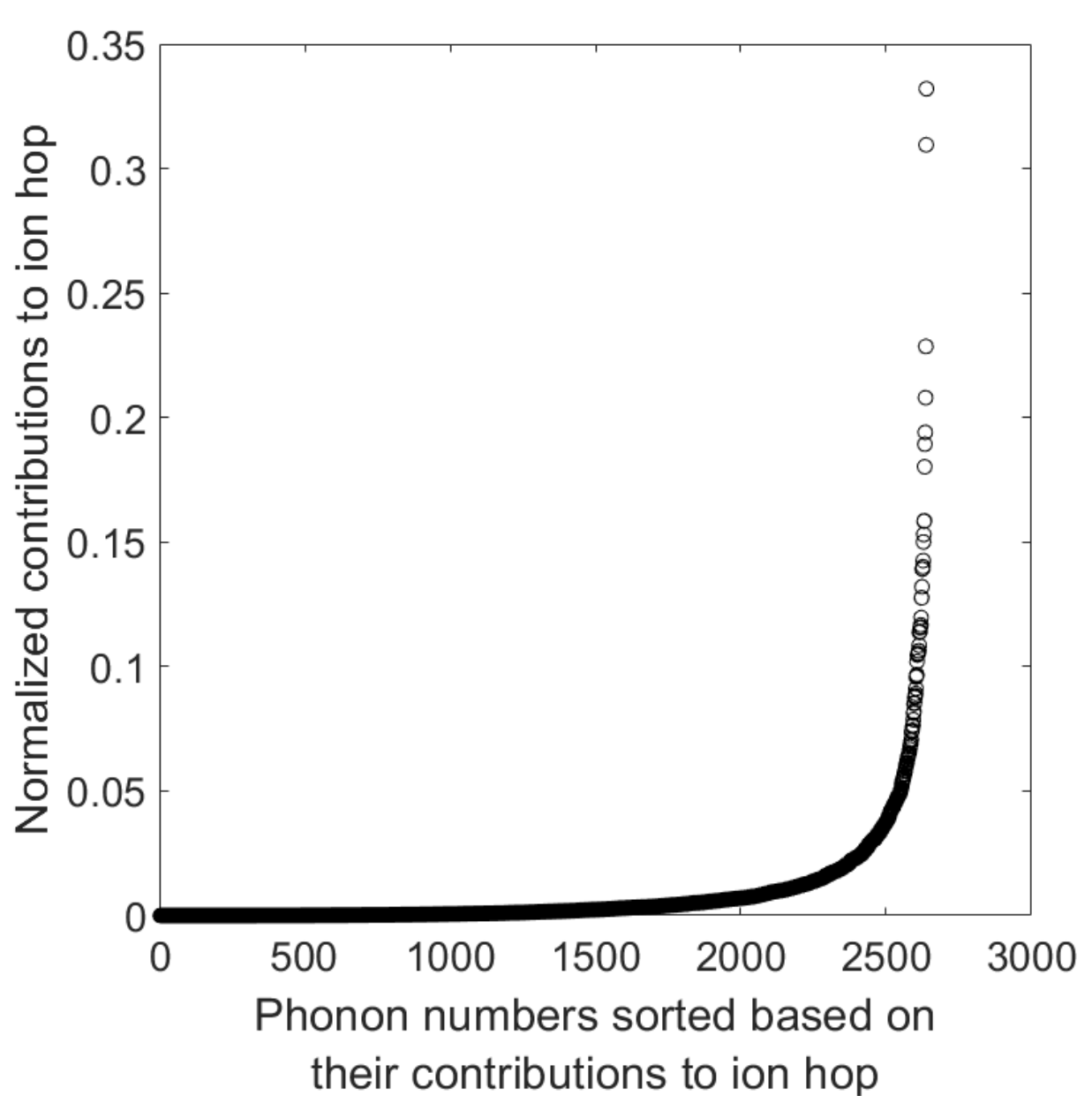

**Figure S9.** Sorted normalized phonon contributions to ion hop. This plot is the same as the scattered plot in **Fig. S6**, except that the phonons are sorted based on their normalized contributions to ion hop. The x-axis goes from 1 to 2640, which is the total number of vibrational modes that the contributions of which were investigated in this study: (22 hops) × (40 atoms) × (3 degrees of freedom for each atom) = 2640.

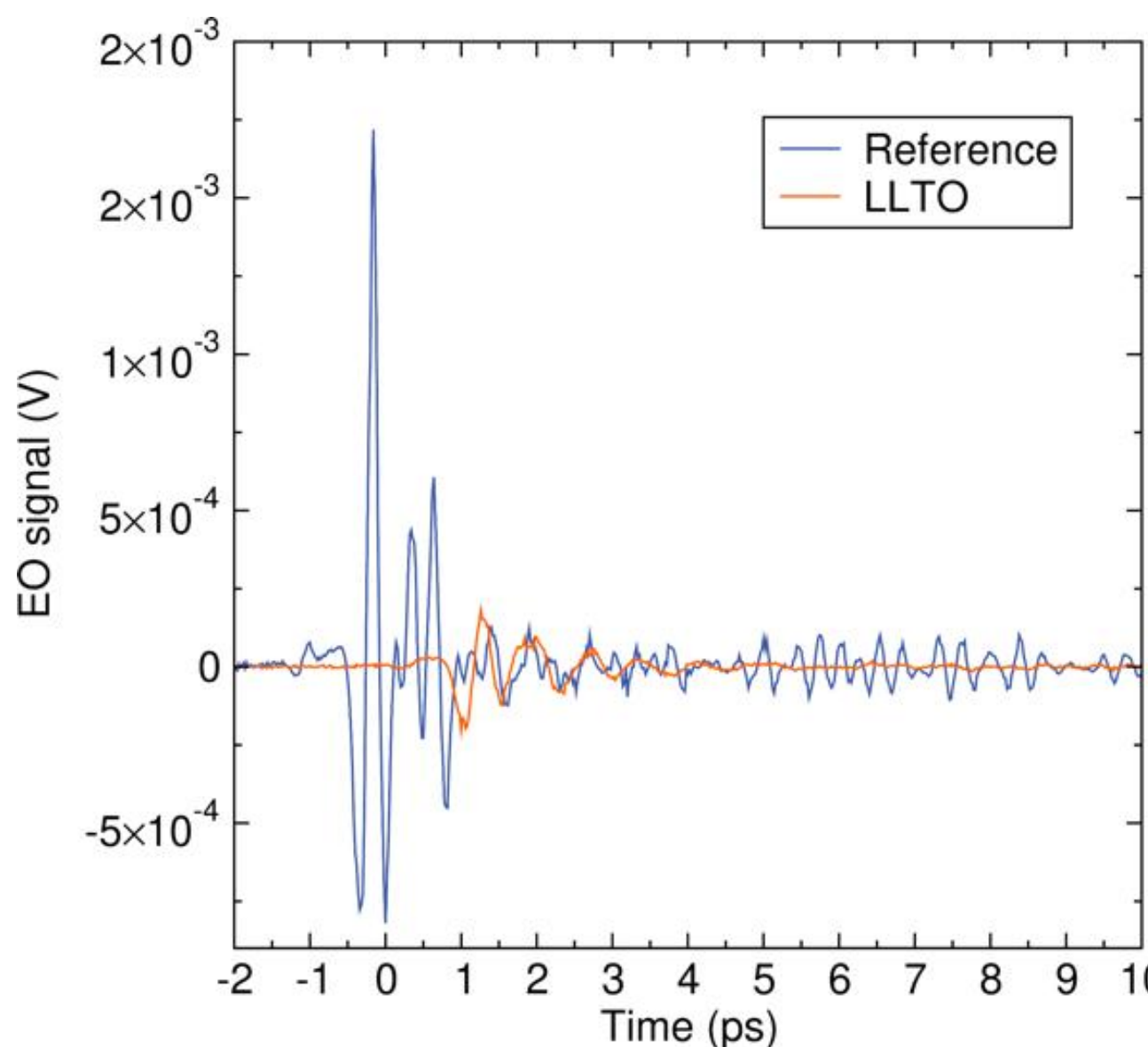

**Figure S10.** THz time domain trace with and without LLTO. The drop in EO signal with the LLTO sample indicates the absorption of the THz field by the sample. The generation process for this signal is described in the **Methods** section.

**Table S2**. The average change and standard deviations for $R_{bulk}$ under THz illumination at each of the measured powers with three trials at each power across four samples. We note the large standard deviations, particularly at low laser power, are due to instability in the laser source coupled with variations in $R_{bulk}$ across different samples and between trials.

| Laser power (mW) | $\Delta R_{bulk}$ (Ω) | Standard deviation in $\Delta R_{bulk}$ (Ω) |
| --- | --- | --- |
| 0.24 | 24.1 | 46.6 |
| 0.57 | 72.8 | 55.6 |
| 0.90 | 114 | 24.7 |
| 1.3 | 195 | 54.9 |
| 1.6 | 258 | 126 |

**Table S3**. The average change and standard deviations for $R_{bulk}$ under 800 nm laser illumination at each of the measured powers with four trials at each power across three samples. We note the large standard deviations stem from instabilities in the laser power and variations in the absolute value of in $R_{bulk}$ across different samples and between trials.

| Laser power (mW) | $\Delta R_{bulk}$ (Ω) | Standard deviation in $\Delta R_{bulk}$ (Ω) |
| --- | --- | --- |
| 5 | 144 | 61.7 |
| 10 | 335 | 192 |
| 15 | 373 | 197 |
| 20 | 582 | 258 |
| 25 | 625 | 306 |

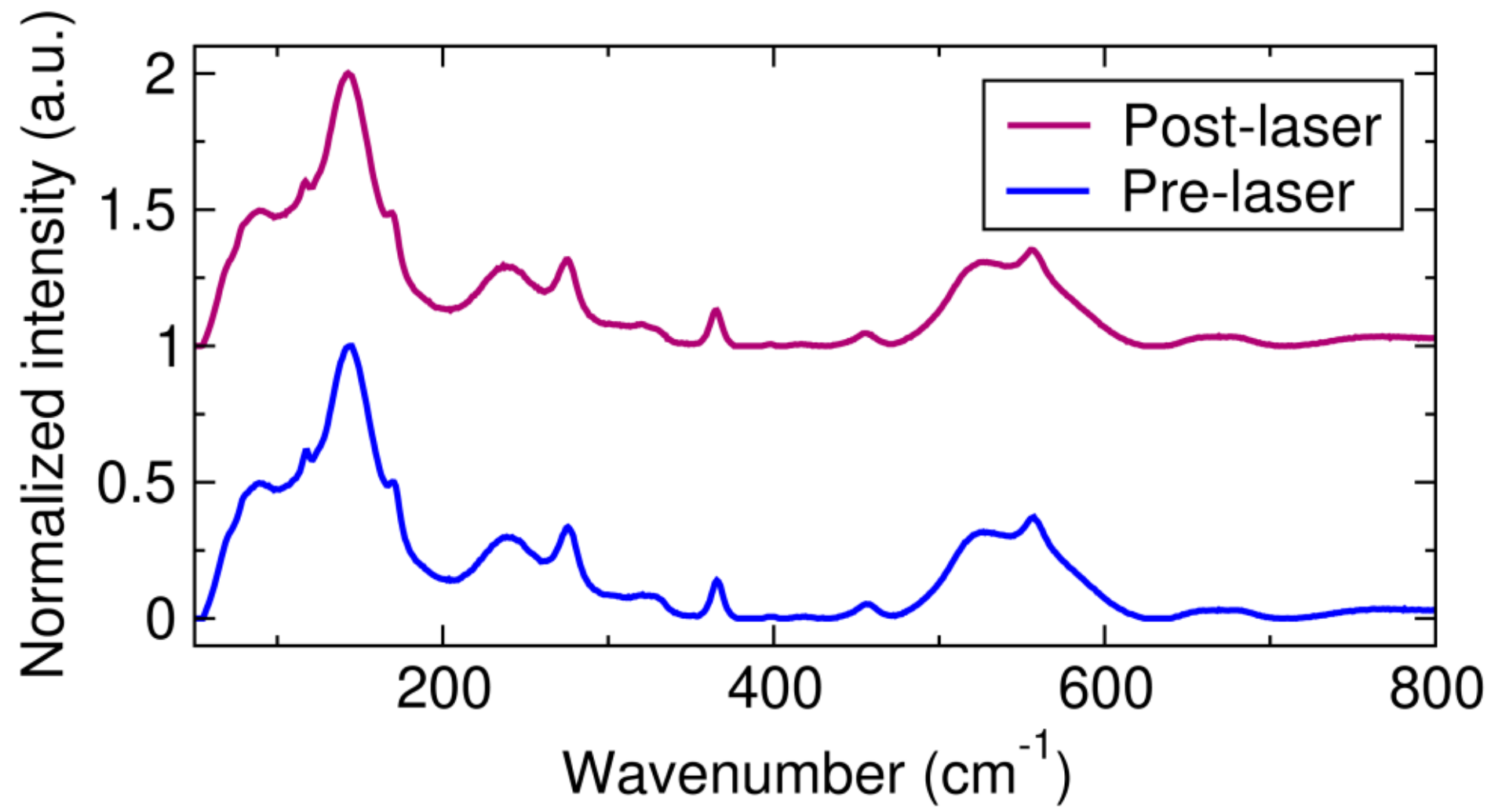

**Figure S11.** Raman spectra of LLTO taken before and after laser excitation on the same spot. One can see that there is no change in the Raman before and after excitation, ruling out sample damage.

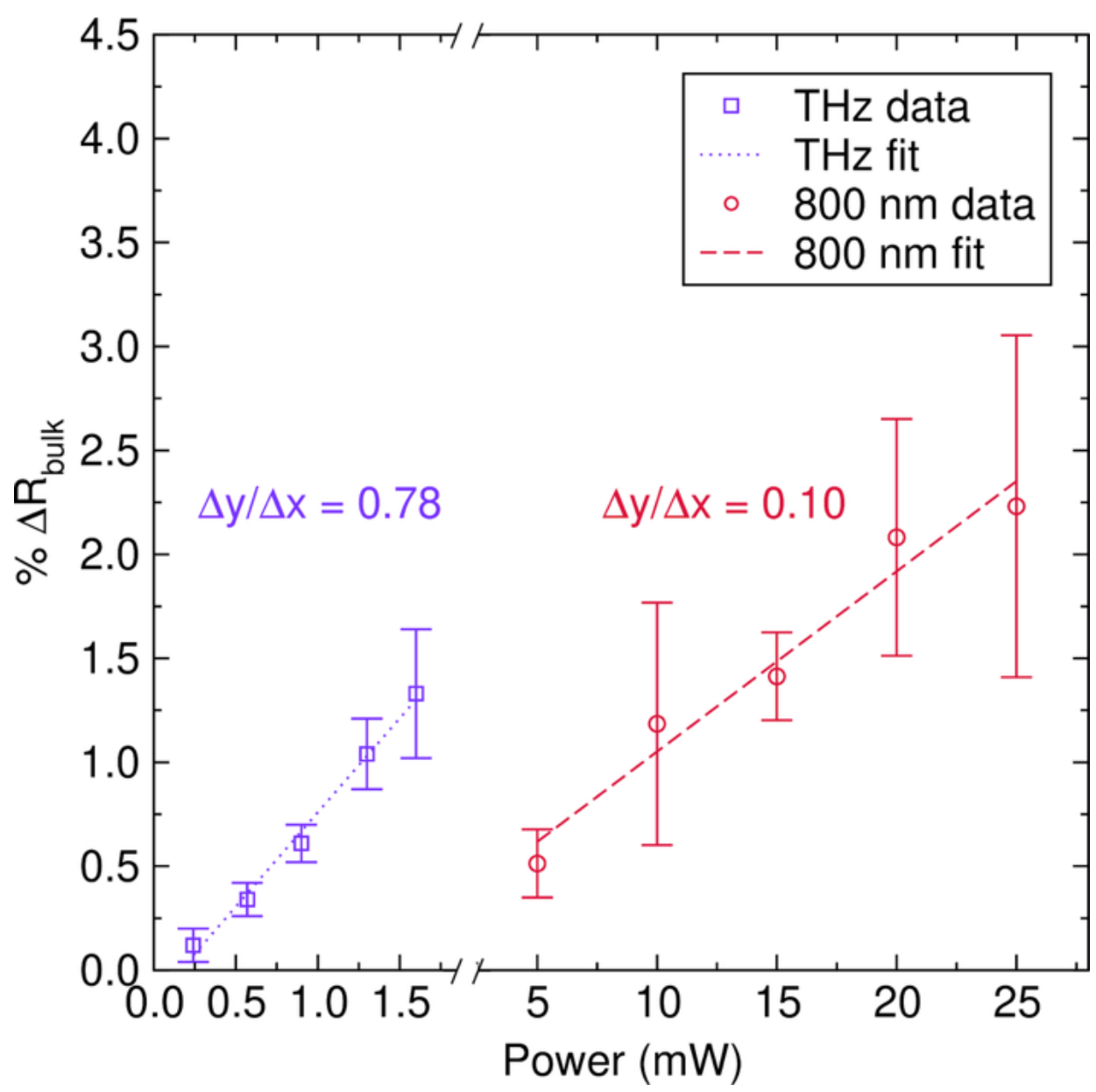

**Figure S12.** The shift in $R_{bulk}$ for 800 nm and THz light excitation. The shifts in impedance from **Fig. 3** plotted against power with equal spot sizes of 250 μm, showing a 0.78% percent decrease in bulk resistance per mW of power for the THz light excitation ($R^2 = 0.9918$) and a 0.10% change in bulk resistance per mW of power for the 800 nm light excitation ($R^2 = 0.9905$).

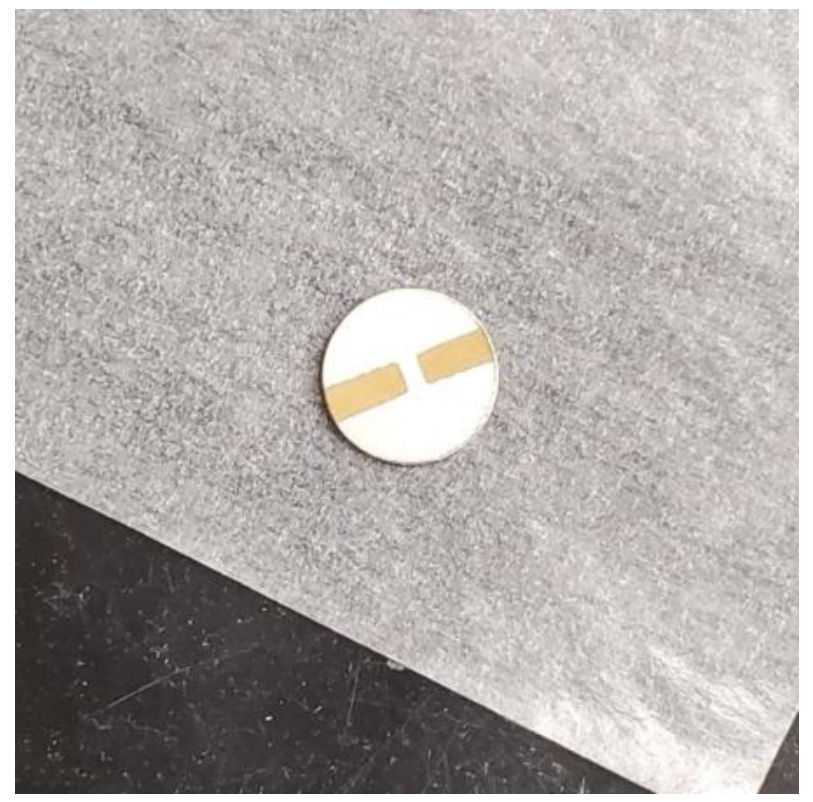

**Figure S13**. The pellets used for laser-driven EIS experiments. The sputtered gold electrodes demonstrate the in-plane electrode geometry where the light excitation was focused directly between the two electrodes.

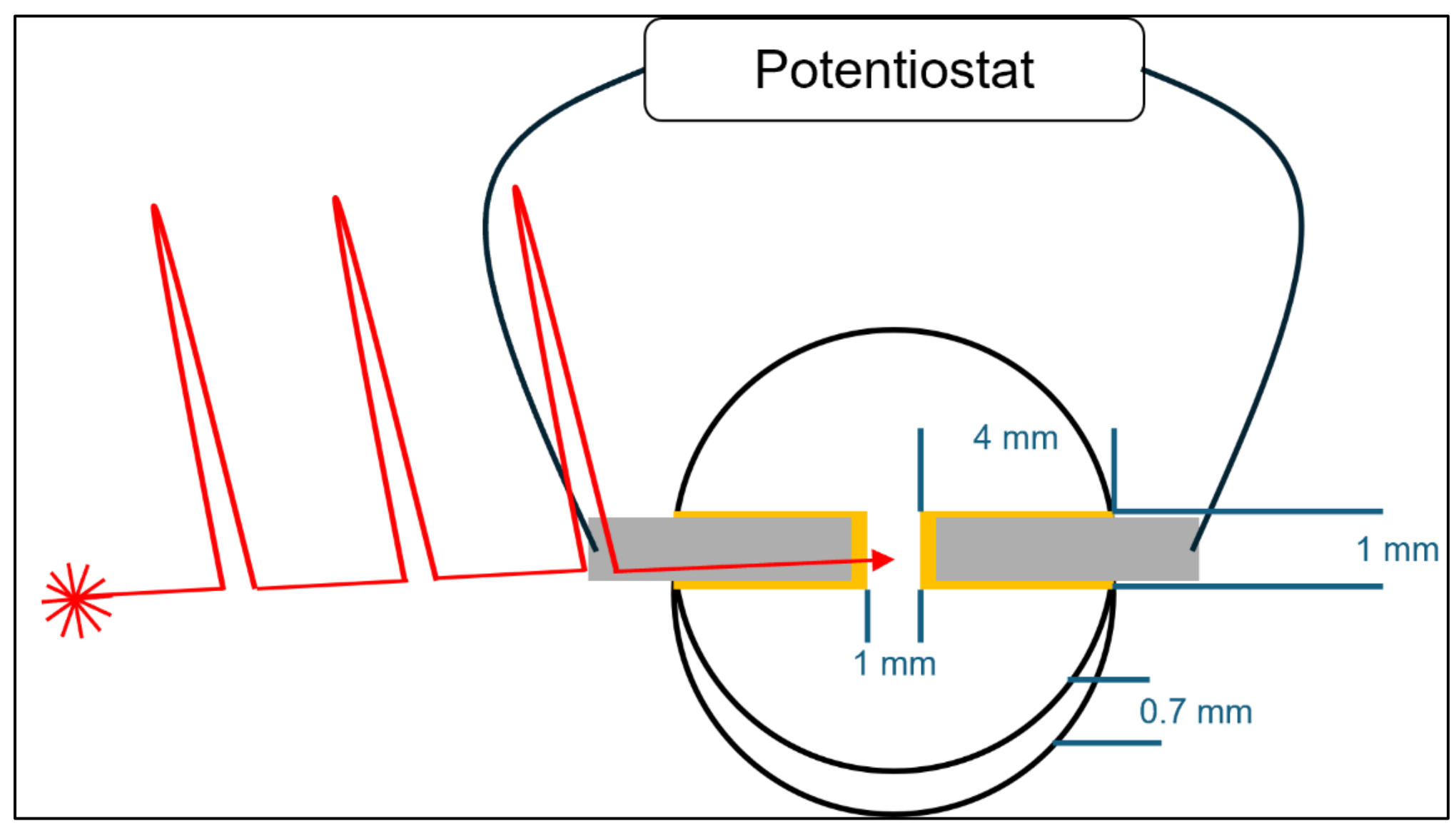

**Figure S14**. A graphical depiction of the laser-driven EIS experiments where the laser (in red) excites a 1 mm gap between two Au-sputtered electrodes which are 4 mm in diameter and 1 mm in width. The sample itself is 0.7 mm thick with a 9 mm radius. The Au-sputtered electrodes make electrochemical contact with stainless steel electrodes that are connected to a potentiostat which records the laser-modulated EIS signal.

**Table S4**. The summary of laser excitation and alternating current (AC) field parameters used for each experiment presented in this work. This incorporates differences in repetition rate and spot size for each experiment.

| Experiment | Wavelength | Laser fluence ($mJ/cm^2$) | Average power density ($mW/cm^2$) | AC Frequency (Hz) | AC Sinus Amplitude (mV) |
|---|---|---|---|---|---|
| Laser-driven EIS | 800 nm | 10.2-50.9 | $1.01*10^4$-$5.09*10^4$ | $1\text{-}32*10^6$ | 50-100 |
| Laser-driven EIS | 0.7-4.5 THz (67-428 μm) | 1.96-13.0 | $488\text{-}3.26*10^3$ | $1\text{-}32*10^6$ | 50-100 |
| LUIS | 800 nm | 16.3 | $8.15*10^3$ | $19*10^9$ | 178 |
| LUIS | 0.5-5.5 THz (54.5-599 μm) | 2.70 | $1.25*10^3$ | $19*10^9$ | 178 |

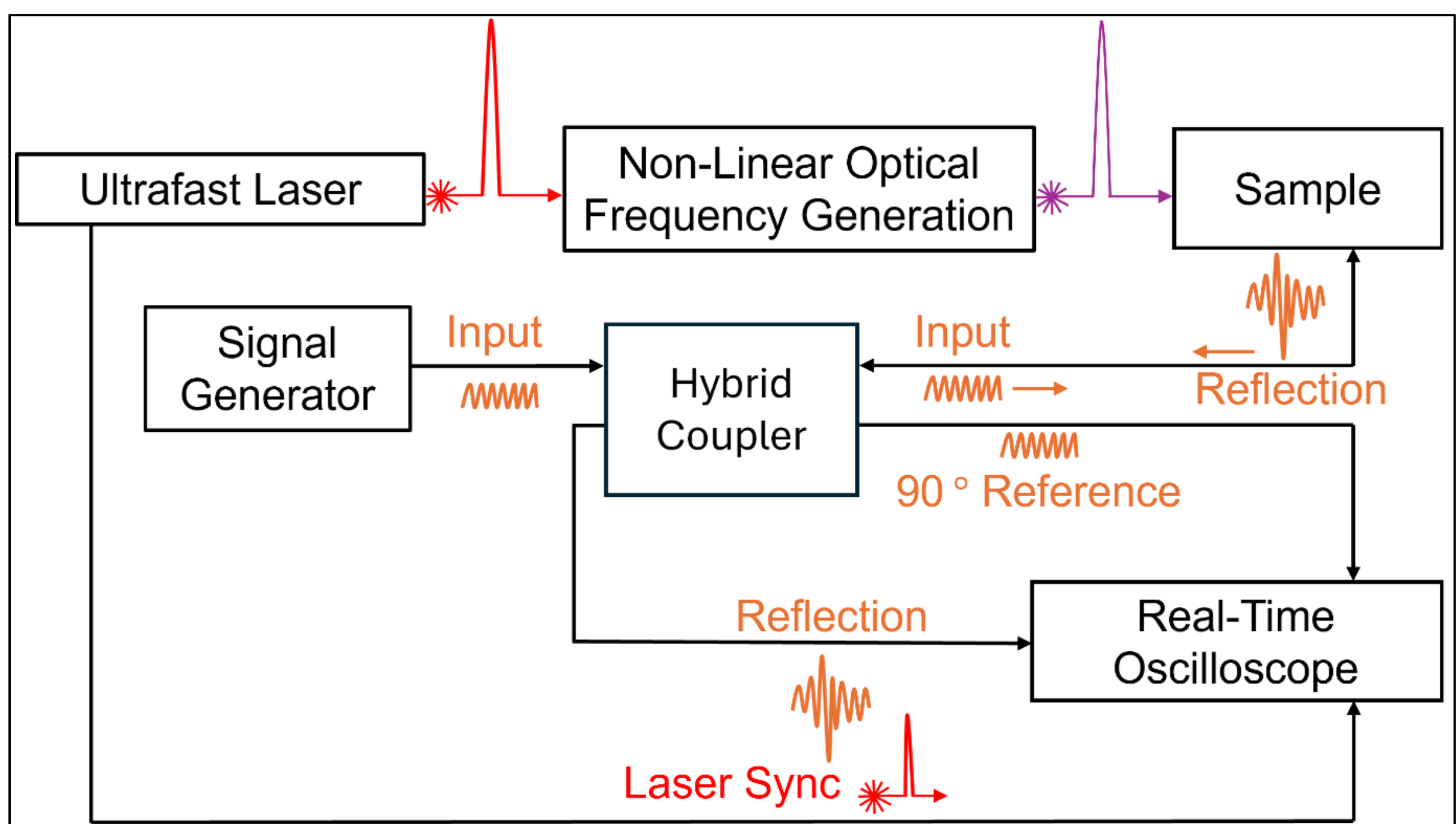

**Figure S15**. A diagram of the LUIS technique. A narrowband laser is generated by an ultrafast laser and, using non-linear optical techniques, a wide array of wavelengths can be generated. This light is then used to excite different dynamics within the LLTO, perturbing a GHz signal, which is being transmitted to the sample-pin interface via coaxial cables. The GHz signal is generated by a signal generator, routed through a hybrid coupler, reflected off the sample, which is perturbed by the laser, and then routed to the real-time oscilloscope, which is synced via a laser trigger to the time base of the ultrafast laser.

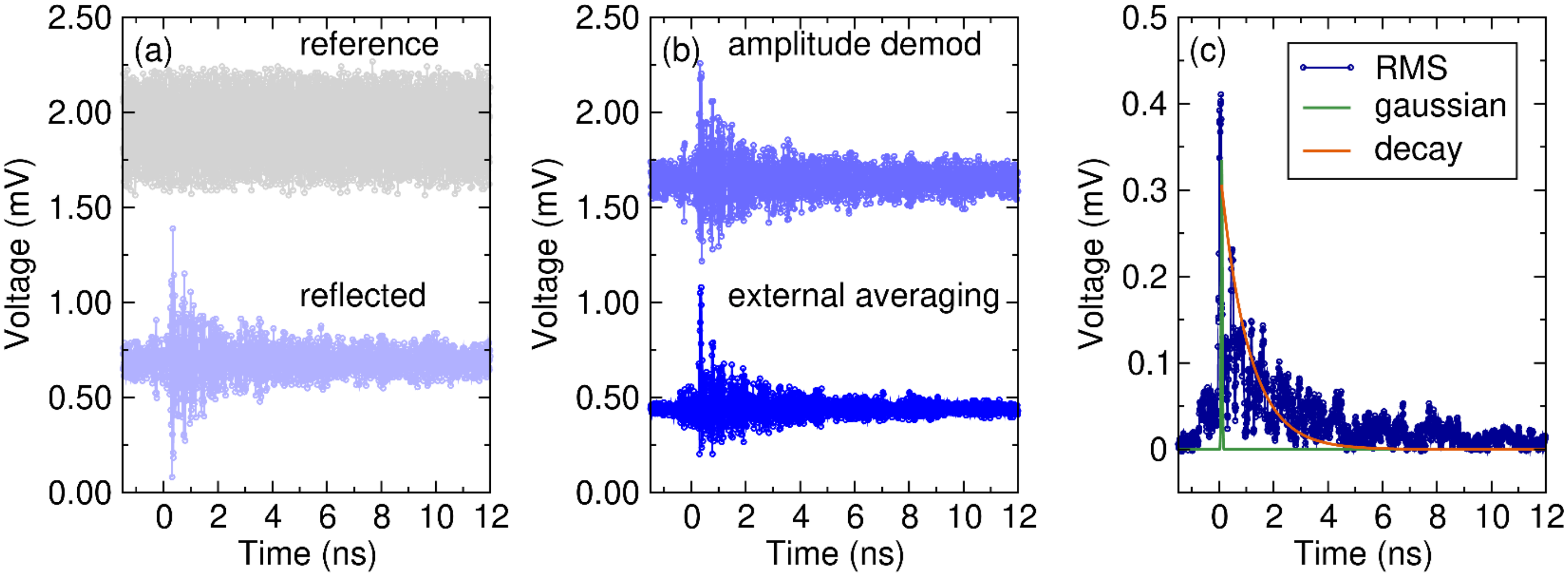

**Figure S16**. The processing steps for the extraction of the THz-LUIS signal beginning with (a) the reference signal collected from the 90° reference port in **Fig. S15** and the reflected signal routed to the oscilloscope from the perturbed sample, both averaged 1024 times. (b) The application of the oscilloscope's amplitude demodulation function and 1024 averages via the oscilloscope's math function. (c) The application of a root-mean-squared envelope function to the externally averaged signal in (b), as well as the gaussian rise and exponential decay components from the fitting model.

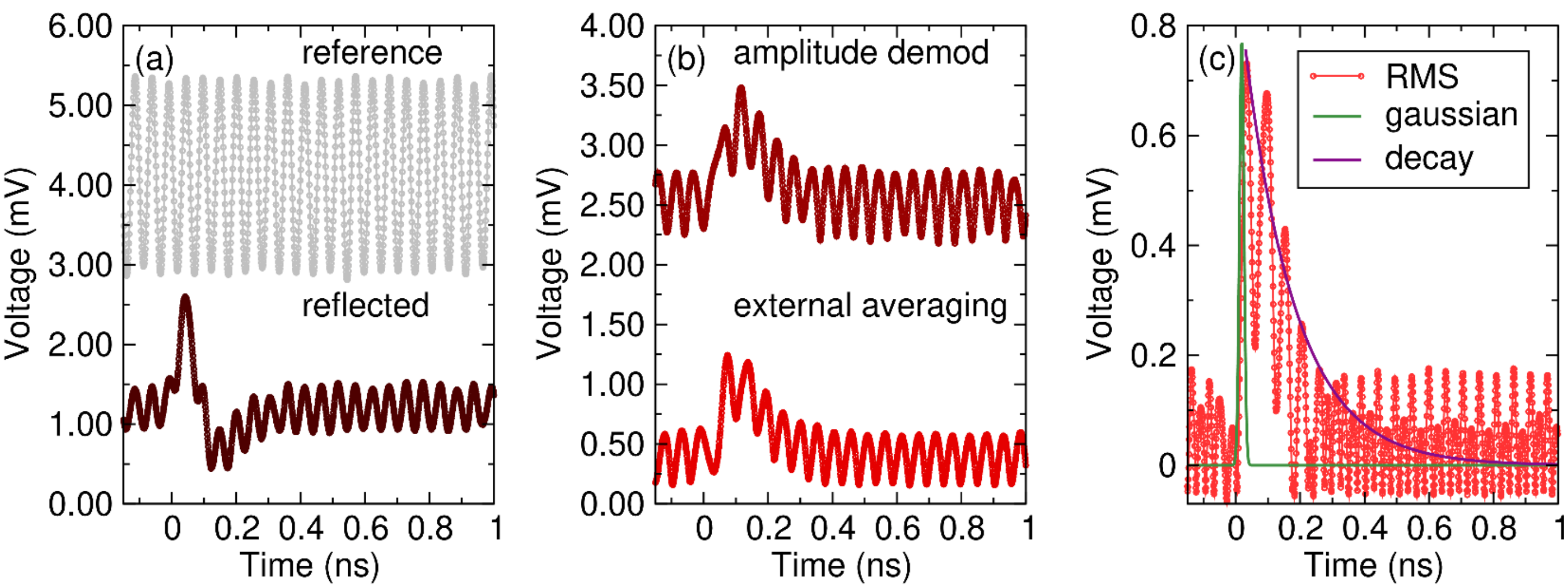

**Figure S17**. The processing steps for the extraction of the 800 nm-LUIS signal beginning with (a) the reference signal collected from the 90° reference port in **Fig. S15** and the reflected signal routed to the oscilloscope from the perturbed sample, both averaged 1024 times. (b) The application of the oscilloscope's amplitude demodulation function and 1024 averages via the oscilloscope's math function. (c) The application of a root-mean-squared envelope function to the externally averaged signal in (b), as well as the gaussian rise and exponential decay components from the fitting model.

**Table S5**. A summary of the rise time, or gaussian width ($\sigma$), and decay times ($\tau_d$), reported with a 95% confidence interval, from the LUIS measurements. The fitting model is described in Equations 30-33 of the **Methods**.

| Wavelength | $\sigma$ (ps) | $\sigma$ Confidence Interval (ps) | $\tau_d$ (ps) | $\tau_d$ Confidence Interval (ps) |
|---|---|---|---|---|
| 800 nm | 7.06 | 5.16-8.97 | 159 | 151-167 |
| 0.5-5.5 THz (54.5-599 μm) | 20.5 | 19.0-22.6 | 1030 | 1000-1060 |

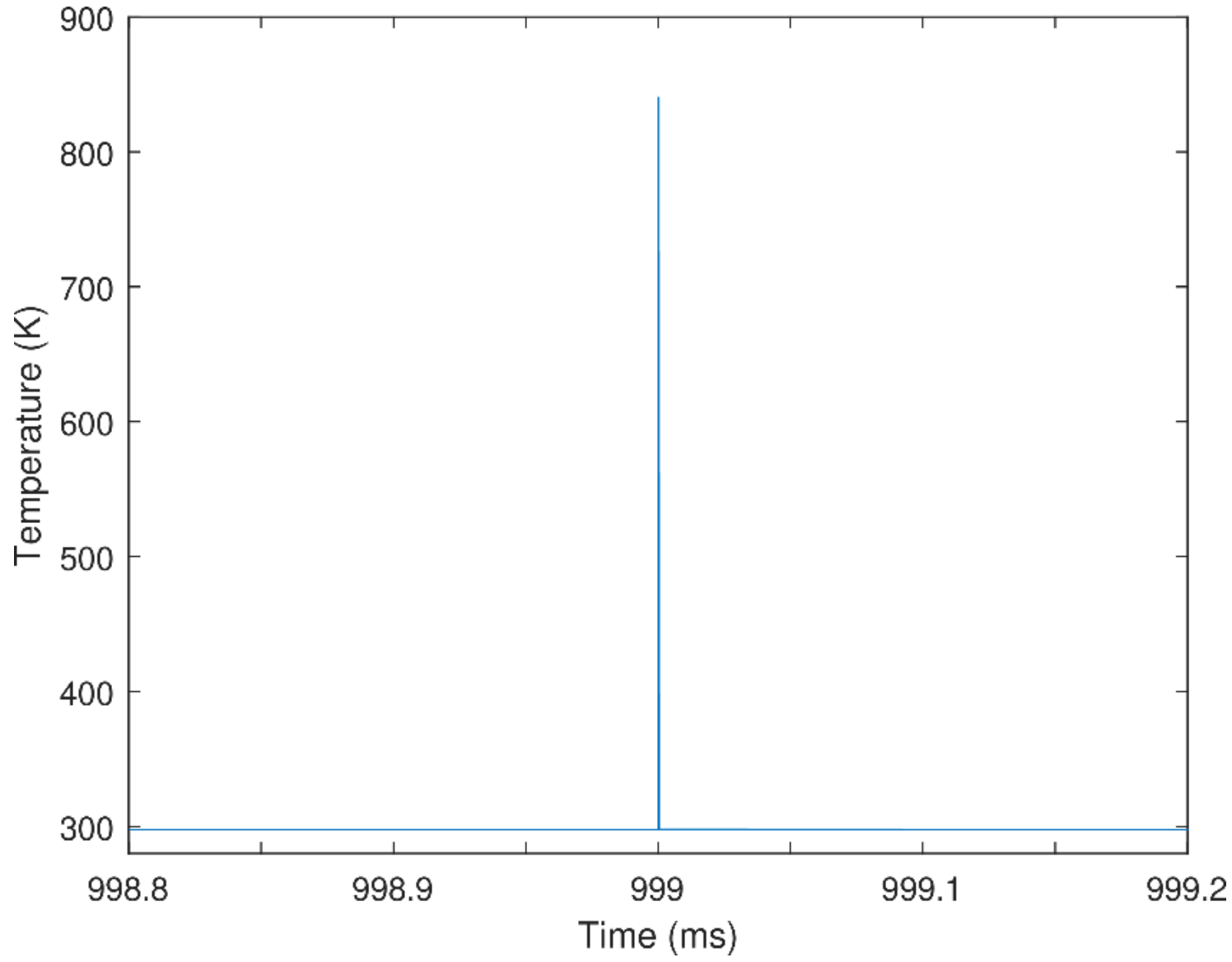

**Figure S18.** The peak temperature in the sample during and after the final pulse for the 800 nm, 20 mW light simulation. The nearly instantaneous rise and decay of heat occurs in less than 100 ns and is indicative of the extremely fast response induced by an ultrafast laser pulse.

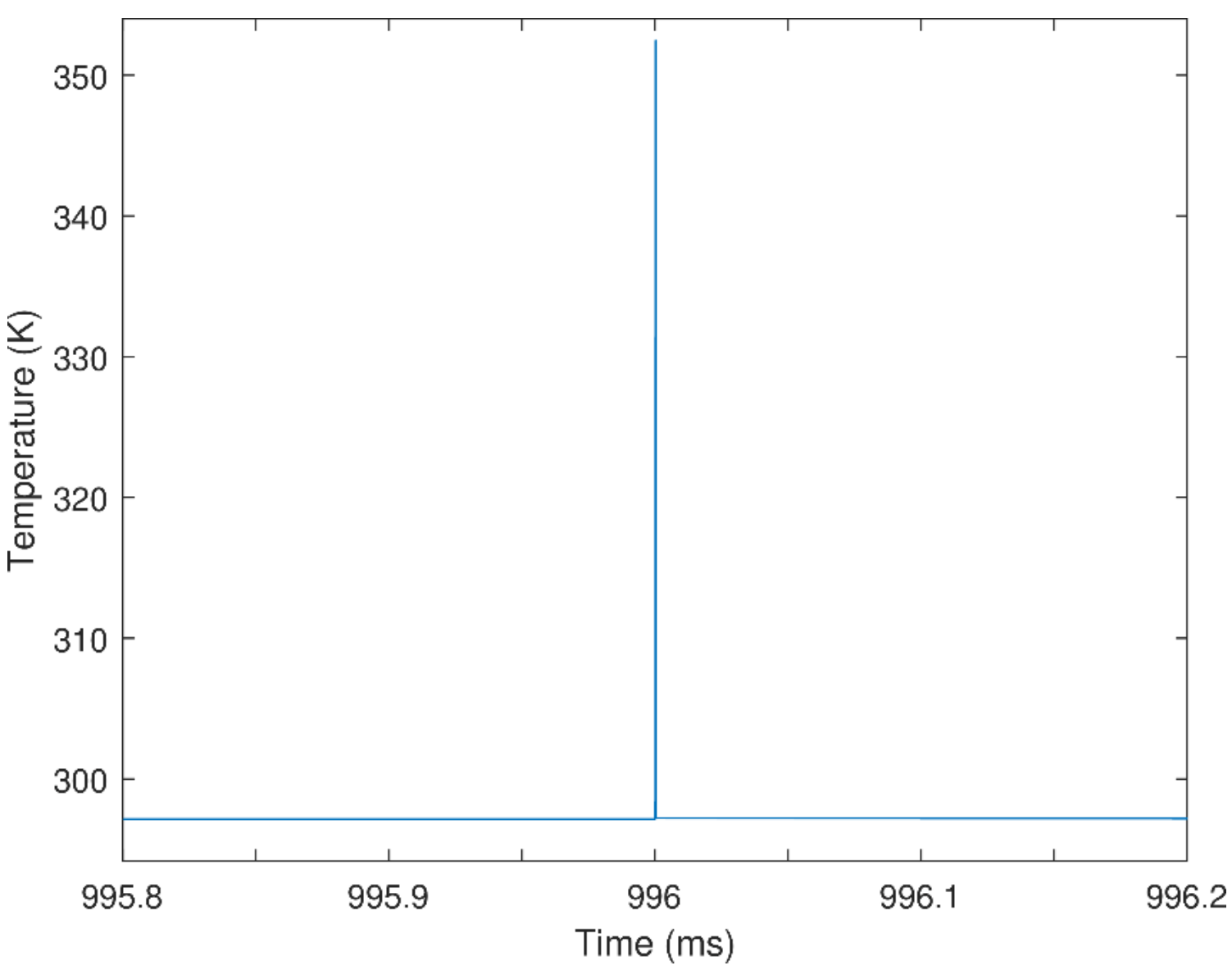

**Figure S19.** The peak temperature in the sample during and after the final pulse for the THz light simulation with 1 mW of laser power. The nearly instantaneous rise and decay of heat occurs in less than 100 ns and is indicative of the extremely fast response induced by an ultrafast laser pulse.

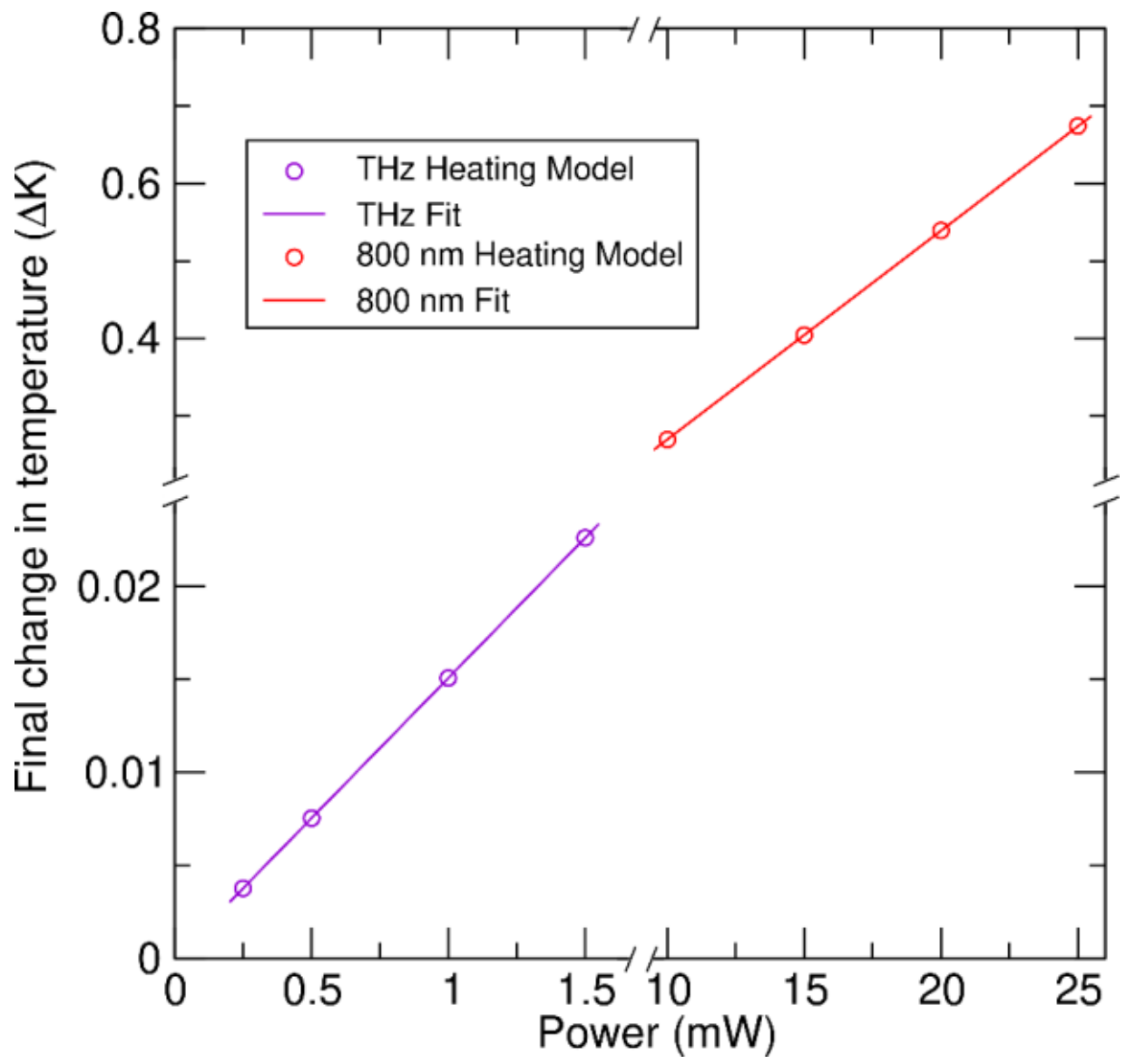

**Figure S20.** The final temperature of the sample due to optical heating as a function of average laser power for both the 800 nm and THz excitation in the range of excitation powers used experimentally. The final temperature is determined by taking the average maximum temperature at all time points during and after the final pulse. Our results suggest that the 800 nm excitation raises the sample temperature by 0.03 K per mW of power while the THz light raises the sample temperature by 0.02 K per mW of power.

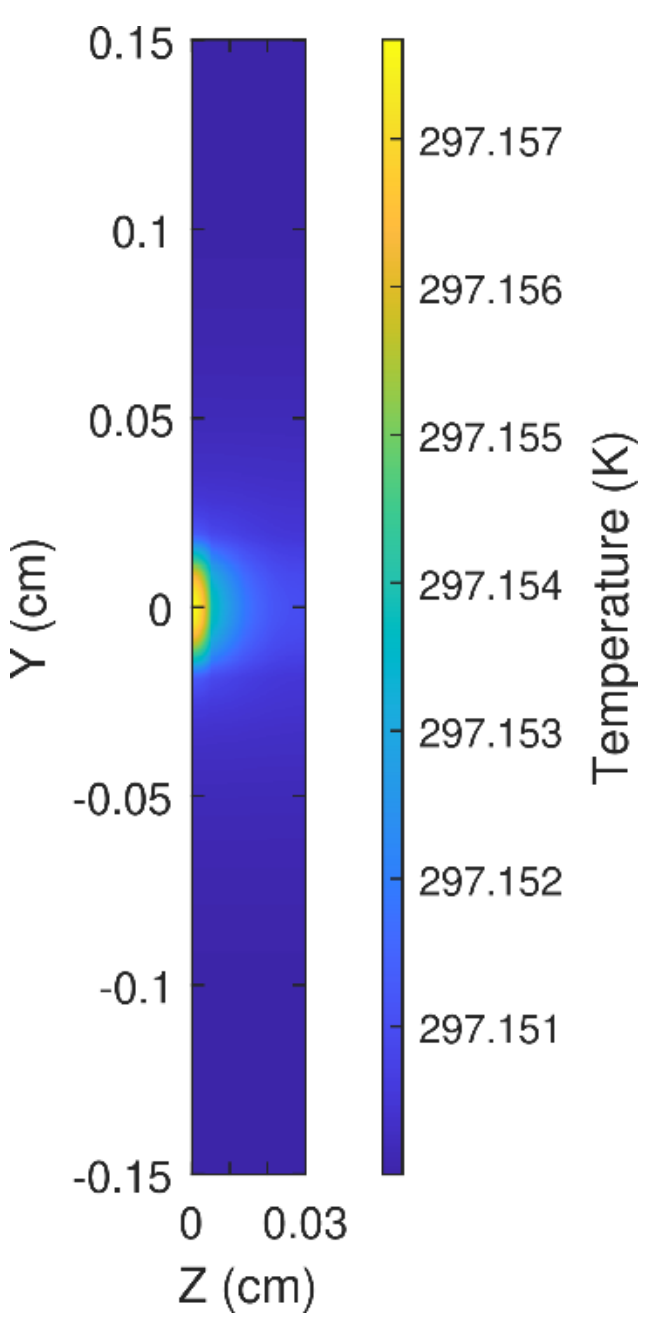

**Figure S21.** Cross-section of the sample after 1 s of ultrafast laser heating, modeled with a 1 mW THz light excitation with a penetration depth of 6 µm. The color bar corresponds to the temperature of the sample in Kelvin, with the initial temperature at 297.15 K. The sample and excitation are centered at x = 0 and y = 0.

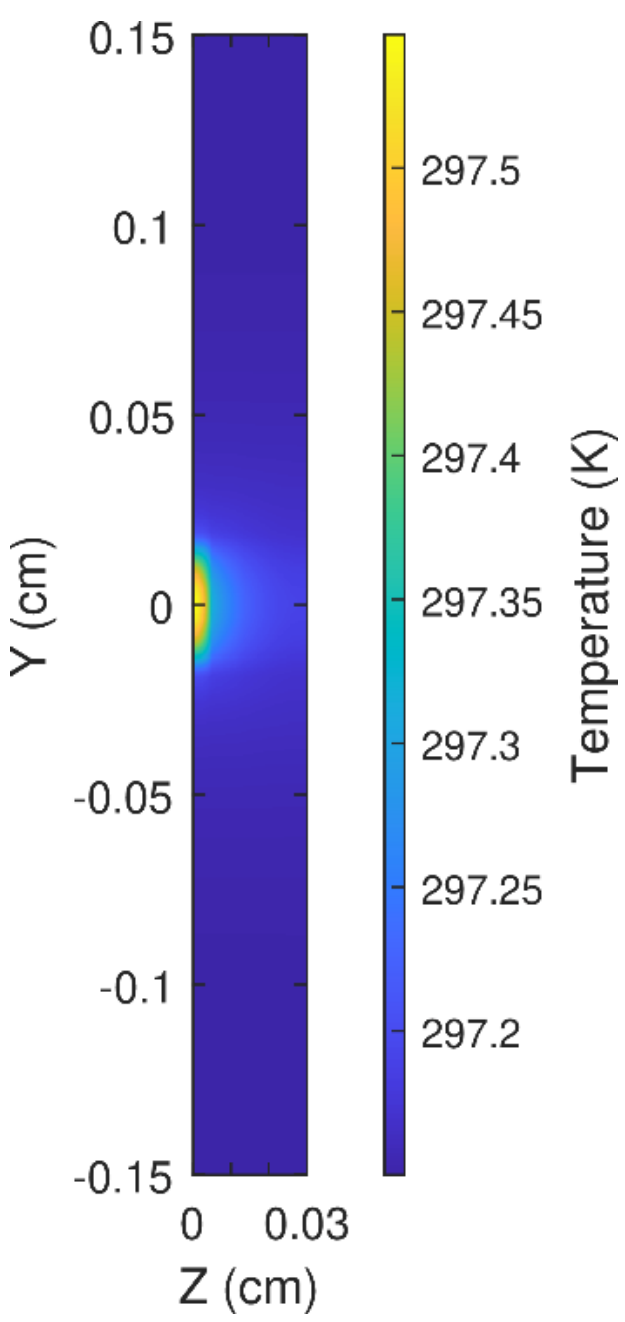

**Figure S22.** Cross-section of the sample after 1 s of ultrafast laser heating, modeled with a 20 mW 800 nm light excitation with a penetration depth of 7 µm. The color bar corresponds to the temperature of the sample in Kelvin, with the initial temperature at 297.15 K. The sample and excitation are centered at x = 0 and y = 0.

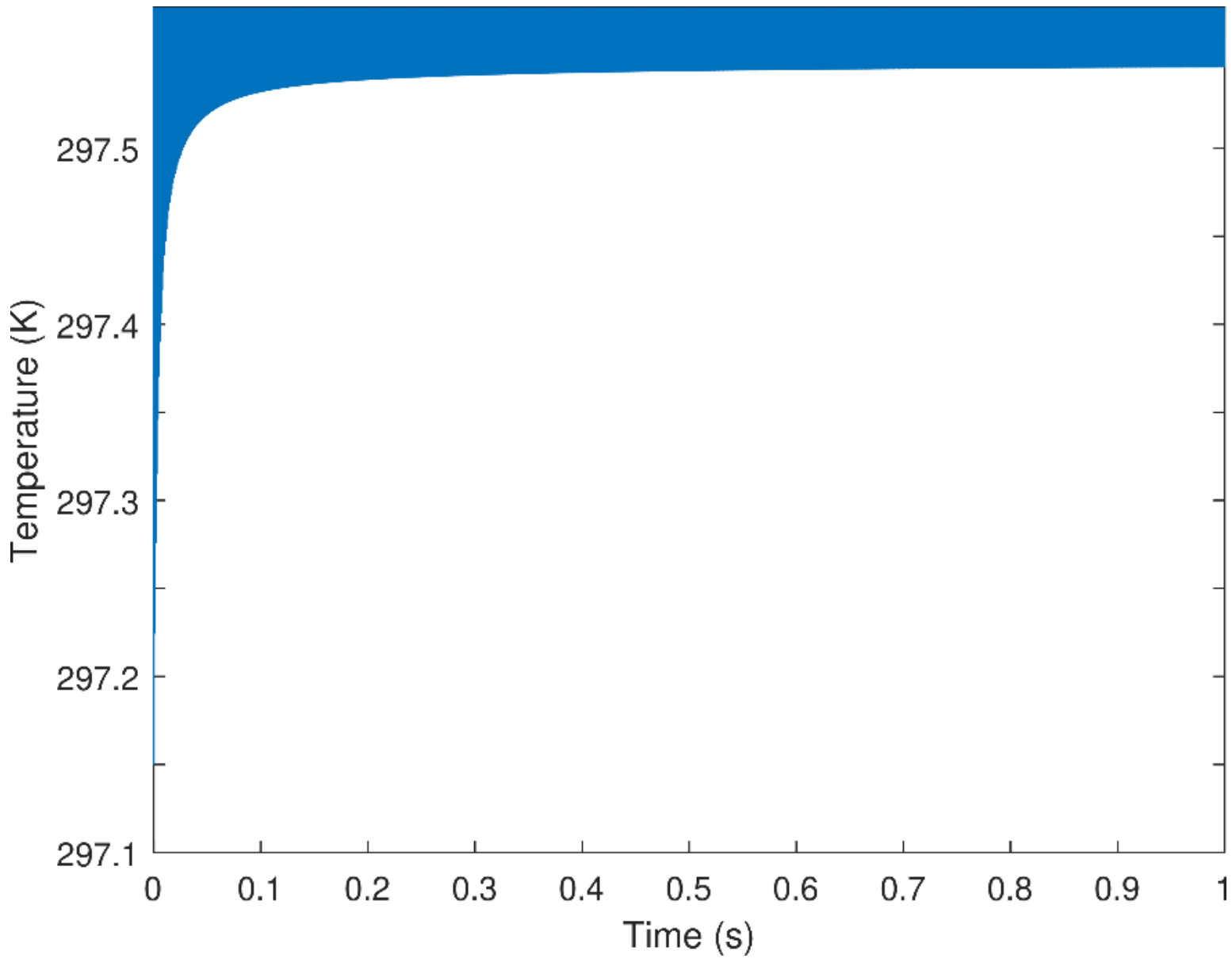

**Figure S23.** The baseline maximum temperature in the sample for 800 nm, 20 mW light. The simulation demonstrates that the sample's baseline temperature reaches steady state within one second. Each pulse still causes a nearly instantaneous rise and decay, but a steady baseline temperature of 297.55 K is reached in several tenths of a second.

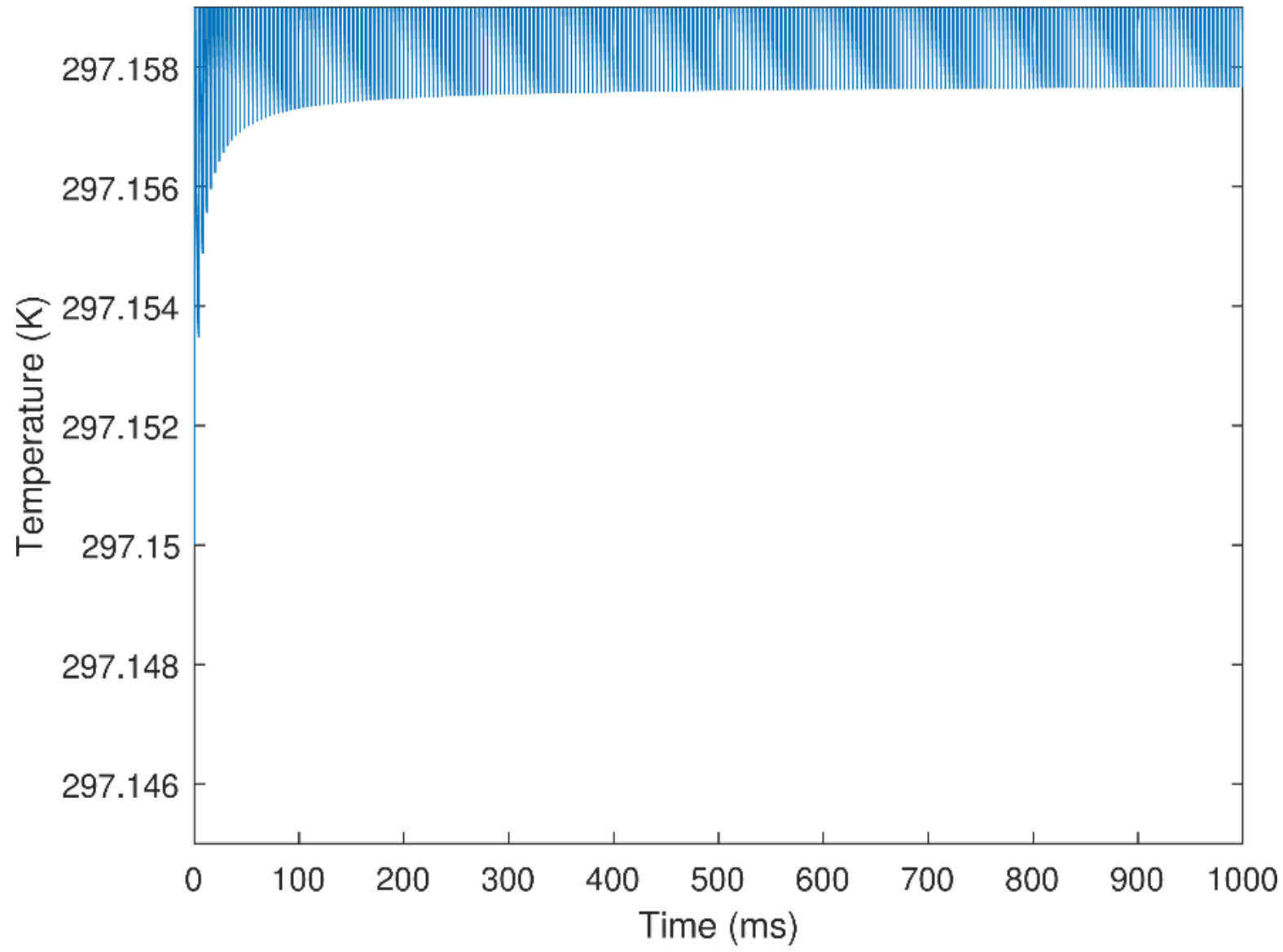

**Figure S24.** The baseline maximum temperature in the sample for THz light. The simulation demonstrates that the sample's baseline temperature reaches steady state within one second. Each pulse still causes a nearly instantaneous rise and decay, but a steady baseline temperature of ~297.158 K is reached in several tenths of a second.

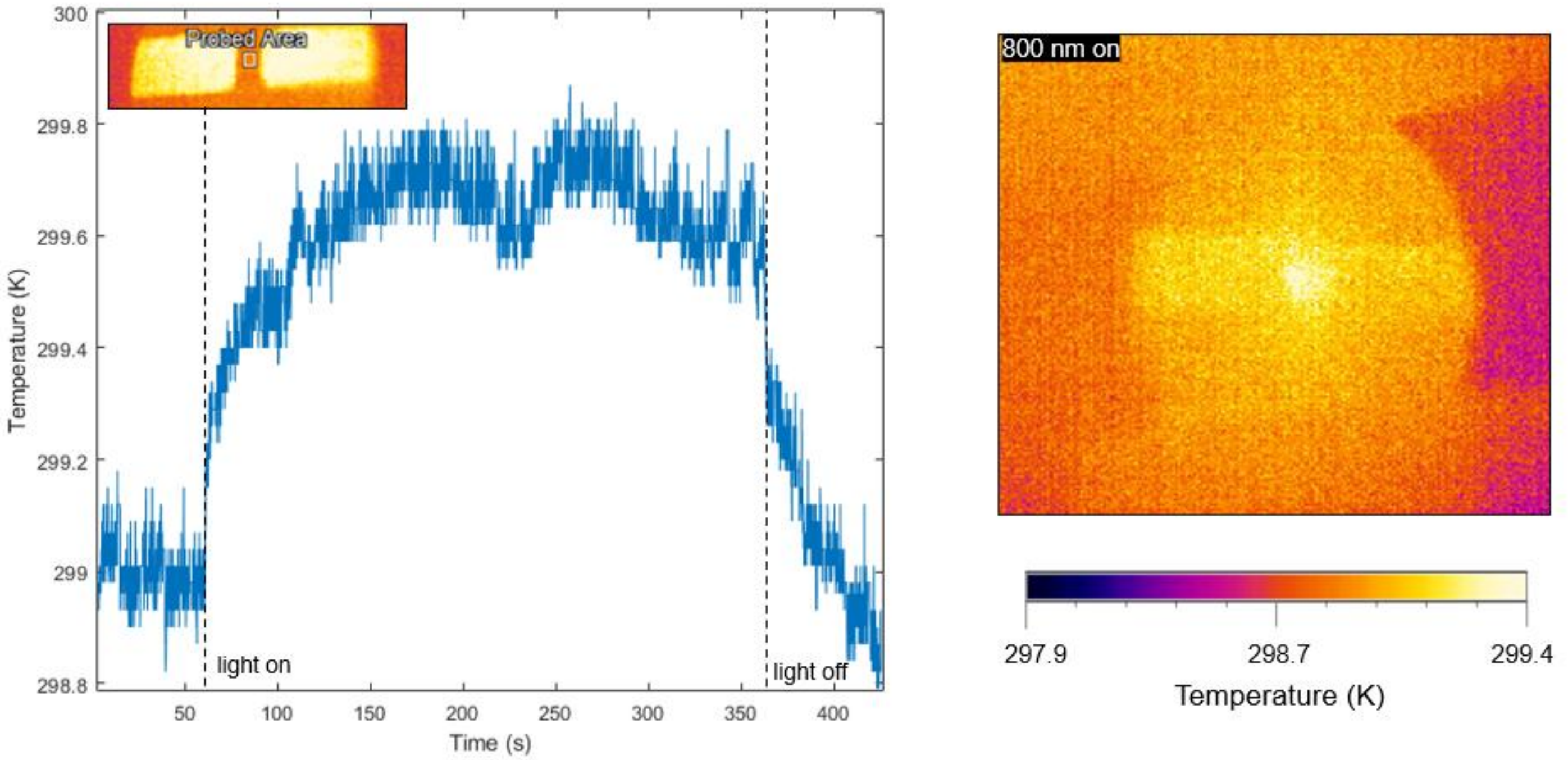

**Figure S25.** Experimental characterization of the sample's surface temperature under 800 nm, 20 mW light using an infrared thermal gun. Left: Surface temperature of the probed area upon illumination between 60 and 360 seconds. Right: Surface map of the sample with the 800 nm laser beam focused between the gold electrodes.

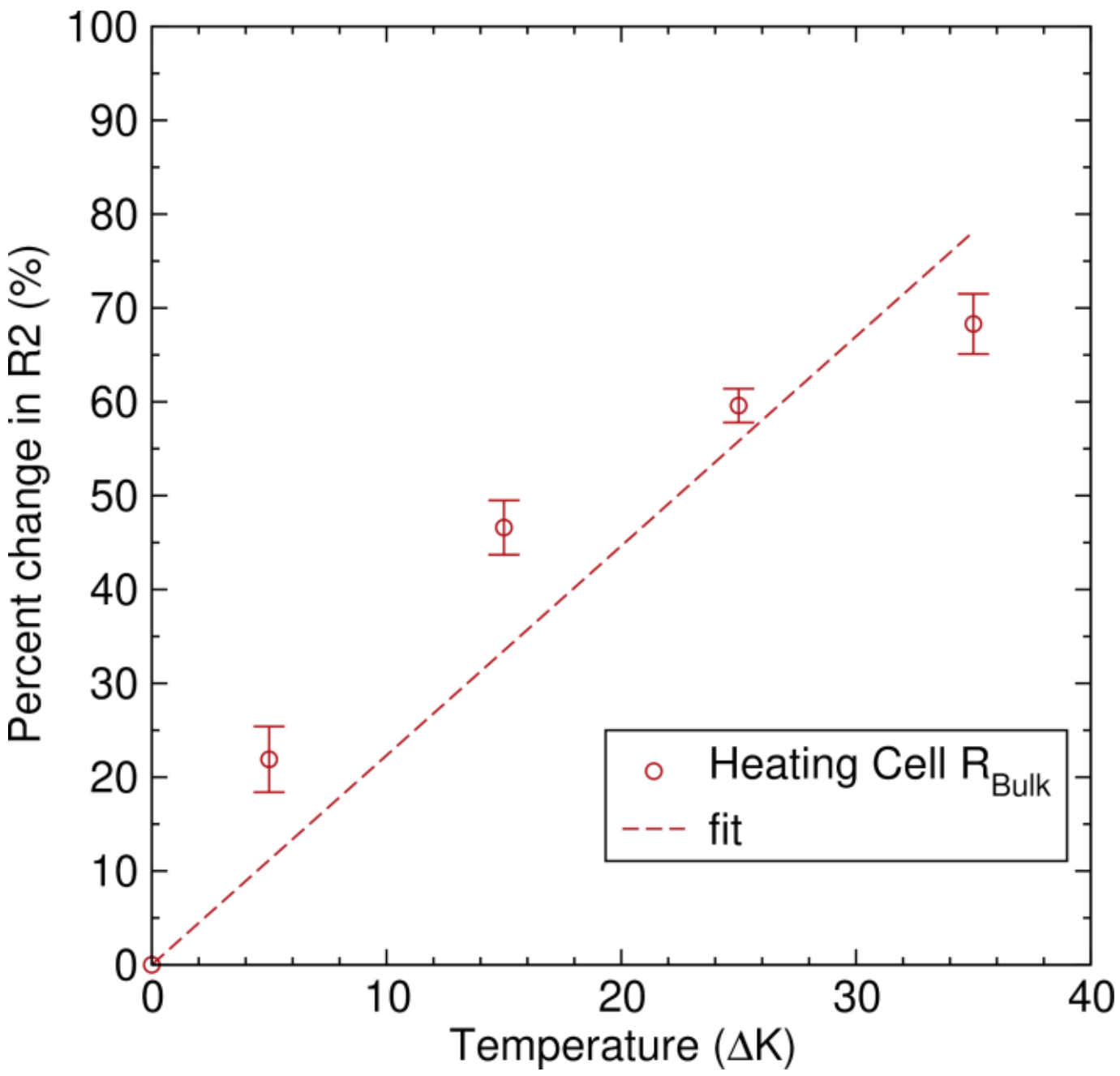

**Figure S26.** The percent change in impedance as a function of temperature. A 2.23% change in $R_{bulk}$ correlates to a 1 K change ($R^2 = 0.9634$).

**Table S6**. Changes in the characteristic final temperature as the thermal conductivity is changed for the 20 mW 800 nm excitation.

| Thermal conductivity ratio ($\kappa/\kappa_{nominal}$) | ΔT (K) |
|---|---|
| 0.5 | 0.533 |
| 0.75 | 0.537 |
| 1 | 0.539 |
| 1.25 | 0.539 |
| 1.5 | 0.540 |

**Table S7.** Changes in the characteristic final temperature as the thermal conductivity is changed for the 1 mW THz light excitation.

| Thermal conductivity ratio ($\kappa/\kappa_{nominal}$) | ΔT (K) |
|---|---|
| 0.5 | 0.0150 |
| 0.75 | 0.0151 |
| 1 | 0.0150 |
| 1.25 | 0.0150 |
| 1.5 | 0.0151 |

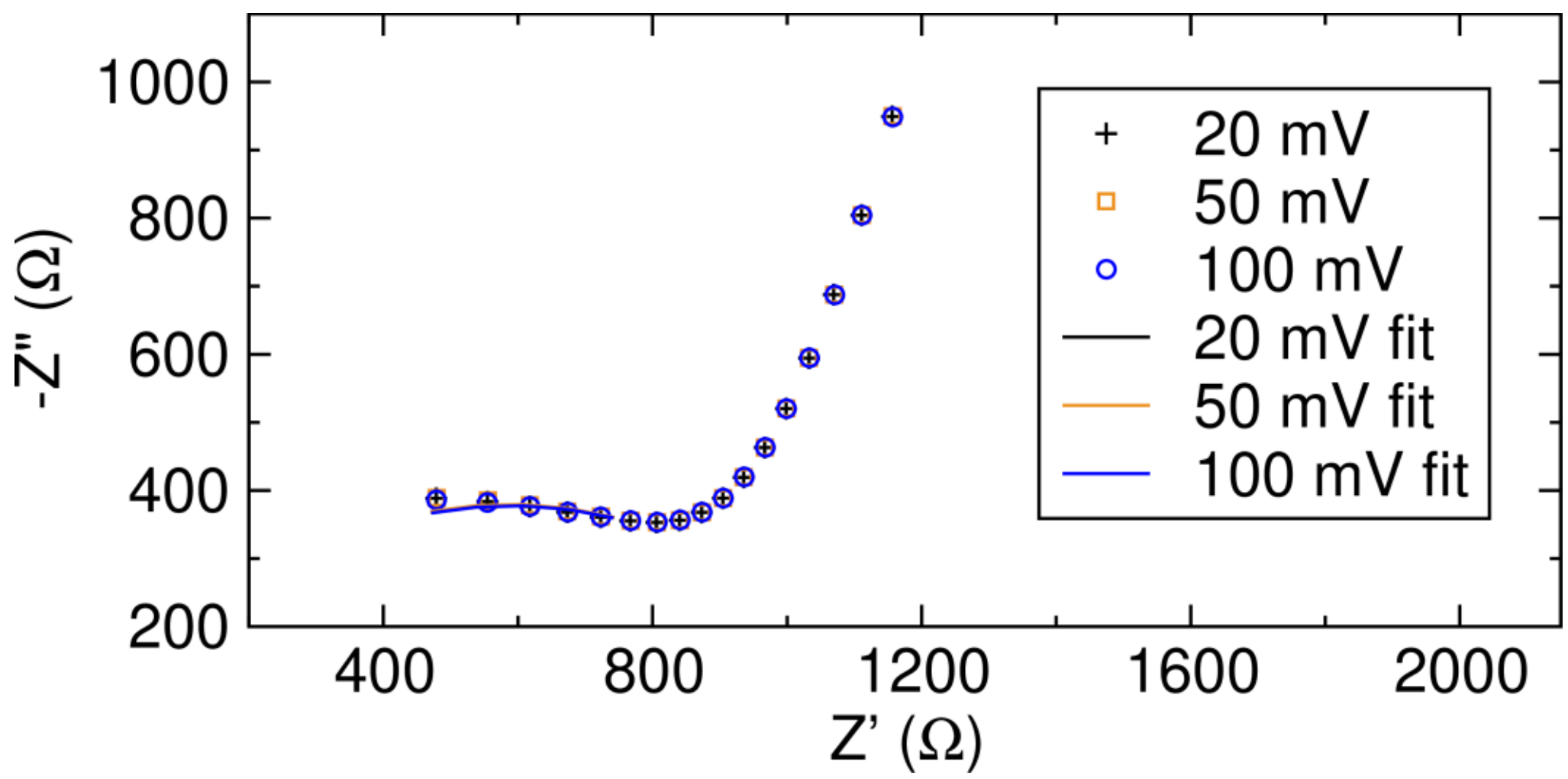

**Figure S27.** EIS performed at varied sinus amplitude between 20 and 100 mV, zoomed in to the bulk feature, demonstrating linearity due to the overlapping nature of the curves for each sinus amplitude.

**Table S8.** $R_{bulk}$ values taken across three samples and three trials for each sinus amplitude in **Fig. S28**.

| Sinus amplitude (mV) | $R_{bulk}$ (Ω) | Standard deviation in $R_{bulk}$ (Ω) |
|---|---|---|
| 20 | 1132 | 55 |
| 50 | 1134 | 56 |
| 100 | 1142 | 55 |

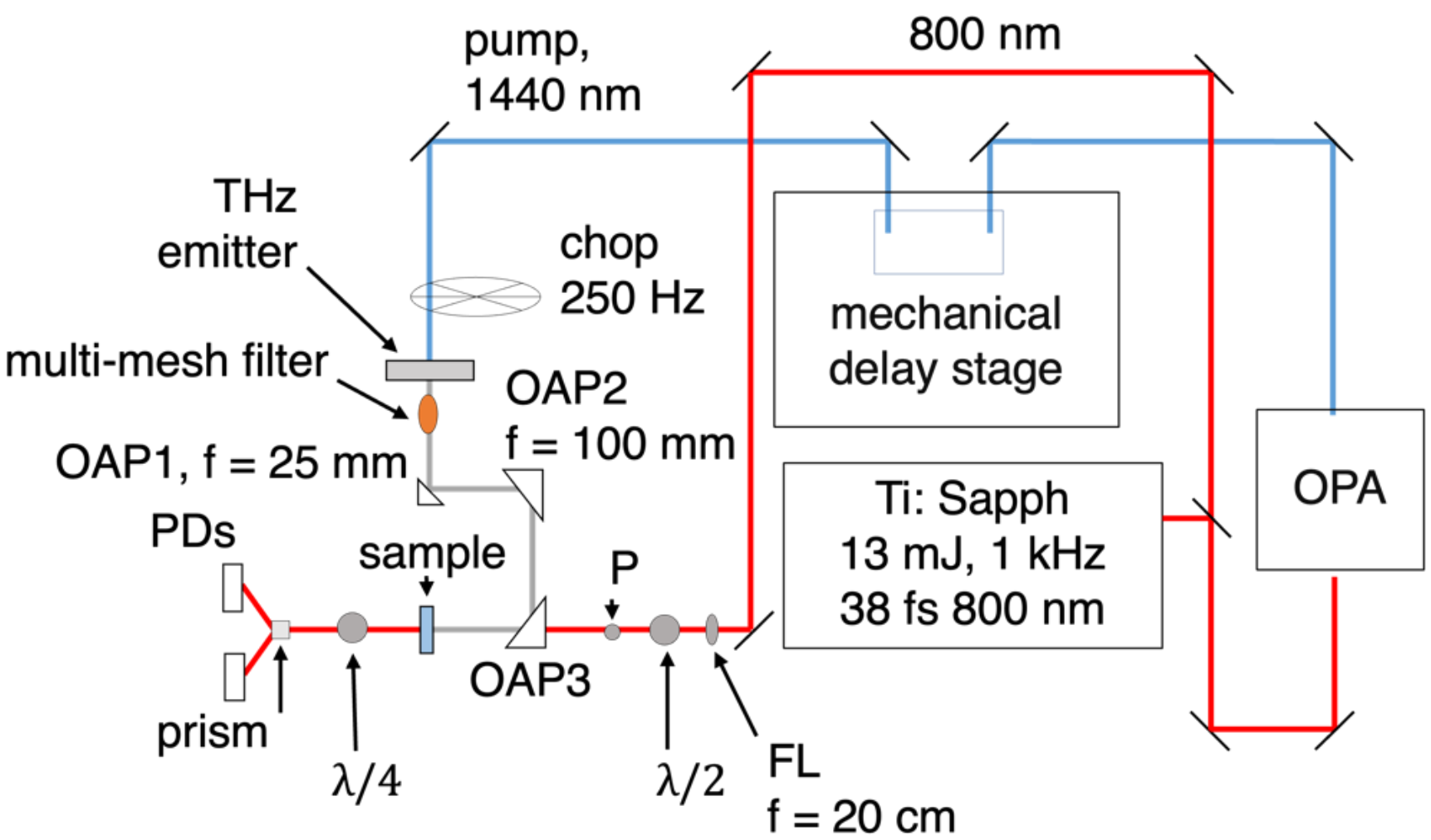

**Figure S28.** Diagram of the THz field set up. OPA: Optical Parameter Amplifier, OAP: Off-Axis Parabolic, FL: focal lens, P: polarizer, PDs: Photodiodes.

**Table S9.** IR power to THz field strength calibration yielding the average power calculation.

| IR Pump Power (mW) | Field Strength (keV/cm) | Peak to Peak Voltage (V) | Average THz Power (mW) |
|---|---|---|---|
| 1000 | 162.5 | 2.0 | 1.6 |
| 800 | 141.22 | 1.6 | 1.3 |
| 700 | 128.68 | 1.3 | 1.1 |
| 600 | 114.86 | 1.1 | 0.90 |
| 500 | 96.61 | 0.9 | 0.73 |
| 400 | 83.42 | 0.7 | 0.57 |
| 200 | 46.88 | 0.3 | 0.24 |

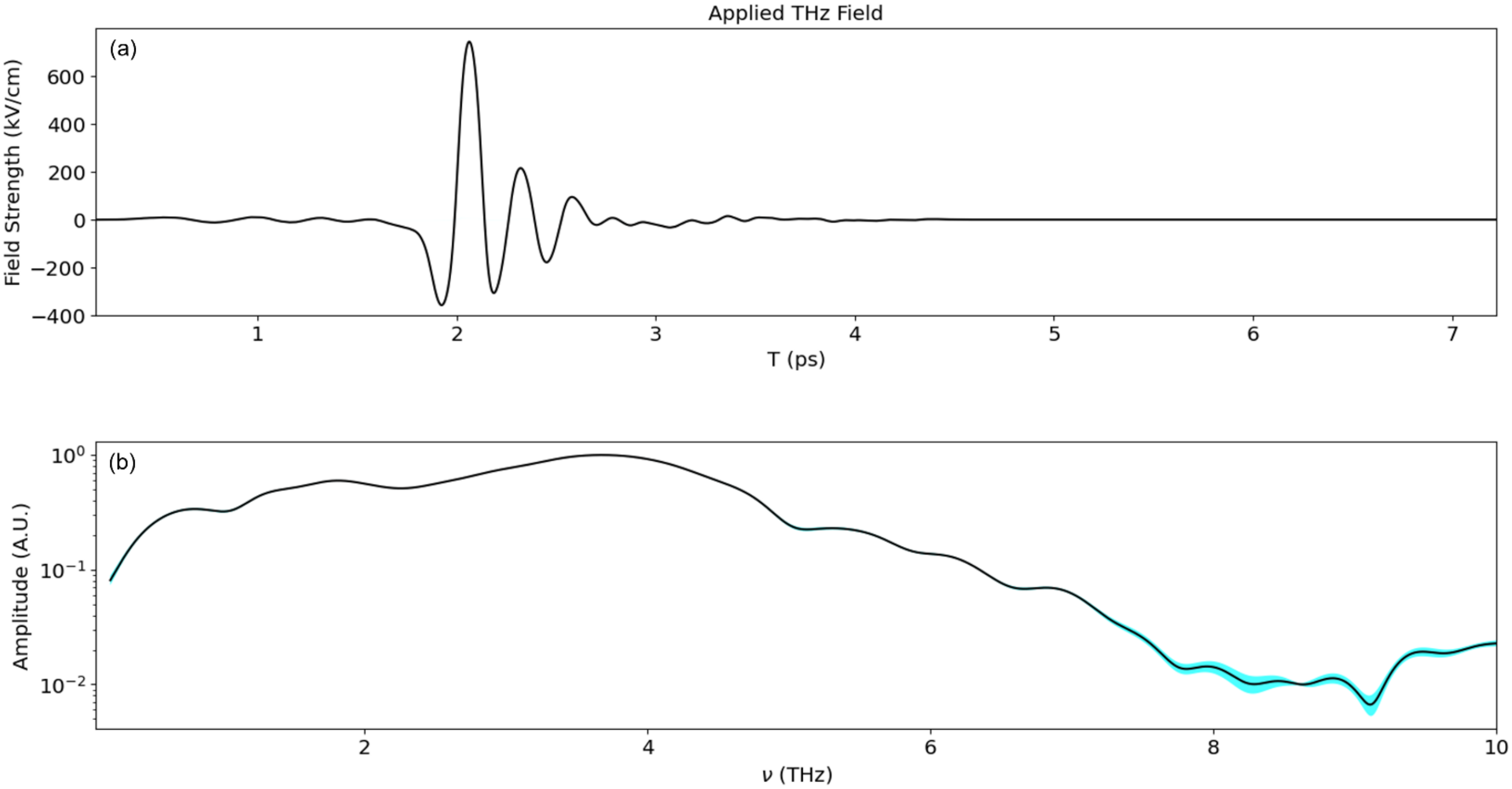

**Figure S29**. The (a) time domain trace of the applied THz field and (b) the frequency content of the same light, with uncertainties in the amplitude of the field shown, which was used to perturb the sample in the THz-LUIS experiment and was generated as described in the **Methods** section.